\documentclass[review]{elsarticle}
\usepackage{}
\usepackage[latin1]{inputenc} 
\usepackage{listings, tcolorbox}
\usepackage{lineno,hyperref}
\usepackage{amssymb}
\usepackage{color}
\usepackage{amsmath}
\usepackage{tikz}
\usepackage{amsfonts}
\usepackage{mathrsfs}
\usepackage{amsthm}

\usepackage{bbding}
\usepackage{mathrsfs}
\usepackage{amsmath, amsfonts, amssymb, mathrsfs, txfonts}
\usepackage{graphicx, subfigure}
\usepackage{wasysym}
\usepackage{color}
\usepackage{bm}
\usepackage{enumerate}

\usepackage{pifont}
\usepackage{txfonts}
\usepackage{amsmath}
\usepackage{graphicx}
\usepackage{amsfonts}
\usepackage{amssymb}
\usepackage{mathrsfs,psfrag,eepic,epsfig}
\usepackage{epstopdf}

\newtheorem{thm}{Theorem}[section]
 
 \newtheorem{lem}[thm]{Lemma}
 
 \theoremstyle{definition}
 
 \theoremstyle{remark}

\biboptions{numbers,sort&compress}

\modulolinenumbers[5]

\journal{Journal of \LaTeX\ Templates}

\makeatletter \@addtoreset{equation}{section}
\renewcommand{\theequation}{\arabic{section}.\arabic{equation}}

\begin{document}

\begin{frontmatter}
\title{Inverse scattering transform of the coupled modified Korteweg-de Vries equation with nonzero boundary conditions}
\tnotetext[mytitlenote]{Project supported by the Fundamental Research Fund for the Central Universities under the grant No. 2019ZDPY07.\\
\hspace*{3ex}$^{*}$Corresponding author.\\
\hspace*{3ex}\emph{E-mail addresses}:  sftian@cumt.edu.cn and
shoufu2006@126.com (S. F. Tian)}

\author{Xiao-Fan Zhang, Shou-Fu Tian$^{*}$ and Jin-Jie Yang}
\address{
School of Mathematics , China University of Mining and Technology,\\ Xuzhou 221116, People's Republic of China\\
}

\begin{abstract}
In this work, we extend the Riemann-Hilbert (RH) method in order to study the coupled modified Korteweg-de Vries equation (cmKdV) under nonzero boundary conditions (NZBCs), and successfully find its solutions with their various dynamic propagation behaviors. In the process of spectral analysis, it is necessary to introduce Riemann surface to avoid the discussion of multi-valued functions, and to obtain the analytical and asymptotic properties needed to establish the RH problem. The eigenfunction have a column that is not analytic in a given region, so we introduce the auxiliary eigenfunction and the adjoint matrix, which is necessary to derive the analytical eigenfunctions. The eigenfunctions have three kinds of symmetry, which leads to three kinds of symmetry of the scattering matrix, and the discrete spectrum is also divided into three categories by us. The asymptoticity of the modified eigenfunction is derived. Based on the analysis, the RH problem with four jump matrices in a given area is established, and the relationship between the cmKdV equation and the solution of the RH problem is revealed. The residue condition of reflection coefficient with simple pole is established. According to the classification of discrete spectrum, we discuss the soliton solutions corresponding to three kinds of discrete spectrum classification and their propagation behaviors in detail.
\end{abstract}


\end{frontmatter}


\section{Introduction}
It is well-known that the modified Korteweg-de Vries (mKdV) equation with a real constant $a$
\begin{align}\label{mkdv}
u_{t}+au^{2}u_{x}+u_{xxx}=0,
\end{align}
plays an important role in soliton equations,  which has  far-reaching significance in physical applications \cite{Athorne-1987, Miure-1987, Wadati-1973, Yajima-1975}. Very recently, the coupled systems applied in the field of mathematical physics have been widely studied. Many methods have been thoroughly applied to these models, and many useful conclusions have been given in the application field \cite{Dodd-1982, Levi-1983, Hirota-1981, Wu-1999, Tam-2000,Hu-2003}. Moreover, one of the most typical coupled systems is coupled modified Korteweg-de Vries equation (cmKdV) equation, which is
\begin{align}\label{QQ1}
\begin{split}
u_{t}=-u_{xxx}+6u^{2}u_{x}+3v^{2}u_{x}+3uvv_{x}=0,\\
v_{t}=-v_{xxx}+6v^{2}v_{x}+3u^{2}v_{x}+3uvu_{x}=0,
\end{split}
\end{align}
where $u$ and $v$ are two real-valued functions of $x$ and $t$. Obviously, if $v=0$, the cmKdV equation  \eqref{QQ1} can be reduced to  the mKdV  equation \eqref{mkdv} with $a=6$.  Four kind of soliton solutions and new explicit exact solutions for a new generalized Hirota-Satsuma coupled KdV equation were obtained in  \cite{Fan-2001, chen-2003}, respectively.  Algebro-geometric solutions of the cmKdV hierarchy were studied in \cite{geng-2014}.  Initial-boundary value problem of the cmKdV equation was studied  by the Fokas method in \cite{tian}. In addition, there are some excellent results about cmKdV equation \cite{wu-2017, Xue-2015, geng-2019-JGP, Ma-2018-jgp}.

In this work, we study the cmKdV equation \eqref{QQ1} with $3\times3$ spectrum problem, which has the following nonzero boundary conditions (NZBCs)
\begin{align}\label{NZBCs}
\lim_{x\rightarrow\pm\infty}\mathbf{q}(x,t)=\mathbf{q}_{\pm}=\mathbf{q}_{0}(x,t)e^{i\theta^{\pm}},
\end{align}
where $\mathbf{q}$ and $\mathbf{q}_{0}$ are the two-component vectors, $\|\cdot\|$ is the standard Euclidean
norm, $\theta^{\pm}$ are real numbers and $\|\mathbf{q}_{0}\|^{2}=q_{0}^{2}$. It is worth noting that although the NZBCs of cmKdV equation was also studied in \cite{tian, Tian-CPAA-2018}, there are some differences in the following aspects: (i) The initial boundary value problem in their work is based on half line and interval, while our work is on the real axis. (ii) Their research is based on Dirichlet boundary value, first Neumann boundary values and second Neumann boundary values, while our work is based on a nonzero background wave with parallel boundary values.  (iii) The results of the research are different. The solution given in \cite{tian, Tian-CPAA-2018} is a general solution with integral form, but we give an exact solution.  Our idea is to establish a relationship between the eigenfunction and the Riemann-Hilbert (RH) problem by constructing the RH problem, so that the solutions of the initial value problem of cmKdV equation can be expressed by the solution of the RH problem, and finally the exact solutions can be restored via the solution of the RH problem.

As a more effective method for solving integrable systems than inverse scattering transformation (IST) proposed by Gardner, Green, Kruskal and Miura for solving the initial value problems of nonlinear evolution equations \cite{GGKM-1967, GGKM-1974}, RH method  not only can be used to construct exact solutions \cite{ WDS-JMP-2010, Guo-JMAA-2018, Zhang-2019-pd,  Yangjj, Geng-WM-2016, ZZC-ZAMP-GI} and long-time asymptotic solutions of higher-order spectral problems \cite{huang-jde-2019, LN-JMP-2019, Lenells}, but also to study initial boundary value problems \cite{Biondini-2014-jmp, Pichler-2017-ijam, TSF-PAMS-2018, TSF-JDE-2017, wen-ss, YJK-SIAM-2010, YZY-JNS-2020, yjj-TMP-2020}, orthogonal polynomials and random matrices \cite{ LMP-2020,  CY-JMAA-2019,  FEG-SAM-2019, xu-cmp-2020}. The core idea of the RH problem is to establish the eigenfunction and the scattering matrix related to the initial value conditions, and to study the basic properties required for the RH problem in the direct scattering process. Then the reflection coefficient appearing in the RH problem is studied, and its residue condition at the simple pole is established to regularize the established RH problem. Finally, the plemelj formula is used to solve the regular RH problem to reconstruct the exact solution of nonlinear integrable equations without reflection potential.


On the other hand, there have been a lot of related literatures on the study of nonlinear integrable equations with $2\times2$ spectrum problems under NZBCs \cite{YZY-JNS-2020, JMP-2014-B}, but the conversion from $2\times2$ to $3\times3$ matrix Lax pairs involves many novelties \cite{SIAM-2015, DK-N-2015}, the main reasons are as follows: (i) one of the columns of the eigenfunction is not analytic in the given area; (ii) the analytic of the scattering matrix cannot be directly given according to the Wronskian determinant (the reason is that the second column of the eigenfunction is not analytic in the given region); (iii) establishment of the asymptotic of eigenfunctions; (iv) the complexity of the analysis area makes it difficult to establish jumping conditions; (v) according to the corresponding Wronskian determinant, the discrete spectrum has three classifications. All of the problems mentioned above, which  will appear in this work and will be solved one by one. Finally, we give the exact solution expression of cmKdV equation without reflection potential.

The structure of this work is as follows. In section $2$, we study the spectral problem of cmKdV equation with nonzero boundary conditions. In the process of spectral analysis, the coefficient matrix $X$ of spectral problem has two eigenvalues, which are multi-valued functions. In order to solve this problem, Riemann surface is introduced. Because of the complexity with the $3\times3$ spectral problem, we also use the cross product to construct an appropriate analytic function and obtain the analytic properties of the auxiliary eigenfunction. In section $3$, we study three symmetries of eigenfunction, scattering matrix and auxiliary eigenfunction. In section $4$,  the distribution of discrete spectrum and the asymptotic solution of eigenfunctions are studied in detail to establish the appropriate RH problem. In section $5$, the RH problem with four jump matrices is successfully established. In section $6$, the trace formula by using reflection coefficient and discrete spectral points can be obtained, and the reconstruction formula without reflection potential are successfully derived. In section $7$,  we transform the original RH problem into a regularized RH problem via removing the asymptotics and poles, and  the soliton solutions of cmKdV equation by using Cramer's rule are successfully obtained. In section $8$, we mainly discuss the dynamic behaviors of single soliton solutions under NZBCs, including $N_{1}=1$ and $N_{2}=1$.
\section{Direct scattering}
\subsection{Some notations}
In this subsection, we fix some notations used  this work. $\sigma_{3}$ is Pauli matrix, and the expression of $Q$ is
\begin{align}
\sigma_{3}=\left(
             \begin{array}{cc}
               \textbf{I} & \textbf{0} \\
               \textbf{0}^{\dag} & -1 \\
             \end{array}
           \right)
,\quad~
Q=\left(
    \begin{array}{cc}
      \textbf{0} & \textbf{q} \\
      -\textbf{q}^{\dag} & 0 \\
    \end{array}
  \right).
\end{align}
\begin{itemize}
  \item The left side of the equation of \eqref{Lax} are called scattering problem, $k$ ia the scattering parameter, $\textbf{q}(x,t)=(u~~v^{*})^{T}$ denotes the scattering potential,
where $u$, $v$ are two potentials, superscript $T$ represents the transposition of a matrix and $v^{*}$ is the complex conjugate of $v$.
  \item The matrix $\textbf{I}$ represents the identity matrix, $\textbf{0}$ represents the zero vector or a zero matrix.
  \item The superscript $\dag$ represents Hermitian matrix, the function $\Phi(x,t)=(\Phi_{ij}(x,t))_{3\times3}$ $(i,j=1,2,3)$.
\end{itemize}
\subsection{Lax pair, Riemann surface and uniformization}
In this section, we shall study coupled modified Korteweg-de Vries (cmKdV) equation  with $3\times3$ spectral problem. In order to construct the Jost function solution of the cmKdV equation under NZBCs, we introduce the  new parameter to transform the spectral parameter from the $k$ plane to the $z$ plane, the purpose is to avoid the appearance of multi-valued function.

The $3\times3$ Lax pair associated with the cmKdV equation  is
\addtocounter{equation}{1}\label{Lax}
\begin{align}
\Phi_{x}(x,t,k)=U\Phi(x,t,k),\tag{\theequation a}\\
\Phi_{t}(x,t,k)=V\Phi(x,t,k),\tag{\theequation b}
\end{align}
where
\addtocounter{equation}{1}
\begin{align*}
U&=ik\sigma_{3}+Q,\tag{\theequation a}\label{Q4}\\
V=4ik^{3}\sigma_{3}+4k^{2}Q&-2ik\sigma_{3}(Q_{x}-Q^{2})+2Q^{3}-Q_{xx}+[Q_{x},Q],\tag{\theequation b}\label{Q5}
\end{align*}
and  $k$ is a constant spectral parameter. It is noted that the cmKdV equation can be derived from the zero curvature equation
\begin{align}
U_{t}-V_{x}+[U,V]=0,
\end{align}
which is the compatibility condition of the two equations in \eqref{Lax}.

We can expect that when the  boundary condition \eqref{NZBCs} is satisfied, the solution of the scattering problems can be approximately expressed by the solution of the asymptotic scattering problems
\addtocounter{equation}{1}
\begin{align}
\Phi_{x}(x,t,k)=U_{\pm}\Phi(x,t,k),\tag{\theequation a}\label{Q6}\\
\Phi_{t}(x,t,k)=V_{\pm}\Phi(x,t,k),\tag{\theequation b}\label{Q7}
\end{align}
where
\addtocounter{equation}{1}
\begin{align}
\lim_{x\rightarrow\pm\infty}U&=ik\sigma_{3}+Q_{\pm}\triangleq U_{\pm},\tag{\theequation a}\label{Q8}\\
\lim_{x\rightarrow\pm\infty}V=4ik^{3}\sigma_{3}&+4k^{2}Q_{\pm}+2ik\sigma_{3}Q^{2}+2Q_{\pm}^{3}
\triangleq V_{\pm}.
\tag{\theequation b}\label{Q9}
\end{align}
The eigenvalues of $U_{\pm}$ are $ik$ and $\pm i\lambda$, similarly,  the eigenvalues of $V_{\pm}$ are
$4ik^{2}$ and $\pm 2i\lambda(2k^{2}-q_{0}^{2})$, where
\begin{align}\label{1}
\lambda^{2}=k^{2}+q_{0}^{2},
\end{align}
$\lambda(k)$ is a double-valued function.
In order to solve this problem, we introduce a Riemann surface defined by \eqref{1}, in which two complex $k$-planes cut by branch secant $[-iq_{0}, iq_{0}]$, that is  $S_{1}$ and $S_{2}$ are glued together, where the fulcrum is $k=\pm iq_{0}$. On such the Riemann surface, $\lambda$ is a single valued function of $k$, which is composed of two single valued analytic branches.

\centerline{\begin{tikzpicture}[scale=0.5]
\filldraw (-9,1) -- (-9,9) to (-1,9) -- (-1,1);
\filldraw (9,1) -- (9,9) to (1,9) -- (1,1);
\filldraw (9,-1) -- (9,-9) to (1,-9) -- (1,-1);
\filldraw (-9,-1) -- (-9,-9) to (-1,-9) -- (-1,-1);
\path [fill=gray] (1,-5) -- (9,-5) to (9,-9) -- (1,-9);
\filldraw[white, line width=0.5](-1,-5)--(3,-5) arc (-180:0:2);
\path [fill=white] (1,-1) -- (9,-1) to (9,-5) -- (1,-5);
\filldraw[gray, line width=0.5](3,-5)--(7,-5) arc (0:180:2);
\path [fill=white] (-9,5)--(-9,9) to (-1,9) -- (-1,5);
\path [fill=gray] (-9,1)--(-9,5) to (-1,5) -- (-1,1);
\path [fill=white] (1,1)--(1,5) to (9,5) -- (9,1);
\path [fill=gray] (1,5)--(1,9) to (9,9) -- (9,5);
\path [fill=white] (-1,-1)--(-1,-5) to (-9,-5) -- (-9,-1);
\path [fill=gray] (-1,-5)--(-1,-9) to (-9,-9) -- (-9,-5);
\filldraw[red, line width=0.5] (2,2) to (-2,-2)[->];
\filldraw[red, line width=0.5] (-2,-8) to (2,-8)[->];
\draw[fill] (-5,5)node[below]{} circle [radius=0.035];
\draw[fill] (5,5)node[below]{} circle [radius=0.035];
\draw[fill] (-5,-5)node[below]{} circle [radius=0.035];
\draw[fill] (5,-5)node[below]{} circle [radius=0.035];
\draw[-][thick](-9,5)--(-8,5);
\draw[-][thick](-8,5)--(-7,5);
\draw[-][thick](-7,5)--(-6,5);
\draw[-][thick](-6,5)--(-5,5);
\draw[-][thick](-5,5)--(-4,5);
\draw[-][thick](-4,5)--(-3,5);
\draw[-][thick](-3,5)--(-2,5);
\draw[-][thick](-2,5)--(-1,5)[->][thick]node[above]{$Rek$};;
\draw[-][thick](-5,1)--(-5,2);
\draw[-][thick](-5,2)--(-5,3);
\draw[-][thick](-5,3)--(-5,4);
\draw[-][thick](-5,4)--(-5,5);
\draw[-][thick](-5,5)--(-5,6);
\draw[-][thick](-5,6)--(-5,7);
\draw[-][thick](-5,7)--(-5,8);
\draw[-][thick](-5,8)--(-5,9)[->] [thick]node[above]{$Imk$};
\draw[-][thick](1,5)--(2,5);
\draw[-][thick](2,5)--(3,5);
\draw[-][thick](3,5)--(4,5);
\draw[-][thick](4,5)--(5,5);
\draw[-][thick](5,5)--(6,5);
\draw[-][thick](6,5)--(7,5);
\draw[-][thick](7,5)--(8,5);
\draw[-][thick](8,5)--(9,5)[->][thick]node[above]{$Rek$};
\draw[-][thick](5,1)--(5,2);
\draw[-][thick](5,2)--(5,3);
\draw[-][thick](5,3)--(5,4);
\draw[-][thick](5,4)--(5,5);
\draw[-][thick](5,5)--(5,6);
\draw[-][thick](5,6)--(5,7);
\draw[-][thick](5,7)--(5,8);
\draw[-][thick](5,8)--(5,9);
\draw[-][thick](-9,-5)--(-8,-5);
\draw[-][thick](-8,-5)--(-7,-5);
\draw[-][thick](-7,-5)--(-6,-5);
\draw[-][thick](-6,-5)--(-5,-5);
\draw[-][thick](-5,-5)--(-4,-5);
\draw[-][thick](-4,-5)--(-3,-5);
\draw[-][thick](-3,-5)--(-2,-5);
\draw[-][thick](-2,-5)--(-1,-5)[->][thick]node[above]{$Re\lambda$};
\draw[-][thick](-5,-1)--(-5,-2);
\draw[-][thick](-5,-2)--(-5,-3);
\draw[-][thick](-5,-3)--(-5,-4);
\draw[-][thick](-5,-4)--(-5,-5);
\draw[-][thick](-5,-5)--(-5,-6);
\draw[-][thick](-5,-6)--(-5,-7);
\draw[-][thick](-5,-7)--(-5,-8);
\draw[-][thick](-5,-8)--(-5,-9);
\draw[-][thick](1,-5)--(2,-5);
\draw[-][thick](2,-5)--(3,-5);
\draw[-][thick](3,-5)--(4,-5);
\draw[-][thick](4,-5)--(5,-5);
\draw[-][thick](5,-5)--(6,-5);
\draw[-][thick](6,-5)--(7,-5);
\draw[-][thick](7,-5)--(8,-5);
\draw[-][thick](8,-5)--(9,-5)[->][thick]node[above]{$Rez$};
\draw[-][thick](5,-1)--(5,-2);
\draw[-][thick](5,-2)--(5,-3);
\draw[-][thick](5,-3)--(5,-4);
\draw[-][thick](5,-4)--(5,-5);
\draw[-][thick](5,-5)--(5,-6);
\draw[-][thick](5,-6)--(5,-7);
\draw[-][thick](5,-7)--(5,-8);
\draw[-][thick](5,-8)--(5,-9);
\draw[->](5,9)[thick]node[above]{$Imk$};
\draw[->](-5,-1)[thick]node[above]{$Im\lambda$};
\draw[->](5,-1)[thick]node[above]{$Imz$};
\draw[fill] (-5,7) circle [radius=0.055]node[left]{\footnotesize$iq_{0}$};
\draw[fill] (-5,3) circle [radius=0.055]node[left]{\footnotesize$-iq_{0}$};
\draw[fill] (5,7) circle [radius=0.055]node[left]{\footnotesize$iq_{0}$};
\draw[fill] (5,3) circle [radius=0.055]node[left]{\footnotesize$-iq_{0}$};
\draw[fill] (-7,-5) circle [radius=0.055]node[below]{\footnotesize$-q_{0}$};
\draw[fill] (-3,-5) circle [radius=0.055]node[below]{\footnotesize$q_{0}$};
\draw(5,-5) [red, line width=1] circle(2);
\filldraw[red, line width=1.5] (-5,7) to (-5,3);
\filldraw[red, line width=1.5] (5,7) to (5,3);
\filldraw[red, line width=1.5] (-7,-5) to (-3,-5);
\draw[fill][black] (-8,7) [thick]node[right]{\footnotesize$S_{1}$};
\draw[fill][black] (2,7) [thick]node[right]{\footnotesize$S_{2}$};
\draw[fill][black] (-4,7) [thick]node[right]{\footnotesize$Imk>0$};
\draw[fill][black] (-4,3) [thick]node[right]{\footnotesize$Imk<0$};
\draw[fill][black] (6,7) [thick]node[right]{\footnotesize$Imk>0$};
\draw[fill][black] (6,3) [thick]node[right]{\footnotesize$Imk<0$};
\draw[fill][black] (-4,-7) [thick]node[right]{\footnotesize$Im\lambda<0$};
\draw[fill][black] (-4,-3) [thick]node[right]{\footnotesize$Im\lambda>0$};
\draw[fill][black] (7,-7) [thick]node[right]{\footnotesize$D_{2}$};
\draw[fill][black] (7,-3) [thick]node[right]{\footnotesize$D_{1}$};
\draw[fill][black] (5,-4) [thick]node[right]{\footnotesize$D_{4}$};
\draw[fill][black] (5,-6) [thick]node[right]{\footnotesize$D_{3}$};
\draw[fill][red] (0,5) node[]{\footnotesize$+$};
\draw[fill][black] (-2,0) [thick]node[right]
{\footnotesize$\lambda=\sqrt{k^{2}+q_{0}^{2}}$};
\draw[fill][black] (0,-8) [thick]node[below]
{\footnotesize$\lambda=(z+q_{0}^{2}/z)/2$};
\end{tikzpicture}}
\quad\quad\quad~~~~Figure 1: The transformation relationship from $k-$  plane to $\lambda-$ complex plane and $z-$ complex plane.
\\

The value of the function differs by one sign, so the local polar coordinates can be introduced on $S_{1}$
\begin{align}
\left\{\begin{aligned}
k+iq_{0}=r_{1}e^{i\theta_{1}},\\
k-iq_{0}=r_{2}e^{i\theta_{2}},
\end{aligned}\right.
\end{align}
where $\theta_{1}$, $\theta_{2}$ $\in [-\frac{\pi}{2},\frac{3\pi}{2}]$, then the two single-value analytic branch functions on the Riemann surface  can be written as
\begin{align}\label{2}
\lambda(k)=\left\{\begin{aligned}
(r_{1}r_{2})^{\frac{1}{2}}e^{\frac{i(\theta_{1}+\theta_{2})}{2}},  ~~on~~S_{1},\\
-(r_{1}r_{2})^{\frac{1}{2}}e^{\frac{i(\theta_{1}+\theta_{2})}{2}},  ~~on~~S_{2}.
\end{aligned}\right.
\end{align}
The next step is to define a single valued variable
\begin{align*}
z=k+\lambda.
\end{align*}
Therefore, the transformation \eqref{2} maps the secants $[-iq_{0}, iq_{0}]$ of $S_{1}$ and $S_{2}$ to the secant of $[-q_{0}, q_{0}]$; mapping $\Im k>0$ of $S_{1}$ and $\Im k<0$ of $S_{2}$ to $\Im \lambda>0$ and mapping $\Im k<0$ of $S_{1}$ and $\Im k>0$ of $S_{2}$ to $\Im \lambda<0$.
Thus, the transformation relationship from $k-$  plane to $\lambda-$ complex plane and $z-$ complex plane is shown in the figure 1.

The continuous spectrum is composed of all the values of $k$, such that $\lambda(k)\in\mathbb{R}$, namely $k\in\mathbb{R}\backslash(-q_{0}, q_{0})$, which corresponding to the entire real axis in the complex $z$-plane.  We can clearly see that sheet$-S_{1}$ and sheet$-S_{2}$ can be mapped to the outside and inside of $C_{0}$,  respectively. Therefore, $z$ has two different asymptotic states when $k\rightarrow\infty$, that is $z\rightarrow\infty$ and $z\rightarrow0$, which will be involved in the following study of the asymptotic behavior of eigenfunction and scattering matrix. In the complex $z-$plane, the jump contour is defined as $\Sigma=\mathbb{R}\cup \mathbb{C}_{0}$, and the analytic region can be expressed as
\begin{align}
\begin{split}
D_{1}&=\{z\in \mathbb{C}: |z|^{2}-q_{0}^{2}>0 \cap\Im z>0\},
~D_{2}=\{z\in \mathbb{C}: |z|^{2}-q_{0}^{2}>0~\cap\Im z<0\},\\
D_{3}&=\{z\in \mathbb{C}: |z|^{2}-q_{0}^{2}<0~\cap\Im z<0\},
~D_{4}=\{z\in \mathbb{C}: |z|^{2}-q_{0}^{2}<0~\cap\Im z>0\},
\end{split}
\end{align}
and $\bigcup_{j=1}^{4}\bar{D}_{j}=\mathbb{C}_{0}$. In what follows, we derive the eigenvector corresponding to the eigenvalue. To simplify the calculation, the orthogonal vector with the following lemma can be defined as follows.
\begin{lem}
For any two-component complex valued vector $\omega=(\omega_{1}\quad \omega_{2})^{T}$,
then its orthogonal vector is $\omega^{\bot}=(\omega_{2}\quad -\omega_{1})^{\dag}$ , and it has the property that
$\omega^{\bot}\omega=\omega^{T}\omega^{\bot}=0$.
\end{lem}
Obviously, if $U_{\pm}$ and $V_{\pm}$ satisfy this relation $[U_{\pm}, V_{\pm}]=0$, we can find an invertible matrix $\Gamma_{\pm}$ so that they can be diagonalized at the same time. Then we can write the eigenvalues and the corresponding eigenvector matrix of the asymptotic scattering problem as
\addtocounter{equation}{1}
\begin{align}
U_{\pm}\Gamma_{\pm}&=i\Gamma_{\pm}J,\tag{\theequation a}\label{10}\\
V_{\pm}\Gamma_{\pm}&=i\Gamma_{\pm}\Omega,\tag{\theequation b}\label{11}
\end{align}
where
\begin{align*}
&J=\text{diag}(\lambda,~k,~-\lambda),\\
\Omega=\text{diag}&\left(2\lambda(2k^{2}-q_{0}^{2}),~4k^{3},~-2\lambda(2k^{2}-q_{0}^{2})\right),\\
&\Gamma_{\pm}=\left(
               \begin{array}{ccc}
                 \frac{\textbf{q}_{\pm}}{q_{0}} & \frac{\textbf{q}_{\pm}^{\bot}}{q_{0}} & \frac{i\textbf{q}_{\pm}}{k+\lambda} \\
                 \frac{iq_{0}}{k+\lambda} & 0 & 1 \\
               \end{array}
             \right).
\end{align*}
Therefore,  the invertible matrix of $\Gamma_{\pm}$  is
\begin{align}
(\Gamma_{\pm})^{-1}=-\frac{1}{\gamma(z)}D(z)\left(
                                              \begin{array}{cc}
                                                (\textbf{q}_{\pm})^{\dag} & -\frac{iq_{0}}{z} \\
                                                (\textbf{q}_{\pm}^{\bot})^{T} & 0 \\
                                                -\frac{i(\textbf{q}_{\pm})^{\dag}}{z} & 1 \\
                                              \end{array}
                                            \right),
\end{align}
where
$D(z)=\text{diag}(-1, -\gamma(z), -1)$, $\det\Gamma_{\pm}=-(1+
\frac{q_{0}^{2}}{z^{2}})\triangleq-\gamma(z)$.
For the sake of strict definition, we decompose the asymptotic behavior of potential energy and rewrite Lax pair as the polynomial form, that is
\addtocounter{equation}{1}
\begin{align}
\Phi_{\pm,x}(x,t,z)=U_{\pm}\Phi_{\pm}(x,t,z)+\Delta Q_{\pm}\Phi_{\pm}(x,t,z),\tag{\theequation a}\label{Q10}\\
\Phi_{\pm,t}(x,t,z)=V_{\pm}\Phi_{\pm}(x,t,z)+\Delta \hat{Q}_{\pm}\Phi_{\pm}(x,t,z),\tag{\theequation b}\label{Q11}
\end{align}
where
\begin{align*}
\Delta Q_{\pm}&=Q-Q_{\pm},\\
\Delta \hat{Q}_{\pm}=4k^{2}Q-2ik\sigma_{3}(Q_{x}-Q^{2})&+2Q^{3}-Q_{xx}+[Q_{x},Q]-(4k^{2}Q_{\pm}+2ik\sigma_{3}
Q_{\pm}^{2}+2Q_{\pm}^{3}).
\end{align*}
\subsection{Analysis property}
For all $z\in\Sigma$, we can define the  solutions $\Phi_{\pm}(x,t,z)$ as the simultaneous solution of the Lax pair satisfying the boundary conditions
\begin{align}\label{3}
\Phi_{\pm}(x,t,z)\sim\Gamma_{\pm}(z)e^{i\Theta(x,t,z)},~~x\rightarrow\pm\infty,
\end{align}
where \textbf{$\theta$}$(x,t,z)$ is the $3\times3$ diagonal matrix,
\begin{align}
\mathbf{\theta}(x,t,z)=J(z)x+\Gamma(z) t=\text{diag}(\theta_{1}(x,t,z), \theta_{2}(x,t,z), -\theta_{1}(x,t,z)),\\
\theta_{1}(x,t,z)=\lambda[x+2(2k^{2}-q_{0}^{2})t], ~\theta_{2}(x,t,z)=k(x+4k^{2}t).
\end{align}
The purpose of introducing the simultaneous solution of Lax pairs is to show that the scattering coefficient is independent of time. In order to eliminate the oscillation of asymptotic exponent, we introduce the following modified Jost eigenfunctions
\begin{align}\label{16}
\mu_{\pm}(x,t,z)=\Phi_{\pm}(x,t,z)e^{-i\Theta}{(x,t,z)},
\end{align}
that means
\begin{align}
\lim_{x\rightarrow\pm\infty}\mu_{\pm}(x,t,z)=\Gamma_{\pm}(x,t,z).
\end{align}
As a result, the Lax pair of the modified eigenfunctions $\mu_{\pm}(x,t,z)$ can be derived
\addtocounter{equation}{1}
\begin{align}
(\Gamma^{-1}_{\pm}\mu_{\pm})_{x}=[iJ, \Gamma^{-1}_{\pm}\mu_{\pm}]+\Gamma^{-1}_{\pm}\Delta Q_{\pm}\mu_{\pm},\tag{\theequation a}\label{Q12}\\
(\Gamma^{-1}_{\pm}\mu_{\pm})_{t}=[i\Omega, \Gamma^{-1}_{\pm}\mu_{\pm}]+\Gamma^{-1}_{\pm}\Delta \hat{Q}_{\pm}\mu_{\pm},\tag{\theequation b}\label{Q13}
\end{align}
which can be written in full differential form, to this end,  the following two solutions of modified Jost eigenfunctions can be obtained by integrating along two special paths,
\begin{align}
\begin{split}
\mu_{-}(x,t,z)=\Gamma_{-}+\int_{-\infty}^{x}\Gamma_{-}e^{i(x-y)J(z)}\Gamma_{-}^{-1}\Delta Q_{-}\mu_{-}e^{-i(x-y)J(z)}dy,\\
\mu_{+}(x,t,z)=\Gamma_{+}-\int_{x}^{+\infty}\Gamma_{+}e^{i(x-y)J(z)}\Gamma_{+}^{-1}\Delta Q_{+}\mu_{-}e^{-i(x-y)J(z)}dy.
\end{split}
\end{align}
\begin{thm}
If $q(\cdot, t)-q_{+}\in \mathbb{L}^{1}(a, +\infty)~~(q(\cdot, t)-q_{-}\in \mathbb{L}^{1}(-\infty, a))$, for any constants $a\in \mathbb{R}$, the modified eigenfunctions $\mu_{\pm, j}(x,t,z)$ can be analyzed in the corresponding region of the complex $z$ plane
\begin{align}
\mu_{+,1}: D_{3}\qquad~\mu_{+,2}: Im~z>0 \qquad~\mu_{+,3}: D_{2},\\
\mu_{-,1}: D_{4}\qquad ~\mu_{-,2}: Im~z<0 \qquad ~\mu_{-,3}: D_{1},
\end{align}
where $\mu_{\pm,j}(x,t,z)(j=1,2,3)$ denote the   $j-th$ column of the modified eigenfunctions $\mu_{\pm}(x,t,z)$.
\end{thm}
The equation \eqref{16} shows that the $\Phi_{\pm}(x,t,z)$ column also have the same analytical and boundedness. Because of the relationship between $\Phi_{\pm}(x,t,z)$ and $\mu_{\pm}(x,t,z)$, which have the same analytic properties. On the basis of boundary conditions \eqref{NZBCs}, employing the Abel's theorem to derive the determinant of $\Phi_{\pm}(x,t,z)$
\begin{align}\label{4}
\det \Phi_{\pm}(x,t,z)=-\gamma(z)e^{i\theta_{2}(x,t,z)}.
\end{align}
It is obvious that $\Phi_{\pm}$ are the fundamental matrix solutions of Lax pair, so they must exist an invertible matrix $A(z)$ such that
\begin{align}\label{5}
\Phi_{-}(x,t,z)=\Phi_{+}(x,t,z)A(z),
\end{align}
where the scattering matrix $A(z)=(a_{ij}(z))_{3\times3}$, which  is a matrix that does not depend on time. From the above relationships \eqref{4} and \eqref{5}, we can deduce $\det A(z)=1$. For convenience, one introduces $A(z)^{-1}=B(z)=(b_{ij}(z))_{3\times3}$. In the scalar case, the analytical property of the diagonal scattering coefficient comes from the Wronskians, which is expressed as the analytic eigenfunction. However, this method is not suitable for this work.

\begin{thm}
If $q(\cdot, t)-q_{+}\in \mathbb{L}^{1}(a, +\infty)(q(\cdot, t)-q_{-}\in \mathbb{L}^{1}(-\infty, a))$, for any constants $a\in \mathbb{R}$, the diagonal elements of the scattering matrix  can be analytic in the corresponding region of the complex $z$ plane
\begin{align}
a_{11}: D_{4},\qquad a_{22}:Im~ z<0, \qquad a_{33}:D_{1},\\
b_{11}: D_{3},\qquad b_{22}:Im~ z>0, \qquad b_{33}:D_{2}.
\end{align}
\end{thm}
\subsection{Auxiliary eigenfunctions and adjoint matrix}
In the $2\times2$ spectral problem, the analysis of the scattering matrix can be derived from the Wronskian determinant of \eqref{5}, but this method is no longer applicable in this work. The reason is that in the spectral problem with $3\times3$ Lax pair, the certain column of the Jost eigenfunction is not analytic in all given region. To this end, the so-called `adjoint' Lax pair is introduced to overcome the defect of analyticity.
\begin{align}\label{XF1}
\begin{split}
\widetilde{\Phi}_{x}(x,t,z)=\widetilde{U}\widetilde{\Phi}(x,t,z),\\
\widetilde{\Phi}_{t}(x,t,z)=\widetilde{V}\widetilde{\Phi}(x,t,z),
\end{split}
\end{align}
where $\widetilde{U}=\overline{U}(x,t,\overline{z})$, $\widetilde{V}=\overline{V}(x,t,\overline{z})$ and $Q(x,t)$ satisfies the following relations
\begin{align*}
\overline{Q}=-Q^{T},~~~~Q=-Q^{\dag},~~~~Q\sigma_{3}=-\sigma_{3}Q,~~~~Q^{T}\sigma_{3}=-\sigma_{3}Q^{T}.
\end{align*}
It is easy to verify that the eigenvalues of the $\widetilde{U}$ are $ik$ and $\mp i\lambda$, in the same way the eigenvalues of the $\widetilde{V}$ are $-4ik^{3}$ and $\mp2i\lambda(2k^{2}-q_{0}^{2})$. The asymptotic spectral problem with $3\times3$ Lax pair  can be obtained,
\addtocounter{equation}{1}
\begin{align}
\widetilde{\Phi}_{\pm,x}(x,t,z)=\widetilde{U_{\pm}}\widetilde{\Phi_{\pm}}(x,t,z),
\tag{\theequation a}\label{Q16}\\
\widetilde{\Phi}_{\pm,t}(x,t,z)=\widetilde{V_{\pm}}\widetilde{\Phi_{\pm}}(x,t,z),\tag{\theequation b}\label{Q17}
\end{align}
which also have simultaneous solutions, we have
\begin{align}\label{29}
\widetilde{\Phi}_{\pm,x}(x,t,z)\sim\widetilde{\Gamma}_{\pm}(z)e^{-i\Theta(x,t,z)},\qquad x\rightarrow\infty.
\end{align}
Obviously, the relation of $\widetilde{\Gamma}_{\pm}(x,t,z)=\overline{\Gamma_{\pm}(\overline{z})}$ and $\det \widetilde{\Gamma}_{\pm}(z)=-\gamma(z)$ can be deduced. Based on the transformation between $\Phi_{\pm}(x,t,z)$ and $\mu_{\pm}(x,t,z)$, we can also get the analytic properties of $\mu_{\pm,j}(x,t,z)$,
\begin{align}
\widetilde{\mu}_{+,1}: D_{4},\qquad \widetilde{\mu}_{+,2}:Im ~z<0, \qquad \widetilde{\mu}_{+,3}:D_{1},\\
\widetilde{\mu}_{-,1}: D_{3},\qquad \widetilde{\mu}_{-,2}:Im~ z>0, \qquad \widetilde{\mu}_{-,3}:D_{2}.
\end{align}
It is obvious that $\widetilde{\Phi}_{\pm}(x,t,z)$ are the fundamental matrix solutions of Lax pair, so they must exist an invertible matrix $\widetilde{A(z)}$, such that
\begin{align}\label{5-1}
\widetilde{\Phi}_{-}(x,t,z)=\widetilde{\Phi}_{+}(x,t,z)\widetilde{A(z)}.
\end{align}
Similarly, we define the scattering matrix $\widetilde{B}(z)$=$\widetilde{A}^{-1}(z)$. At the same time, we find the analytic property of the following diagonal scattering coefficient in the scattering matrix
\begin{align}
a_{11}: D_{3},\qquad a_{22}:Im~ z>0, \qquad a_{33}:D_{2},\\
b_{11}: D_{4},\qquad b_{22}:Im~ z<0, \qquad b_{33}:D_{1}.
\end{align}

In order to reconstruct the solution of the Lax pair \eqref{Lax} from the modified Lax pair \eqref{XF1}, one can  introduce the following Lemma.
\begin{lem}\label{6}
For any vectors $\textbf{u}, \textbf{v}\in \mathbb{C}^{3}$, the following equalities can be derived and the notation ``$\times$" represents the cross product,
\begin{align*}
[(\sigma_{3}\textbf{u})\times \textbf{v}]+[\textbf{u}\times(&\sigma_{3}\textbf{v})]-[\textbf{u}\times \textbf{v}]-[(\Lambda \textbf{u})\times(\Lambda \textbf{v})]=\textbf{0},\\
\Lambda[\textbf{u}\times \textbf{v}]&=-[(\Lambda \textbf{u})\times(\Lambda \textbf{v})],\\
Q[\textbf{u}\times \textbf{v}]&+[(Q^{T}\textbf{u})\times \textbf{v}]+[\textbf{u}\times(Q^{T}\textbf{v})]=\textbf{0},\\
\Lambda Q^{2}[\textbf{u}\times \textbf{v}]+&[\Lambda(Q^{T})^{2}\textbf{u}\times \textbf{v}]+[\textbf{u}\times\Lambda(Q^{T})^{2}\textbf{v}]=\textbf{0}.
\end{align*}
\end{lem}
By Lemma \ref{6} and direct calculation, we can get the theorem
\begin{thm}\label{7}
If $\widetilde{v}(x,t,z)$ and $\widetilde{\omega}(x,t,z)$ are any two solutions of the adjoint Lax pair \eqref{XF1}, then $u(x,t,z)=e^{i\theta_{2}(x,t,z)}[\widetilde{v}\times \widetilde{\omega}](x,t,z)$ is  the solution of \eqref{Lax}.
\end{thm}

By the transformation relation between the modified eigenfunction and the matrix function, which is composed of eigenvectors related to eigenvalues, we find that there exist relations between the modified eigenfunctions and the eigenfunctions
\begin{align}\label{17}
\begin{split}
\Phi_{\pm,j}(x,t,z)=\frac{e^{i\theta_{2}}[\widetilde{\Phi}_{\pm,\ell}
\times\widetilde{\Phi}_{\pm,m}]}{\gamma_{j}(z)},\\
\widetilde{\Phi}_{\pm,j}(x,t,z)=\frac{e^{-i\theta_{2}}[\Phi_{\pm,\ell}
\times\Phi_{\pm,m}]}{\gamma_{j}(z)},
\end{split}
\end{align}
where $\gamma_{1}(z)=\gamma_{3}(z)=-1$ and $\gamma_{2}(z)=-\gamma(z)$.
\begin{proof}
The theorem \ref{7} and equation \ref{29} yield
\begin{align}
\mathbf{u}(x,t,z)=-e^{-i\theta_{1}}\Gamma_{\pm,3}, \quad x\rightarrow\pm\infty.
\end{align}
Moreover, $\mathbf{u}_{\pm}(x,t,z)$ must be a linear combination of the column of $\Phi_{\pm}(x,t,z)$, so there exist the scalar functions $m_{\pm}(z), n_{\pm}(z)$ and $p_{\pm}(z)$ such that they are a linear simultaneous equations of $\mathbf{u}(x,t,z)$, then the corresponding scalar function is obtained by the method of comparison coefficient.
\end{proof}

Since a column of the Jost eigenfunction is not analytic in a given region, in light of these results, we define the auxiliary eigenfunctions $\chi_{j}(x,t,z), (j=1,2,3,4)$, which are also the solution of the Lax pair \eqref{Lax}
\addtocounter{equation}{1}\label{2.35}
\begin{align}
\chi_{1}(x,t,z)=e^{i\theta_{2}}[\widetilde{\Phi}_{-,2}\times\widetilde{\Phi}_{+,3}](x,t,z),\tag{\theequation a}\\
\chi_{2}(x,t,z)=e^{i\theta_{2}}[\widetilde{\Phi}_{+,2}\times\widetilde{\Phi}_{-,3}](x,t,z),\tag{\theequation b}\\
\chi_{3}(x,t,z)=e^{i\theta_{2}}[\widetilde{\Phi}_{-,1}\times\widetilde{\Phi}_{+,2}](x,t,z),\tag{\theequation c}\\
\chi_{4}(x,t,z)=e^{i\theta_{2}}[\widetilde{\Phi}_{+,1}\times\widetilde{\Phi}_{-,2}](x,t,z).\tag{\theequation d}
\end{align}
It is obviously to deduce the analytic property of $\chi_{j}(x,t,z)$ from the analytic property of modified eigenfunctions $\widetilde{\Phi}_{\pm}(x,t,z)$.
In the previous section, we know that because of the linear  dependence of $\Phi_{\pm}(x,t,z)$, the eigenfunction is associated with the scattering matrix. Now we can get the relationship between the scattering matrices $A(z)$  and $\widetilde{A}(z)$ from \eqref{17}
\begin{align}
\widetilde{A(z)}=D(z)(A^{-1}(z))^{T}D^{-1}(z).
\end{align}
\begin{thm}
For $z\in \Sigma$, the Jost eigenfunctions have the relations as follows,
\begin{align}\label{8}
\begin{split}
\Phi_{-,1}(x,t,z)=\frac{a_{21}(z)\Phi_{-,2}(x,t,z)-\chi_{2}(x,t,z)}{a_{22}(z)}=
\frac{a_{31}(z)\Phi_{-,3}(x,t,z)-\chi_{1}{(x,t,z)}}{a_{33}(z)},\\
\Phi_{+,1}(x,t,z)=\frac{b_{21}(z)\Phi_{+,2}(x,t,z)-\chi_{1}(x,t,z)}{b_{22}(z)}=
\frac{b_{31}(z)\Phi_{+,3}(x,t,z)-\chi_{2}{(x,t,z)}}{b_{33}(z)},\\
\Phi_{-,3}(x,t,z)=\frac{a_{23}(z)\Phi_{-,2}(x,t,z)-\chi_{3}(x,t,z)}{a_{22}(z)}=
\frac{a_{13}(z)\Phi_{-,1}(x,t,z)-\chi_{4}{(x,t,z)}}{a_{11}(z)},\\
\Phi_{+,3}(x,t,z)=\frac{b_{23}(z)\Phi_{+,2}(x,t,z)-\chi_{4}(x,t,z)}{b_{22}(z)}=
\frac{b_{13}(z)\Phi_{+,1}(x,t,z)-\chi_{3}{(x,t,z)}}{b_{11}(z)}.
\end{split}
\end{align}
\end{thm}
The advantage of the adjoint eigenfunctions will be instrumental in obtaining many valid results in the following sections. Furthermore, in order to eliminate the oscillation term of the auxiliary eigenfunctions $\chi_{j}(z)$, we define the modified auxiliary eigenfunctions
\begin{align}\label{25}
m_{j}&=e^{-i\theta_{1}}\chi_{j}(x,t,z), \qquad j=1,2,\\
m_{j}&=e^{i\theta_{1}}\chi_{j}(x,t,z),  ~~\qquad j=3,4.
\end{align}
These results will play a crucial role in the characteristics of the following discrete spectrum and the construction of the RH problem.
\section{Symmetries }
For the system with ZBCs, the symmetry of the scattering problem is mainly generated by mapping $k\mapsto k^{*}$. For the system with NZBCs, in the process of solving the asymptotic spectral problem, in addition to the usual spectral parameter $k$, the spectral matrix is diagonalized and resulting in another parameter $\lambda$, which is a multivalued function of $k$. For this case, Zakharov and Shabat introduced a Riemann surface $S$ to deal with this case, which makes the symmetry under this condition more complicated. Because after removing the asymptotic oscillation, the Jost solution does not tend to the identity matrix. In this section, we will discuss the symmetries of $z\mapsto z^{*}$ and $z\mapsto \frac{q_{0}^{2}}{z}$ for the Jost eigenfunctions, scattering matrix and auxiliary eigenfunctions, respectively.

\subsection{The first symmetries}

In this section, we mainly study the case of $z\mapsto z^{*}$, that is, mapping the upper half plane to the lower half plane. Let's first introduce the Lemma
\begin{lem}
If $\Phi(x,t,z)$ is the nonsingular solution of Lax pair \eqref{Lax}, then $\omega(x,t,z)=(\Phi^{\dag}(x,t,z^{*}))^{-1}$ is also a solution.
\end{lem}
\begin{lem}\label{18}
For all $z\in \Sigma$, the Jost eigenfunctions have the symmetry as follows
\begin{align}\label{9}
(\Phi^{\dag}(x,t,z^{*}))^{-1}C(z)=\Phi_{\pm}(x,t,z),
\end{align}
where $C(z)=\text{diag}(\gamma(z), 1, \gamma(z))$.
\end{lem}
Using the relation between adjoint matrix and inverse matrix, it will also be convenient to note also that
\begin{align}
[\Phi_{\pm}^{-1}(x,t,z)]^{T}=\frac{[\Phi_{\pm,2}\times\Phi_{\pm,3}, \Phi_{\pm,3}\times\Phi_{\pm,1},  \Phi_{\pm,1}\times\Phi_{\pm,2}](x,t,z)}{\det\Phi_{\pm}(x,t,z)}.
\end{align}
Then using Lemma \ref{18} in the scattering relation \eqref{5} yields the following.
\begin{lem}
The scattering matrix $A(z)$ satisfies the following symmetry
\begin{align}
A(z)=C(z)B(z)C^{-1}(z).
\end{align}
\end{lem}
Componentwise, it is not difficult to get the relation of scattering coefficient
\begin{align*}
b_{11}(z)=a_{11}^{*}(z^{*}),~~~b_{12}(z)=\frac{1}{\gamma(z)}a_{21}^{*}(z^{*}),~~~b_{31}(z)=a_{31}^{*}(z^{*}),\\
b_{21}(z)=\gamma(z)a_{21}^{*}(z^{*}),~~~b_{22}(z)=a_{22}^{*}(z^{*}),~~~b_{23}(z)=\gamma(z)a_{32}^{*}(z^{*}),\\
b_{31}(z)=a_{13}^{*}(z^{*}),~~~b_{32}(z)=\frac{1}{\gamma(z)}a_{23}^{*}(z^{*}),~~~b_{33}(z)=a_{33}^{*}(z^{*}).
\end{align*}
Using \eqref{8} and \eqref{9}, we can find the relationship between eigenfunctions and auxiliary  eigenfunctions and conclude as follows.
\addtocounter{equation}{1}\label{19}
\begin{lem}
\begin{align}
\Phi_{+,1}^{*}(z^{*})&=-\frac{e^{-i\theta_{2}(x,t,z)}}{b_{22}(z)}[\Phi_{+,2}\times\chi_{4}](x,t,z),
\tag{\theequation a}\\
\Phi_{-,1}^{*}(z^{*})&=-\frac{e^{-i\theta_{2}(x,t,z)}}{a_{22}(z)}[\Phi_{-,2}\times\chi_{3}](x,t,z),
\tag{\theequation b}\\
\Phi_{+,2}^{*}(z^{*})&=-\frac{e^{-i\theta_{2}(x,t,z)}}{\gamma(z)b_{11}(z)}[\chi_{3}\times\Phi_{+,1}](x,t,z)
=-\frac{e^{-i\theta_{2}(x,t,z)}}{\gamma(z)b_{33}(z)}[\Phi_{+,3}\times\chi_{2}](x,t,z),
\tag{\theequation c}\\
\Phi_{-,2}^{*}(z^{*})&=-\frac{e^{-i\theta_{2}(x,t,z)}}{\gamma(z)a_{11}(z)}[\chi_{4}\times\Phi_{-,1}](x,t,z)
=-\frac{e^{-i\theta_{2}(x,t,z)}}{\gamma(z)a_{33}(z)}[\Phi_{-,3}\times\chi_{1}](x,t,z),
\tag{\theequation d}
\end{align}
\begin{align}
\Phi_{+,3}^{*}(z^{*})&=-\frac{e^{-i\theta_{2}(x,t,z)}}{b_{22}(z)}[\chi_{1}\times\Phi_{+,2}](x,t,z),
\tag{\theequation e}\\
\Phi_{-,3}^{*}(z^{*})&=-\frac{e^{-i\theta_{2}(x,t,z)}}{a_{22}(z)}[\chi_{2}\times\Phi_{-,2}](x,t,z)
\tag{\theequation f}.
\end{align}
\end{lem}
The auxiliary eigenfunctions satisfy the similar symmetry relations.
\begin{lem}\label{20}
\begin{align}
\begin{split}
\chi_{1}^{*}(z^{*})=e^{-i\theta_{2}}[\Phi_{-,2}\times\Phi_{+,3}](z),\qquad\qquad z\in D_{2},\\
\chi_{2}^{*}(z^{*})=e^{-i\theta_{2}}[\Phi_{+,2}\times\Phi_{-,3}](z),\qquad\qquad z\in D_{1},\\
\chi_{3}^{*}(z^{*})=e^{-i\theta_{2}}[\Phi_{-,1}\times\Phi_{+,2}](z),\qquad\qquad z\in D_{4},\\
\chi_{4}^{*}(z^{*})=e^{-i\theta_{2}}[\Phi_{+,1}\times\Phi_{-,2}](z),\qquad\qquad z\in D_{3}.
\end{split}
\end{align}
\end{lem}
In addition, the proof of  Lemma \ref{20} and \eqref{17} yields
\begin{align}
\Phi_{\pm,j}^{*}(x,t,z^{*})=\frac{e^{-i\theta_{2}(z)}[\Phi_{\pm,\ell}\times\Phi_{\pm,m}](x,t,z)}{\gamma_{j}(z)},
\end{align}
where $j, \ell, m, (j, \ell, m=1,2,3)$ still meet the rotation relationships.
\subsection{The second symmetries}
In this section, we mainly study the case of $z\mapsto-\frac{q_{0}^{2}}{z}$, that is, the inside of the circle $C_{0}$ with radius $q_{0}$ is mapped to the outside of the circle. We use this symmetry to relate the values of the eigenfunctions on the two sheets. Since $\Phi_{\pm}(-\frac{q_{0}^{2}}{z})$ is the solution of Lax pair and the linear correlation of the solutions, it is not difficult to deduce that the Jost eigenfunctions have the following symmetries.
\begin{align}
\Phi_{\pm}(x,t,z)=\Phi_{\pm}(x,t,-\frac{q_{0}^{2}}{z})\pi(z),
\end{align}
where
\begin{align*}
\pi(z)=\left(
         \begin{array}{ccc}
           0 & 0 & \frac{iq_{0}}{z} \\
           0 & 1 & 0 \\
           \frac{iq_{0}}{z} & 0 & 0 \\
         \end{array}
       \right).
\end{align*}

As before, the analytic properties of eigenfunctions allow us to extend some of the above relationships:
\begin{align}\label{12}
\begin{split}
\Phi_{\pm,1}(z)&=\frac{iq_{0}}{z}\Phi_{\pm,3}(-\frac{q_{0}^{2}}{z}),\\
\Phi_{\pm,2}(z)&=\Phi_{\pm,2}(-\frac{q_{0}^{2}}{z}),\\
\Phi_{\pm,3}(z)&=\frac{iq_{0}}{z}\Phi_{\pm,1}(-\frac{q_{0}^{2}}{z}).
\end{split}
\end{align}
We use \eqref{5} again to obtain the following lemma.
\begin{lem}
\begin{align}
A(-\frac{q_{0}^{2}}{z})=\pi(z)A(z)\pi^{-1}(z),\qquad~
B(-\frac{q_{0}^{2}}{z})=\pi(z)B(z)\pi^{-1}(z),
\end{align}
\end{lem}
which lead to the relationships between the scattering coefficients
\begin{align*}
a_{11}(z)&=a_{33}(-\frac{q_{0}^{2}}{z}),\qquad a_{12}(z)=\frac{z}{iq_{0}}a_{32}(-\frac{q_{0}^{2}}{z}),\qquad
a_{13}(z)=a_{31}(-\frac{q_{0}^{2}}{z}),\\
a_{21}(z)&=\frac{iq_{0}}{z}a_{23}(-\frac{q_{0}^{2}}{z}),\qquad
a_{22}(z)=a_{22}(-\frac{q_{0}^{2}}{z}),\qquad
a_{23}(z)=\frac{iq_{0}}{z}a_{21}(-\frac{q_{0}^{2}}{z}),\\
a_{31}(z)&=a_{13}(-\frac{q_{0}^{2}}{z}),\qquad
a_{32}(z)=\frac{z}{iq_{0}}a_{12}(-\frac{q_{0}^{2}}{z}),\qquad
a_{33}(z)=a_{11}(-\frac{q_{0}^{2}}{z}).
\end{align*}
The same set of equations applies to the elements of $B(z)$. Finally, we combine \eqref{20} with  \eqref{12} to conclude the following.
\begin{lem}
The auxiliary eigenfunctions satisfy these symmetries
\begin{align}
\chi_{1}(z)=\frac{iq_{0}}{z}\chi_{4}(-\frac{q_{0}^{2}}{z}),\qquad z\in D_{1},\\
\chi_{2}(z)=\frac{iq_{0}}{z}\chi_{3}(-\frac{q_{0}^{2}}{z}),\qquad z\in D_{2}.
\end{align}
\end{lem}
\subsection{The third symmetries}
In this section,   we can combine the above two symmetries to obtain the relationship between the eigenfunction and the scattering coefficient, we will discuss the symmetries of the reflection coefficients, which will simplify the following process
\begin{align*}
&\rho_{1}(z)=\frac{a_{21}(z)}{a_{11}(z)}=\frac{\gamma(z)b_{12}^{*}(z^{*})}{b_{11}^{*}(z^{*})},\qquad
\rho_{2}(z)=\frac{a_{31}(z)}{a_{11}(z)}=\frac{b_{13}^{*}(z^{*})}{b_{11}^{*}(z^{*})},\\
&\rho_{3}(z)=\frac{a_{32}(z)}{a_{22}(z)}=\frac{b_{23}^{*}(z^{*})}{\gamma(z)b_{22}^{*}(z^{*})},\qquad
\rho_{1}(-\frac{q_{0}^{2}}{z})=\frac{z}{iq_{0}}\frac{a_{23}(z)}{a_{33}(z)}
=\frac{z}{iq_{0}}\frac{\gamma(z)b_{32}^{*}(z^{*})}{b_{33}^{*}(z^{*})},\\
&\rho_{2}(-\frac{q_{0}^{2}}{z})=\frac{a_{13}(z)}{a_{33}(z)}=\frac{b_{31}^{*}(z^{*})}{b_{33}^{*}(z^{*})},\qquad
\rho_{3}(-\frac{q_{0}^{2}}{z})=\frac{iq_{0}}{z}\frac{a_{12}(z)}{a_{22}(z)}
=\frac{iq_{0}}{z}\frac{b_{21}^{*}(z^{*})}{\gamma(z)b_{22}^{*}(z^{*})}.
\end{align*}

In addition, we also want to imply that the symmetry of the scattering coefficients do not lead to any symmetrical relationship between these reflection coefficients. We will obtain the trace formula for analyzing the scattering coefficient in sec. $6$, and then we can combine all the above symmetries to reconstruct the entire scattering matrix.
\section{Discrete spectrum and asymptotic behavior}
Recall that in the $2\times2$ spectral problem with NZBC, there is a one-to-one correspondence between the zero point of the analytical scattering coefficient and the discrete eigenvalues. In addition, the self-adjointness of the scattering problem shows that this discrete eigenvalue $k$ must be real, and it can be proved that discrete eigenvalues cannot appear in the continuous spectrum. In this work, the scattering problem in \eqref{Lax} is also self-adjointness. For solving the RH problem, we need to transform the original RH problem into a standard RH problem via removing the asymptotics and poles. The purpose of this section is to find the poles by studying the discrete spectrum and residue conditions, and discuss the asymptotic property of modified eigenfunctions $\mu_{\pm}(x,t,z)$.
\subsection{Discrete spectrum}
Because of the complexity of the analytic property of the eigenfunction, the auxiliary eigenfunctions are introduced. The discrete spectrum of $3\times3$ matrix under NZBCs is more complex than that of the $2\times2$ matrix. Aiming at the analytic property of eigenfunctions, the discrete function is characterized. In order to study the discrete spectrum in this case, we first conclude as follows.
\begin{lem}\label{21}
\begin{align}
\mho_{1}(x,t,z)=(\chi_{1}(x,t,z), \Phi_{+,2}(x,t,z), \Phi_{-,3}(x,t,z)),\\
\mho_{2}(x,t,z)=(\chi_{2}(x,t,z), \Phi_{-,2}(x,t,z), \Phi_{+,3}(x,t,z)),\\
\mho_{3}(x,t,z)=(\Phi_{+,1}(x,t,z), \Phi_{-,2}(x,t,z), \chi_{3}(x,t,z)),\\
\mho_{4}(x,t,z)=(\Phi_{-,1}(x,t,z), \Phi_{+,2}(x,t,z), \chi_{4}(x,t,z)),
\end{align}
which are analytic in four regions $D_{j}(j=1,2,3,4)$, respectively.
\end{lem}
Taking \eqref{8} into account and by the direct calculations, the Wronskian determinant of $\mho_{j}(z)(j=1,2,3,4)$ can be derived
\begin{align*}
Wr(\mho_{1}(z))=-a_{33}(z)b_{22}(z)\gamma(z)e^{i\theta_{2}(z)},\\
Wr(\mho_{2}(z))=-a_{22}(z)b_{22}(z)\gamma(z)e^{i\theta_{2}(z)},\\
Wr(\mho_{3}(z))=-b_{11}(z)a_{22}(z)\gamma(z)e^{i\theta_{2}(z)},\\
Wr(\mho_{4}(z))=-b_{22}(z)a_{11}(z)\gamma(z)e^{i\theta_{2}(z)}.
\end{align*}

It is not difficult to find that the  columns of the eigenfunction $\mho_{1}(x,t,z)$ is linearly dependent at the zeros of $a_{33}(z)$ and $b_{22}(z)$, the columns of the eigenfunction $\mho_{2}(x,t,z)$ is linearly dependent at the zeros of $a_{22}(z)$ and $b_{33}(z)$, the columns of the eigenfunction $\mho_{3}(x,t,z)$ is linearly dependent at the zeros of $a_{22}(z)$ and $b_{11}(z)$ and the column of the eigenfunction $\mho_{4}(x,t,z)$ is linearly dependent at the zeros of $a_{11}(z)$ and $b_{22}(z)$. In addition, the symmetry of the scattering coefficients mean that these zeros are not independent of each other.
\begin{lem}
Let $Im~ z_{0}> 0$, and $z=z_{0}$, we have
\begin{align}
b_{22}(z_{0})=0\Leftrightarrow a_{22}(z_{0}^{*})=0\Leftrightarrow a_{22}(-\frac{q_{0}^{2}}{z_{0}^{*}})=0 \Leftrightarrow b_{22}(-\frac{q_{0}^{2}}{z_{0}})=0.
\end{align}
\end{lem}
\begin{lem}
Let $Im~ z_{0}> 0$, and $|z|\geq q_{0}$, we have
\begin{align}
a_{33}(z_{0})=0\Leftrightarrow b_{33}(z_{0}^{*})=0\Leftrightarrow b_{11}(-\frac{q_{0}^{2}}{z_{0}^{*}})=0 \Leftrightarrow a_{11}(-\frac{q_{0}^{2}}{z_{0}})=0.
\end{align}
\end{lem}
The set of discrete spectrum is given by
\begin{align*}
Z=\left\{z_{0}, z_{0}^{*}, -\frac{q_{0}^{2}}{z_{0}^{*}}, -\frac{q_{0}^{2}}{z_{0}}\right\}.
\end{align*}
Now we only need to discuss the following three cases of zero points $a_{11}(z)$ and $b_{22}(z)$,
\begin{align*}
(i)~~a_{33}(z_{0})=0,~~ b_{22}(z_{0})\neq0;\\
(ii)~~a_{33}(z_{0})\neq0, ~~b_{22}(z_{0})=0;\\
(iii)~~a_{33}(z_{0})=0, ~~b_{22}(z_{0})=0.
\end{align*}
The following lemma are also useful to describe the discrete spectrum.

\centerline{\begin{tikzpicture}[scale=0.55]
\draw[-][thick](4,4)--(4,-4);
\draw[-][thick](4,-4)--(-4,-4);
\draw[-][thick](-4,-4)--(-4,4);
\draw[-][thick](-4,4)--(4,4);
\draw[fill] (0,-2)node[below]{} circle [radius=0.055];
\draw[fill] (0,2)node[below]{} circle [radius=0.055];
\draw[fill] (2,0)node[below]{} circle [radius=0.055];
\draw[fill] (-2,0)node[below]{} circle [radius=0.055];
\draw[fill] (0,0)node[below]{} circle [radius=0.055];
\path [fill=white] (-4,0)--(-4,4) to (4,4) -- (4,0);
\filldraw[gray, line width=0.5](-4,0)--(-2,0) arc (180:0:2);
\path [fill=gray] (-4,0)--(-4,-4) to (4,-4) -- (4,0);
\filldraw[white, line width=0.5](-4,0)--(-2,0) arc (-180:0:2);
\draw[->][thick](-4,0)--(-3,0);
\draw[-][thick](-3,0)--(-2,0)node[below right]{\footnotesize$0^{-}$};;
\draw[-][thick](-2,0)--(-1,0);
\draw[<-][thick](-1,0)--(0,0);
\draw[-][thick](0,0)--(1,0);
\draw[<-][thick](1,0)--(2,0)node[below right]{\footnotesize$0^{+}$};
\draw[->][thick](2,0)--(3,0);
\draw[->][thick](3,0)--(4,0)[thick]node[right]{$Rez$};
\draw[-][thick](0,0)--(0,1);
\draw[-][thick](0,1)--(0,2)node[below right]{\footnotesize$iq_{0}$};
\draw[-][thick](0,2)--(0,3);
\draw[->][thick](0,3)--(0,4)[thick]node[right]{$Imz$};
\draw[-][thick](0,0)--(0,-1);
\draw[-][thick](0,-1)--(0,-2)node[below right]{\footnotesize$-iq_{0}$};
\draw[-][thick](0,-2)--(0,-3);
\draw[-][thick](0,-3)--(0,-4);
\draw[->][red, line width=0.8] (2,0) arc(0:220:2);
\draw[->][red, line width=0.8] (2,0) arc(0:330:2);
\draw[->][red, line width=0.8] (2,0) arc(0:-330:2);
\draw[->][red, line width=0.8] (2,0) arc(0:-220:2);
\draw[fill][blue] (2.5,2.5) circle [radius=0.035][thick]node[right]{\footnotesize$z_{n}$};
\draw[fill] (2.5,-2.5) circle [radius=0.035][thick]node[right]{\footnotesize$z_{n}^{*}$};
\draw[fill] (-1,1) circle [radius=0.035][thick]node[right]{\footnotesize$-\frac{q_{0}^{2}}{z_{n}}$};
\draw[fill][blue] (-1,-1) circle [radius=0.035][thick]node[right]{\footnotesize$-\frac{q_{0}^{2}}{z_{n}^{*}}$};
\end{tikzpicture}}

\centerline{\noindent {\small \textbf{Figure 2.} The distribution of the set of discrete spectral points on the $z$-plane.} }

\begin{lem}
Suppose that $Im~ z_{0}>0$, and $|z_{0}|>q_{0}$, the following equations are equivalent to each other:
\begin{align*}
&(i)~~\chi_{1}(z_{0})=0;\\
&(ii)~~\chi_{4}(z_{0})=0;\\
&(iii)~~\exists b_{0},~~so~~that~~ \Phi_{-,2}(z_{0}^{*})=b_{0}\Phi_{+,3}(z_{0}^{*});\\
&(iv)~~\exists \widetilde{b_{0}},~~ so ~~ that~~ \Phi_{-,2}(-\frac{q_{0}^{2}}{z_{0}^{*}})=\widetilde{b_{0}}\Phi_{+,1}(-\frac{q_{0}^{2}}{z_{0}^{*}}).
\end{align*}
\end{lem}
\begin{lem}
Suppose that $Im~ z_{0}>0$, and $|z_{0}|>q_{0}$, the following equations are equivalent to each other:
\begin{align*}
&(i)~~\chi_{2}(z_{0}^{*})=0;\\
&(ii)~~\chi_{3}(-\frac{q_{0}^{2}}{z_{0}^{*}})=0;\\
&(iii)~~\exists \hat{b_{0}},~~so~~ that~~ \Phi_{+,2}(z_{0})=\hat{b_{0}}\Phi_{-,3}(z_{0});\\
&(iv)~~\exists \check{b_{0}},~~so~~that~~ \Phi_{+,2}(-\frac{q_{0}^{2}}{z_{0}})=\check{b_{0}}\Phi_{-,1}(-\frac{q_{0}^{2}}{z_{0}}).
\end{align*}
\end{lem}
\begin{lem}
Suppose that $Im~ z_{0}>0$,  $|z_{0}|>q_{0}$, and $a_{22}(z_{0})b_{33}(z_{0})=0$, the following situations are true:

(1)If $z_{0}$ is the first type eigenvalue
\begin{align*}
\Phi_{-,3}(z_{0})&=e_{0}\chi_{1}(z_{0}), \qquad\qquad \chi_{2}(z_{0}^{*})=\hat{e_{0}}\Phi_{+,3}(z_{0}^{*}),\\
\chi_{3}(-\frac{q_{0}^{2}}{z_{0}^{*}})&=\check{e_{0}}\Phi_{+,1}(-\frac{q_{0}^{2}}{z_{0}^{*}}),\qquad
\Phi_{-,1}(-\frac{q_{0}^{2}}{z_{0}})=\widetilde{e_{0}}\chi_{4}(-\frac{q_{0}^{2}}{z_{0}}).
\end{align*}
(2)If $z_{0}$ is the second type eigenvalue
\begin{align*}
\chi_{1}(z_{0})&=f_{0}\Phi_{+,2}(z_{0}), \qquad
\Phi_{-,2}(z_{0}^{*})=\hat{f_{0}}\chi_{2}(z_{0}^{*}),\\
\Phi_{-,2}(-\frac{q_{0}^{2}}{z_{0}^{*}})&=\check{f_{0}}\chi_{3}(-\frac{q_{0}^{2}}{z_{0}^{*}}),\qquad
\chi_{4}(-\frac{q_{0}^{2}}{z_{0}})=\widetilde{f_{0}}\Phi_{+,2}(-\frac{q_{0}^{2}}{z_{0}}).
\end{align*}
(3)If $z_{0}$ is the third type eigenvalue
\begin{align*}
\Phi_{-,3}(z_{0})&=g_{0}\Phi_{+,2}(z_{0}), \qquad \qquad
\Phi_{-,2}(z_{0}^{*})=\hat{g_{0}}\Phi_{+,3}(z_{0}^{*}),\\
\Phi_{-,2}(-\frac{q_{0}^{2}}{z_{0}^{*}})&=\check{g_{0}}\Phi_{+,1}(-\frac{q_{0}^{2}}{z_{0}^{*}}),\qquad
\Phi_{-,1}(-\frac{q_{0}^{2}}{z_{0}})=\widetilde{g_{0}}\Phi_{+,2}(-\frac{q_{0}^{2}}{z_{0}}).
\end{align*}
\end{lem}
In order to easily derive the residue condition, we use the modified eigenfunction to rewrite the above Lemma.
\begin{lem}
(1)~$\{z_{n}\}_{n=1}^{N_{1}}$ are the first type eigenvalue
\begin{align*}
\mu_{-,3}(z_{n})&=e_{n}m_{1}(z_{n})e^{2i\theta_{1}(z_{n})}, \qquad\qquad m_{2}(z_{n}^{*})=\hat{e_{n}}\mu_{+,3}(z_{n}^{*})e^{-2i\theta_{1}(z_{n}^{*})},\\
m_{3}(-\frac{q_{0}^{2}}{z_{n}^{*}})&=\check{e_{n}}\mu_{+,1}(-\frac{q_{0}^{2}}{z_{n}^{*}})
e^{2i\theta_{1}(-\frac{q_{0}^{2}}{z_{n}^{*}})},\qquad
\mu_{-,1}(-\frac{q_{0}^{2}}{z_{n}})=\widetilde{e_{n}}m_{4}
(-\frac{q_{0}^{2}}{z_{n}})e^{-2i\theta_{1}(-\frac{q_{0}^{2}}{z_{n}})}.
\end{align*}
(2)~$\{\zeta_{n}\}_{n=1}^{N_{2}}$ are the second type eigenvalue
\begin{align*}
m_{1}(\zeta_{n})&=f_{n}\mu_{+,2}(\zeta_{n})e^{i(\theta_{2}-\theta_{1})(\zeta_{n})}, \qquad\quad~~~~
\mu_{-,2}(\zeta_{n}^{*})=\hat{f_{n}}m_{2}(\zeta_{n}^{*})e^{i(\theta_{1}-\theta_{2})(\zeta_{n}^{*})},\\
\mu_{-,2}(-\frac{q_{0}^{2}}{\zeta_{n}^{*}})&=\check{f_{n}}m_{3}(-\frac{q_{0}^{2}}{\zeta_{n}^{*}})
e^{-i(\theta_{1}+\theta_{2})(-\frac{q_{0}^{2}}{\zeta_{n}^{*}})},\qquad
m_{4}(-\frac{q_{0}^{2}}{\zeta_{n}})=\widetilde{f_{n}}\mu_{+,2}(-\frac{q_{0}^{2}}{\zeta_{n}})
e^{i(\theta_{1}+\theta_{2})(-\frac{q_{0}^{2}}{\zeta_{n}})}.
\end{align*}
(3)~$\{\omega_{n}\}_{n=1}^{N_{3}}$ are the third type eigenvalue
\begin{align*}
\mu_{-,3}(\omega_{n})&=g_{n}\mu_{+,2}(\omega_{n})e^{i(\theta_{1}+\theta_{2})(\omega_{n})}, \qquad \qquad
\mu_{-,2}(\omega_{n}^{*})=\hat{g_{n}}\mu_{+,3}(\omega_{n}^{*})e^{-i(\theta_{2}+\theta_{1})
(\omega_{n}^{*})},\\
\mu_{-,2}(-\frac{q_{0}^{2}}{\omega_{n}^{*}})&=\check{g_{n}}\mu_{+,1}
(-\frac{q_{0}^{2}}{\omega_{n}^{*}})e^{i(\theta_{1}-\theta_{2})(-\frac{q_{0}^{2}}{\omega_{n}^{*}})},\qquad
\mu_{-,1}(-\frac{q_{0}^{2}}{\omega_{n}})=\widetilde{g_{n}}\mu_{+,2}(-\frac{q_{0}^{2}}{\omega_{n}})
e^{i(\theta_{2}-\theta_{1})(-\frac{q_{0}^{2}}{\omega_{n}})}.
\end{align*}
\end{lem}
\subsection{Asymptotic behavior}
In order to solve the RH problem in the next section, it is necessary to discuss the asymptotic behaviors of the Jost eigenfunction and  the scattering matrix as  $z\rightarrow0$ and $z\rightarrow\infty$. Considering the following formal extensions for $\mu_{+}(x,t,z)$:
\begin{align}\label{22}
\mu_{+}(x,t,z)=\sum_{n=0}^{\infty}\mu_{n}(x,t,z),
\end{align}
where
\begin{align}\label{23}
\mu_{0}(x,t,z)=\Gamma_{+}(z),
\end{align}
\begin{align}\label{24}
\mu_{n+1}(x,t,z)=-\int_{x}^{+\infty}\Gamma_{+}(z)e^{i(x-y)J(z)}\Gamma_{+}^{-1}(z)\Delta Q_{+}(y,t)\mu_{n}(x,t,z)e^{-i(x-y)J(z)}dy.
\end{align}

We will describe the asymptotic behavior of eigenfunctions with \eqref{22}, \eqref{23} and \eqref{24}, namely $z\rightarrow\infty$ and $z\rightarrow0$. As this requires integration by parts, appropriate functional categories must be identified to ensure the effectiveness of the results.
Let us introduce some notations, which are the block diagonal and block off-diagonal terms of $3\times3$ matrix
\begin{align*}
S_{bd}=\left(
         \begin{array}{ccc}
           s_{11} & s_{12} & 0 \\
           s_{21} & s_{22} & 0 \\
           0 & 0 & s_{33} \\
         \end{array}
       \right),
S_{bo}=\left(
         \begin{array}{ccc}
           0 & 0 &  s_{13} \\
           0 & 0 &  s_{23} \\
            s_{31} &  s_{32} & 0 \\
         \end{array}
       \right),
[S_{bd}]_{o}=\left(
               \begin{array}{ccc}
                 0 & s_{12} & 0 \\
                 s_{21} & 0 & 0 \\
                 0 & 0 & 0 \\
               \end{array}
             \right).
\end{align*}
\begin{thm}
The asymptotic behavior of $\mu_{\pm}(x,t,z)$ as $z\rightarrow\infty$ are
\begin{align}
\mu_{\pm}(x,t,z)=\left(
                     \begin{array}{ccc}
                       \frac{\textbf{q}_{\pm}}{q_{0}} &  \frac{\textbf{q}_{\pm}^{\bot}}{q_{0}} & \frac{-i\textbf{q}}{z} \\
                       -\frac{i\textbf{q}^{\dag}q_{\pm}}{zq_{0}} &  -\frac{i\textbf{q}_{\pm}^{\dag}\textbf{q}^{\bot}}{zq_{0}} & 1 \\
                     \end{array}
                   \right)
+\frac{1}{z}(\Gamma_{\pm})_{bd}(E_{1})_{bo}+O(\frac{1}{z^{2}}),\quad z\rightarrow\infty.
\end{align}
Similarly, as $z\rightarrow0$ in the appropriate area of the complex plane,
\begin{align}
\mu_{\pm}(x,t,z)=\left(
                   \begin{array}{ccc}
                     \frac{\mathbf{q}_{\pm}}{q_{0}} & \frac{\mathbf{q}_{\pm}^{\bot}}{q_{0}} &  \frac{i\mathbf{q}_{\pm}}{z} \\
                     \frac{iq_{0}}{z} & 0 & \frac{\textbf{q}_{\pm}^{\dag}\mathbf{q}}{zq_{0}} \\
                   \end{array}
                 \right)
+(\Gamma_{\pm})_{bd}(F_{0})_{bo}+O(1)
,\quad z\rightarrow0.
\end{align}
\end{thm}

We now calculate the asymptotic behavior of the auxiliary eigenfunction $\chi_{j}(x,t,z) (j=1,2,3,4)$. We reviewed the definition of the modified auxiliary eigenfunction \eqref{20}, and combined the above asymptotes with \eqref{25} to get the following results.
\begin{thm}
The asymptotic of $m_{j}$ as $z\rightarrow\infty$
\begin{align}
m_{1}(x,t,z)&=\left(
               \begin{array}{c}
                 -\frac{\mathbf{q}_{-}}{q_{0}} \\
                 \frac{i\mathbf{q}^{\dag}\mathbf{q}_{-}}{zq_{0}} \\
               \end{array}
             \right),\quad
m_{2}(x,t,z)=\left(
               \begin{array}{c}
                 -\frac{\mathbf{q}_{+}}{q_{0}} \\
                 \frac{i\mathbf{q}^{\dag}\mathbf{q}_{+}}{zq_{0}} \\
               \end{array}
             \right),\\
m_{3}(x,t,z)&=\left(
               \begin{array}{c}
                 \frac{i\mathbf{q}_{+}[\mathbf{q}_{-}^{\dag}\mathbf{q}]}{zq_{0}^{2}} \\
                 -\frac{\mathbf{q}_{-}^{\dag}\mathbf{q}_{+}}{q_{0}^{2}} \\
               \end{array}
             \right),~
m_{4}(x,t,z)=\left(
               \begin{array}{c}
                 \frac{i\mathbf{q}_{-}[\mathbf{q}_{+}^{\dag}\mathbf{q}]}{zq_{0}^{2}} \\
                 -\frac{\mathbf{q}_{+}^{\dag}\mathbf{q}_{-}}{q_{0}^{2}} \\
               \end{array}
             \right).
\end{align}
The asymptotic of $m_{j}$ as $z\rightarrow0$
\begin{align}
m_{1}(x,t,z)&=\left(
                \begin{array}{c}
                  -\frac{\mathbf{q}_{-}}{q_{0}^{2}}[\mathbf{q}_{-}^{\dag}\mathbf{q}] \\
                 0 \\
                \end{array}
              \right),~
m_{2}(x,t,z)=\left(
                \begin{array}{c}
                   -\frac{\mathbf{q}_{+}}{q_{0}^{2}}[\mathbf{q}_{+}^{\dag}\mathbf{q}] \\
                 0 \\
                \end{array}
              \right),\\
m_{3}(x,t,z)&=\left(
                \begin{array}{c}
                 0 \\
                 -\frac{i\mathbf{q}_{+}}{z} \\
                \end{array}
              \right),\quad~~~~~
m_{4}(x,t,z)=\left(
                \begin{array}{c}
                 0 \\
                 -\frac{i\mathbf{q}_{-}}{z} \\
                \end{array}
              \right).
\end{align}
\end{thm}

Note that unlike the ZBC case, the off-diagonal terms of the scattering matrix do not all vanish as $z\rightarrow\infty$. However, as we will see, this does not complicate the inverse problem, because all reflection coefficients will still disappear as $z\rightarrow\infty$. For detail see \cite{SIAM-2015}.
\section{Riemann-Hilbert problem}
The starting point of the inverse problem formula is the scattering relationship \eqref{5} and \eqref{5-1}, which will lead to the jump conditions for the RH problem. However, this derivation process is much more complicated than the usual situation we have encountered. The reason is that there are some Jost eigenfunctions that are not analytic, so  \eqref{5} and \eqref{5-1} must be recalculated according to the basic analytic eigenfunction  defined in lemma \ref{21}. In this section, the modified eigenfunction and the solution of the adjoint spectrum problem are used to construct analytic functions in four different regions $\{D_{j}\}_{j=1}^{4}$, so as to construct the corresponding RH problem of the cmKdV equation. It is worth noting that, unlike the $2\times2$ spectral problem, the complex $z-$plane  is divided into four regions $\{D_{j}\}_{j=1}^{4}$, so there are four jump matrices corresponding to the RH problem. For simplicity, we make $D_{+}=D_{1}\cup D_{3}$ and $D_{-}=D_{2}\cup D_{4}$.
\begin{thm}
Define meromorphic functions $M(x,t,z)=M_{j}(x,t,z),z\in D_{j}(j=1,2,3,4)$,
\begin{align*}
M^{+}(x,t,z)=\left\{\begin{aligned}
M_{1}(x,t,z)=\left(\frac{m_{1}}{b_{22}},~~\mu_{+,2},~~\frac{\mu_{-,3}}{a_{33}}\right),\quad~z\in D_{1},\\
M_{3}(x,t,z)=\left(\mu_{+,1},~~\frac{\mu_{-,2}}{a_{22}},~~\frac{m_{3}}{b_{11}}\right),\quad~z\in D_{3},
\end{aligned}\right.
\end{align*}
\begin{align*}
M^{-}(x,t,z)=\left\{\begin{aligned}
M_{2}(x,t,z)=\left(\frac{m_{2}}{b_{33}},~~\frac{\mu_{-,2}}{a_{22}},~~\mu_{+,3}\right),\quad~z\in D_{2},\\
M_{4}(x,t,z)=\left(\frac{\mu_{-,1}}{a_{11}},~~\mu_{+,2},~~\frac{m_{4}}{b_{22}}\right),\quad~z\in D_{4},
\end{aligned}\right.
\end{align*}
where $M_{j}(x,t,z)$ are satifying the jump conditions
\begin{align}
M^{+}(x,t,z)=M^{-}(x,t,z)[I-e^{i\theta(x,t,z)}L(z)e^{-i\theta(x,t,z)}],\quad z\in\Sigma,
\end{align}
where $\Sigma=\Sigma_{1}\cup\Sigma_{2}\cup\Sigma_{3}\cup\Sigma_{4}$, $\Sigma_{j}=\bar{D}_{j}\cap \bar{D}_{(j+1){\text{mod}4}}(j=1,2,3,4)$ and
\begin{align*}
L_{1}&=\left(
        \begin{array}{ccc}
          -\frac{z^{2}}{q_{0}^{2}}\gamma(z)\rho_{3}^{*}(-\frac{q_{0}^{2}}{z^{*}})
          \rho_{3}(-\frac{q_{0}^{2}}{z}) & \frac{z}{iq_{0}}\rho_{3}(-\frac{q_{0}^{2}}{z}) & \rho_{1}(-\frac{q_{0}^{2}}{z})\rho_{3}(-\frac{q_{0}^{2}}{z})-\rho_{2}(-\frac{q_{0}^{2}}{z}) \\
          -\frac{z}{-iq_{0}}\gamma(z)\rho_{3}^{*}(\frac{q_{0}^{2}}{z^{*}}) & 0 & -\frac{iq_{0}}{z}\rho_{1}(-\frac{q_{0}^{2}}{z}) \\
          m_{1} & m_{2} & m_{3} \\
        \end{array}
      \right),~~z\in\Sigma_{1},\\
L_{2}&=\left(
        \begin{array}{ccc}
          0 & 0 & \rho_{2}^{*}(z^{*}) \\
          0 & 0 & 0 \\
          -\rho_{2}^{*}(-\frac{q_{0}^{2}}{z^{*}}) & 0 & -\rho_{2}^{*}(-\frac{q_{0}^{2}}{z^{*}}) \\
        \end{array}
      \right),\qquad\qquad\qquad\qquad\qquad\qquad\qquad\qquad~~ z\in\Sigma_{2},\\
L_{3}&=\left(
        \begin{array}{ccc}
          0 & (-\frac{z}{iq_{0}})\rho_{3}(-\frac{q_{0}^{2}}{z}) & \rho_{2}^{*}(z^{*}) \\
          \rho_{1}(z)+\gamma(z)\rho_{2}(z)\rho_{3}^{*}(z^{*}) & m_{4} & m_{5} \\
          -\rho_{2}(z) & \frac{z}{iq_{0}}\rho_{3}(-\frac{q_{0}^{2}}{z})\rho_{2}(z)-\rho_{3}(z) & -\rho_{2}^{*}({z^{*}})\rho_{2}(z) \\
        \end{array}
      \right),\quad~~ z\in\Sigma_{3},\\
L_{4}&=\left(
        \begin{array}{ccc}
          0 & 0 & -\rho_{2}(-\frac{q_{0}^{2}}{z}) \\
          m_{6} & 1 & m_{7} \\
          \rho_{2}(z) & 0 & \rho_{2}(z)\rho_{2}(-\frac{q_{0}^{2}}{z}) \\
        \end{array}
      \right),\qquad\qquad\qquad\qquad\qquad\qquad\qquad~~~~\qquad z\in\Sigma_{4},
\end{align*}
with
\begin{align*}
m_{1}&=-\rho_{2}^{*}(-\frac{q_{0}^{2}}{z^{*}})-\frac{z^{2}}{q_{0}^{2}}\gamma(z)\rho_{3}^{*}
(-\frac{q_{0}^{2}}{z^{*}})\rho_{3}(-\frac{q_{0}^{2}}{z})\rho_{2}^{*}
(-\frac{q_{0}^{2}}{z^{*}})+\gamma(z)\frac{z}{iq_{0}}\rho_{3}^{*}
(-\frac{q_{0}^{2}}{z^{*}})\rho_{3}(z),\\
m_{2}&=\frac{z}{iq_{0}}\rho_{3}(-\frac{q_{0}^{2}}{z})\rho_{2}^{*}
(-\frac{q_{0}^{2}}{z^{*}})+\rho_{3}(z),\\
m_{3}&=\rho_{1}(-\frac{q_{0}^{2}}{z})\rho_{3}(-\frac{q_{0}^{2}}{z})\rho_{2}^{*}
(-\frac{q_{0}^{2}}{z^{*}})+\frac{iq_{0}}{z}\rho_{1}(-\frac{q_{0}^{2}}{z})\rho_{3}(z)-
\rho_{2}(-\frac{q_{0}^{2}}{z})\rho_{2}^{*}
(-\frac{q_{0}^{2}}{z^{*}}),\\
m_{4}&=-\gamma(z)\rho_{3}^{*}(z^{*})\rho_{3}(z)+\frac{z}{iq_{0}}\rho_{3}(-\frac{q_{0}^{2}}{z})\rho_{1}(z)
+\frac{z}{iq_{0}}\rho_{3}\gamma(z)\rho_{3}^{*}(z^{*})(-\frac{q_{0}^{2}}{z})\rho_{2}(z),\\
m_{5}&=-\gamma(z)\rho_{3}^{*}(z^{*})-\rho_{2}^{*}(z^{*})\rho_{1}(z)-\gamma(z)\rho_{2}^{*}(z^{*})
\rho_{2}(z)\rho_{3}^{*}(z^{*}),\\
m_{6}&=\rho_{1}(z)+\gamma(z)\rho_{3}^{*}(z^{*})\rho_{2}(z)-\frac{z}{iq_{0}}\rho_{3}^{*}
(-\frac{q_{0}^{2}}{z^{*}}),\\
m_{7}&=-\frac{iq_{0}}{z}\rho_{1}(-\frac{q_{0}^{2}}{z})-\gamma(z)\rho_{3}^{*}(z^{*})+\rho_{1}(z)
\rho_{2}(-\frac{q_{0}^{2}}{z})+\gamma(z)\rho_{3}^{*}(z^{*})\rho_{2}(-\frac{q_{0}^{2}}{z})\rho_{2}(z).
\end{align*}
\end{thm}
In order for the above RH problem to recognize the unique solution, appropriate normalization conditions must also be specified. In this case, this condition is provided by the asymptotic behavior of $M^{\pm}$ as $z\rightarrow\infty$ and the contribution of the pole at $0$ to help regularize the  RH problem.  Using  the previously known information, we conclude the following.
\begin{lem}
\begin{align}
M(x,t,z)&=\left(
           \begin{array}{ccc}
             \frac{\mathbf{q}_{+}}{q_{0}} & \frac{\mathbf{q}_{+}^{\bot}}{q_{0}} & 0 \\
             0 & 0 & 1 \\
           \end{array}
         \right)+o(\frac{1}{z}),\qquad z\rightarrow\infty,\\
M(x,t,z)&=\left(
           \begin{array}{ccc}
             0 & 0 & \frac{i\mathbf{q}_{+}}{z} \\
             \frac{iq_{0}}{z} & 0 & 0 \\
           \end{array}
         \right)+o(1),\qquad z\rightarrow 0.
\end{align}
\end{lem}

In order to simplify the calculation of residue conditions, we define the notation $M^{\pm}=(m_{1}^{\pm}, m_{2}^{\pm}, m_{3}^{\pm})$ and $M^{\pm}_{-1,\tau}(x,t)$ denote the residue conditions of $M^{\pm}$ at $z=\tau$. Using the previously known information, we get the following information.
\addtocounter{equation}{1}
\begin{thm}\label{5.2}
The meromorphic functions $M^{\pm}$ satisfy residue conditions under three kinds of zero points
\begin{align}
&M^{+}_{-1,z_{n}}=\left(0,\qquad0,\qquad E_{n}m^{+}_{1}(z_{n})e^{2i\theta_{1}(z_{n})}\right),
\tag{\theequation a}\\
&M^{-}_{-1,z_{n}^{*}}=\left(\widehat{E}_{n}m^{-}_{3}(z_{n}^{*})e^{-2i\theta_{1}(z_{n}^{*})}, \qquad0, \qquad0\right),\tag{\theequation b}\\
&M^{+}_{-1,-\frac{q_{0}^{2}}{z_{n}^{*}}}=\left(0,\qquad0,\qquad\frac{z_{n}^{*}}{iq_{0}}\check{E}_{n}m_{3}^{-}
(z_{n}^{*})e^{2i\theta_{1}(-\frac{q_{0}^{2}}{z_{n}^{*}})}\right),\tag{\theequation c}\\
&M^{-}_{-1,-\frac{q_{0}^{2}}{z_{n}}}=\left(\frac{z_{n}}{iq_{0}}\widetilde{E}_{n}m^{+}_{1}(z_{n})e^{-2i\theta_{1}
(-\frac{q_{0}^{2}}{z_{n}})}, \qquad0, \qquad0\right),\tag{\theequation d}\\
&M^{+}_{-1,\zeta_{n}}=\left(F_{n}m^{+}_{2}(\zeta_{n})e^{i(\theta_{2}-\theta_{1})(\zeta_{n})}, \qquad0, \qquad0\right),\tag{\theequation e}\\
&M^{-}_{-1,\zeta_{n}^{*}}=\left(0,\qquad \hat{F}_{n}m^{-}_{1}(\zeta_{n}^{*})e^{i(\theta_{1}-\theta_{2})(\zeta_{n}^{*})}, \qquad0\right),\tag{\theequation f}\\
&M^{+}_{-1,-\frac{q_{0}^{2}}{\zeta_{n}^{*}}}=\left(0,\qquad\frac{\zeta_{n}^{*}}{iq_{0}}
\check{F}_{n}m^{-}_{1}(\zeta_{n}^{*})e^{-i(\theta_{1}+\theta_{2})(-\frac{q_{0}^{2}}{\zeta_{n}^{*}})}, \qquad0\right),\tag{\theequation g}\\
&M^{-}_{-1,-\frac{q_{0}^{2}}{\zeta_{n}}}=\left(0, \qquad0, \qquad\widetilde{F}_{n}m^{+}_{2}(\zeta_{n})
e^{i(\theta_{1}+\theta_{2})(-\frac{q_{0}^{2}}{\zeta_{n}})}\right),\tag{\theequation h}\\
&M^{+}_{-1,\omega_{n}}=\left(0,\qquad0,\qquad G_{n}m^{+}_{2}(\omega_{n})e^{i(\theta_{1}+\theta_{2})(\omega_{n})}\right),\tag{\theequation i}\\
&M^{-}_{-1,\omega_{n}^{*}}=\left(0,\qquad\hat{G}_{n}m^{-}_{3}(\omega_{n}^{*})e^{-i(\theta_{2}+\theta_{1})
(\omega_{n}^{*})}, \qquad0\right),\tag{\theequation j}\\
&M^{+}_{-1,-\frac{q_{0}^{2}}{\omega_{n}^{*}}}=\left(0,\qquad\frac{\omega^{*}_{n}}{iq_{0}}\check{G}_{n}
m^{-}_{3}(\omega_{n}^{*})e^{i(\theta_{1}-\theta_{2})(-\frac{q_{0}^{2}}{\omega_{n}^{*}})}, \qquad0\right),\tag{\theequation k}\\
&M^{-}_{-1,-\frac{q_{0}^{2}}{\omega_{n}}}=\left(\widetilde{G}_{n}m^{+}_{2}(\omega_{n})e^{i(\theta_{2}-\theta_{1})
(-\frac{q_{0}^{2}}{\omega_{n}})}, \qquad0, \qquad0\right),\tag{\theequation l}
\end{align}
where
\begin{align*}
E_{n}=\frac{e_{n}b_{22}(z_{n})}{a_{33}^{'}(z_{n})},\qquad\hat{E}_{n}=\frac{\hat{e}_{n}}{b_{33}^{'}
(z_{n}^{*})},\qquad\check{E}_{n}=\frac{\check{e}_{n}}{b_{11}^{'}(-\frac{q_{0}^{2}}{z_{n}^{*}})},\qquad
\widetilde{E}_{n}=\frac{\widetilde{e}_{n}b_{22}(z_{n})}{a_{11}^{'}(-\frac{q_{0}^{2}}{z_{n}})},\\
F_{n}=\frac{f_{n}}{b_{22}^{'}(\zeta_{n})},\qquad\hat{F}_{n}=\frac{\hat{f}_{n}b_{33}(\zeta_{n}^{*})}
{a_{22}^{'}(\zeta_{n}^{*})},\qquad\check{F}_{n}=\frac{\check{f}_{n}b_{33}(\zeta_{n}^{*})}
{a_{22}^{'}(-\frac{q_{0}^{2}}{\zeta_{n}^{*}})},\qquad\tilde{F}_{n}=\frac{\tilde{f}_{n}}
{b_{22}^{'}(-\frac{q_{0}^{2}}{\zeta_{n}})},\\
G_{n}=\frac{g_{n}}{a_{33}^{'}(\omega_{n})},\qquad\hat{G}_{n}=\frac{\hat{g}_{n}}{a_{22}^{'}(\omega_{n}^{*})}
,\qquad\check{G}_{n}=\frac{\check{g}_{n}}{a_{22}^{'}(-\frac{q_{0}^{2}}{\omega_{n}^{*}})},\qquad
\widetilde{G}_{n}=\frac{\widetilde{g}_{n}}{a_{11}^{'}(-\frac{q_{0}^{2}}{\omega_{n}})}.
\end{align*}
\end{thm}
\begin{lem}
The elements appear in Theorem \ref{5.2} hold the symmetries
\begin{align*}
\hat{e}_{n}&=\check{e}_{n},\qquad~~ \quad e_{n}=\widetilde{e}_{n},\qquad\qquad\hat{e}_{n}=b_{22}(z_{n}^{*})\bar{e}_{n},\\
\hat{f}_{n}&=\frac{\zeta_{n}^{*}}{iq_{0}}\check{f}_{n},\qquad\widetilde{f}_{n}=\frac{\zeta_{n}}{iq_{0}}f_{n}
,\qquad \bar{f}_{n}=\gamma(\zeta_{n}^{*})a_{33}(\zeta_{n}^{*})\hat{f}_{n},\\
\widetilde{g}_{n}&=\frac{\omega_{n}}{iq_{0}}g_{n},\qquad\check{g}_{n}=\frac{iq_{0}}
{\omega_{n}^{*}}\hat{g}_{n},\qquad \hat{g}_{n}=\frac{\bar{g}_{n}\bar{b}_{22}^{'}(\omega_{n})}
{\bar{\gamma}(\omega_{n})\bar{a}_{33}^{'}(\omega_{n})}.
\end{align*}
\end{lem}
\begin{lem}
The elements appear in Theorem \ref{5.2} hold the symmetries
\begin{align*}
E_{n}&=\frac{z_{n}^{2}}{q_{0}^{2}}\widetilde{E}_{n},\qquad\hat{E}_{n}=\bar{E}_{n},\qquad\qquad\hat{E}_{n}=
\frac{\bar{z}_{n}^{2}}{q_{0}^{2}}\check{E}_{n},\\
F_{n}&=\frac{i\zeta_{n}}{q_{0}}\widetilde{F}_{n},~~~\quad\hat{F}_{n}=\frac{(\zeta_{n}^{*})^{3}}{iq_{0}^{3}}
\check{F}_{n},\qquad \bar{F}_{n}=\gamma(\zeta_{n}^{*})\hat{F}_{n},\\
G_{n}&=\frac{i\omega_{n}}{q_{0}}\widetilde{G}_{n},~~\quad\hat{G}_{n}=\frac{(\omega_{n}^{*})^{3}}
{iq_{0}^{3}}\check{G}_{n},~~\quad \bar{G}_{n}=-\gamma^{*}(\omega_{n})\hat{G}_{n}.
\end{align*}
\end{lem}
\section{Trace formula}
The last work of the inverse problem is to reconstruct the analytic scattering coefficients from the scattering data. Owing to $B=A^{-1}$ and $\det A=\det B=1$, one has
\begin{align}\label{13}
a_{11}b_{11}=\frac{1}{1+\frac{a_{21}b_{12}+a_{31}b_{13}}{a_{11}b_{11}}},~~
a_{22}b_{22}=\frac{1}{1+\frac{a_{12}b_{21}+a_{32}b_{23}}{a_{22}b_{22}}},~~
a_{33}b_{33}=\frac{1}{1+\frac{a_{13}b_{31}+a_{23}b_{32}}{a_{33}b_{33}}}.
\end{align}
Because $a_{22}$ and $b_{22}$ are analytic in the lower half plane and the upper half plane,  respectively, in order to remove the poles generated by the zeros of $a_{22}$ and $b_{22}$, we define two analytic functions
\begin{align}
\beta^{+}(z)=b_{22}(z)e^{i\Delta\theta}\prod_{n=1}^{N_{2}}\frac{z-\bar{\zeta}_{n}}{z-\zeta_{n}}
\frac{z-(-\frac{q_{0}^{2}}{\bar{\zeta}_{n}})}{z-(-\frac{q_{0}^{2}}{\zeta_{n}})}\prod_{n=1}^{N_{3}}
\frac{z-\bar{\omega}_{n}}{z-\omega_{n}}\frac{z-(-\frac{q_{0}^{2}}{\bar{\omega}_{n}})}
{z-(-\frac{q_{0}^{2}}{\omega_{n}})},~~z\in\mathbb{C}^{+},\\
\beta^{-}(z)=\frac{1}{a_{22}}(z)e^{i\Delta\theta}\prod_{n=1}^{N_{2}}\frac{z-\bar{\zeta}_{n}}{z-\zeta_{n}}
\frac{z-(-\frac{q_{0}^{2}}{\bar{\zeta}_{n}})}{z-(-\frac{q_{0}^{2}}{\zeta_{n}})}\prod_{n=1}^{N_{3}}
\frac{z-\bar{\omega}_{n}}{z-\omega_{n}}\frac{z-(-\frac{q_{0}^{2}}{\bar{\omega}_{n}})}
{z-(-\frac{q_{0}^{2}}{\omega_{n}})},~~z\in\mathbb{C}^{-},
\end{align}
where $\Delta\theta=\theta_{+}-\theta_{-}$, $\beta^{\pm}(z)$ are analytic on $\mathbb{C}^{\pm}$, respectively, and there is no zero point on $\mathbb{C}^{\pm}$. Taking \eqref{13} into account and combining with Plemelj formula, we can get
\begin{align}\label{14}
\log b_{11}(z)-\log\frac{1}{a_{11}}&=-\log[1+\rho_{1}(z)\rho_{1}^{*}(z^{*})+\rho_{2}(z)\rho_{2}^{*}(z^{*})],\\
\log b_{22}(z)-\log\frac{1}{a_{22}(z)}&=-\log[1-\frac{z^{2}}{q_{0}^{2}}\rho_{3}(-\frac{q_{0}^{2}}{z})\gamma(z)\rho
_{3}^{*}(-\frac{q_{0}^{2}}{z^{*}})+\gamma(z)\rho_{3}(z)\rho_{3}^{*}(z^{*})],\\
\log a_{33}(z)-\log\frac{1}{b_{33}(z)}&=-\log[1+\rho_{2}(-\frac{q_{0}^{2}}{z})\rho_{2}^{*}(-\frac{q_{0}^{2}}{z^{*}})
-\frac{q_{0}^{2}}{z^{2}}\rho_{1}(-\frac{q_{0}^{2}}{z})\frac{1}{\gamma(z)}\rho_{1}^{*}
(-\frac{q_{0}^{2}}{z^{*}})].
\end{align}
Similarly, we can yield
\begin{align}\label{15}
\log a_{22}(z)-\log \frac{1}{b_{11}}(z)&=\log b_{33}(z)+\log[1-\rho_{2}^{*}(z^{*})\rho_{2}^{*}(-{q_{0}^{2}}{z^{*}})],\\
\log a_{11}(z)-\log \frac{1}{a_{33}}(z)&=\log b_{22}(z)+\log[1-\rho_{2}(z)\rho_{2}(-{q_{0}^{2}}{z})].
\end{align}
Thus, we have
\begin{align}
b_{22}(z)=\exp\left(-i\Delta\theta-\frac{1}{2\pi i}\int_{\mathbb{R}}\frac{J(\zeta)}{\zeta-z}d\zeta\right)\prod_{n=1}^{N_{2}}\frac{z-\zeta_{n}}{z-\bar{\zeta}_{n}}
\frac{z-(-\frac{q_{0}^{2}}{\zeta_{n}})}{z-(-\frac{q_{0}^{2}}{\bar{\zeta}_{n}})}\prod_{n=1}^{N_{3}}
\frac{z-\omega_{n}}{z-\bar{\omega}_{n}}\frac{z-(-\frac{q_{0}^{2}}{\omega_{n}})}
{z-(-\frac{q_{0}^{2}}{\bar{\omega}_{n}})}.
\end{align}
Since $a_{11}(z)$, $b_{11}(z)$, $a_{33}(z)$ and $b_{33}(z)$ are analytic on $D_{4}, D_{3}, D_{1}$ and $D_{2}$, respectively. In order to get a normal RH problem, we first introduce a  sectionally functions, which is analytic on the whole plane, to this end, we define the sectionally functions
\begin{align*}
\dot{\beta}^{+}(z)=\left\{\begin{aligned}&\beta_{1}(z),\quad z\in D_{1},\\
&\beta_{3}(z),\quad z\in D_{3},\end{aligned}\right.
\qquad
\dot{\beta}^{-}(z)=\left\{\begin{aligned}&\beta_{2}(z),\quad z\in D_{2},\\
&\beta_{4}(z),\quad z\in D_{4},\end{aligned}\right.
\end{align*}
where
\begin{align*}
\beta_{1}(z)=\frac{a_{33}(z)}{s_{1}(z)}\prod_{n=1}^{N_{1}}\frac{z-z_{n}}{z-\bar{z}_{n}}
\frac{z-(-\frac{q_{0}^{2}}{\bar{z}_{n}})}{z-(-\frac{q_{0}^{2}}{z_{n}})}\prod_{n=1}^{N_{2}}
\frac{z-\bar{\zeta}_{n}}{z-\zeta_{n}}\frac{z-(-\frac{q_{0}^{2}}{\zeta_{n}})}
{z-(-\frac{q_{0}^{2}}{\bar{\zeta}_{n}})},
\end{align*}
\begin{align*}
\beta_{2}(z)=\frac{s_{1}^{*}(z^{*})}{b_{33}(z)}\prod_{n=1}^{N_{1}}\frac{z-z_{n}}{z-\bar{z}_{n}}
\frac{z-(-\frac{q_{0}^{2}}{\bar{z}_{n}})}{z-(-\frac{q_{0}^{2}}{z_{n}})}\prod_{n=1}^{N_{2}}
\frac{z-\bar{\zeta}_{n}}{z-\zeta_{n}}\frac{z-(-\frac{q_{0}^{2}}{\zeta_{n}})}
{z-(-\frac{q_{0}^{2}}{\bar{\zeta}_{n}})},\\
\beta_{3}(z)=\frac{b_{11}(z)}{s_{2}^{*}(z^{*})}\prod_{n=1}^{N_{1}}\frac{z-z_{n}}{z-\bar{z}_{n}}
\frac{z-(-\frac{q_{0}^{2}}{\bar{z}_{n}})}{z-(-\frac{q_{0}^{2}}{z_{n}})}\prod_{n=1}^{N_{2}}
\frac{z-\bar{\zeta}_{n}}{z-\zeta_{n}}\frac{z-(-\frac{q_{0}^{2}}{\zeta_{n}})}
{z-(-\frac{q_{0}^{2}}{\bar{\zeta}_{n}})},\\
\beta_{4}(z)=\frac{s_{2}(z)}{a_{11}(z)}\prod_{n=1}^{N_{1}}\frac{z-z_{n}}{z-\bar{z}_{n}}
\frac{z-(-\frac{q_{0}^{2}}{\bar{z}_{n}})}{z-(-\frac{q_{0}^{2}}{z_{n}})}\prod_{n=1}^{N_{2}}
\frac{z-\bar{\zeta}_{n}}{z-\zeta_{n}}\frac{z-(-\frac{q_{0}^{2}}{\zeta_{n}})}
{z-(-\frac{q_{0}^{2}}{\bar{\zeta}_{n}})},
\end{align*}
with
\begin{align*}
s_{1}(z)=\prod_{n=1}^{N_{1}}\frac{z-z_{n}}{z-\bar{z}_{n}}\prod_{n=1}^{N_{3}}
\frac{z-\omega_{n}}{z-\bar{\omega}_{n}},\qquad
s_{2}(z)=\prod_{n=1}^{N_{1}}\frac{z-(-\frac{q_{0}^{2}}{z_{n}})}{z-(-\frac{q_{0}^{2}}{\bar{z}_{n}})}
\prod_{n=1}^{N_{3}}\frac{z-(-\frac{q_{0}^{2}}{\omega_{n}})}{z-(-\frac{q_{0}^{2}}{\bar{\omega}_{n}})}.
\end{align*}
The formulas of \eqref{14} can be written as
\begin{align*}
\log\beta_{3}(z)-\log\beta_{4}(z)&=J_{3}(z)=-\log[1+\rho_{1}(z)\rho_{1}^{*}(z^{*})+\rho_{2}(z)
\rho_{2}^{*}(z^{*})],\\
\log\beta_{1}(z)-\log\beta_{2}(z)&=J_{1}(z)=-\log[1+\rho_{2}(-\frac{q_{0}^{2}}{z})\rho_{2}^{*}(-\frac{q_{0}^{2}}{z^{*}})
-\frac{q_{0}^{2}}{z^{2}}\rho_{1}(-\frac{q_{0}^{2}}{z})\frac{1}{\gamma(z)}\rho_{1}^{*}
(-\frac{q_{0}^{2}}{z^{*}})].
\end{align*}
From the formulas of \eqref{15} and the expressions of $\beta_{j}$, we have
\begin{align*}
\log\beta_{3}(z)-\log\beta_{2}(z)&=J_{2}(z)=\frac{1}{2\pi i}\int_{\mathbb{R}}\frac{J_{0}(\zeta)}{\zeta-z}d\zeta-\log[1-\rho_{2}^{*}(z^{*})\rho_{2}^{*}
(-\frac{q_{0}^{2}}{z^{*}})],\\
\log\beta_{1}(z)-\log\beta_{4}(z)&=J_{4}(z)=\frac{1}{2\pi i}\int_{\mathbb{R}}\frac{J_{0}(\zeta)}{\zeta-z}d\zeta-\log[1-\rho_{2}(z)\rho_{2}(-\frac{q_{0}^{2}}{z})],
\end{align*}
where
\begin{align*}
J_{0}(z)=\log[1+\gamma(z)\rho_{3}(z)\rho_{3}^{*}(z^{*})+\gamma(z)\rho_{3}^{*}(-\frac{q_{0}^{3}}{z^{*}})
\rho_{3}(-\frac{q_{0}^{3}}{z})].
\end{align*}
The jump condition of the analytic function $\dot{\beta}^{\pm}(z)$ can be obtained
\begin{align}
\log\dot{\beta}^{+}(z)-\log\dot{\beta}^{-}(z)=J_{i}(z),\quad J_{i}(z)=\Sigma_{i}~(i=1,2,3,4),
\end{align}
Obviously, one has
\begin{align*}
\log\dot{\beta}(z)=\frac{1}{2\pi i}\int_{\Sigma}\frac{J(\xi)}{\xi-z}d\xi, \quad z\in \mathbb{C}\setminus\Sigma.
\end{align*}
Taking logarithm on both sides of $z\in D_{1}$ to derive
\begin{align}
a_{33}(z)=\exp\left(\frac{1}{2\pi i}\int_{\Sigma}\frac{J(\xi)}{\xi-z}d\xi\right)\left(\prod_{n=1}^{N_{1}}
\frac{z-(-\frac{q_{0}^{2}}{z_{n}})}{z-(-\frac{q_{0}^{2}}{\bar{z}_{n}})}\prod_{n=1}^{N_{2}}
\frac{z-\zeta_{n}}{z-\bar{\zeta}_{n}}\frac{z-(-\frac{q_{0}^{2}}{\bar{\zeta}_{n}})}
{z-(-\frac{q_{0}^{2}}{\zeta_{n}})}\prod_{n=1}^{N_{3}}\frac{z-\omega_{n}}{z-\bar{\omega}_{n}}
\right).
\end{align}
Finally, we notice that the whole scattering matrix can be reconstructed by using the trace formula and the reflection coefficients. The expression of asymptotic phase difference of $\Delta\theta=\theta_{+}-\theta_{-}$ is given
\begin{align}
\Delta\theta=\frac{1}{2\pi}\int_{\Sigma}\frac{J(\xi)}{\xi-z}d\xi+2\sum_{n=1}^{N_{1}}arg z_{n}-
4\sum_{n=1}^{N_{2}}arg\xi_{n}-2\sum_{n=1}^{N_{3}}arg \omega_{n}.
\end{align}
\section{Solution of Riemann-Hilbert problem}
The RH problem defined in the previous section includes finding a piecewise meromorphic matrix $M^{\pm}$, which satisfies the jump conditions, asymptotic properties and residue conditions. The solution of this RH problem can be expressed by a mixed system of algebraic integral equations, which is obtained by subtracting the asymptotic behavior of infinity, then regularizing and applying Cauchy operator. In this section, we transform the original RH problem into the regularized RH problem via removing the asymptotics and poles. The Cramer's rule is used to solve the equations, and the expression of the solution is obtained.
\begin{thm}
Let $\tau$ be the set of all discrete spectral points, and by removing the asymptoticity and pole points, the jumping condition can be written as
\begin{align}
\begin{split}
M^{+}-M_{\infty}-M_{o}-\sum_{n=1}^{N}\frac{M_{-1,\tau_{n}}^{+}}{z-\tau_{n}}-\sum_{n=1}^{N}
\frac{M_{-1,-\frac{q_{0}^{2}}{\tau_{n}}}^{+}}{z-\frac{q_{0}^{2}}{\tau_{n}}}-
\sum_{n=1}^{N}\frac{M_{-1,\tau_{n}^{*}}^{-}}{z-\tau_{n}^{*}}-\sum_{n=1}^{N}
\frac{M_{-1,-\frac{q_{0}^{2}}{\tau_{n}^{*}}}^{-}}{z-\frac{q_{0}^{2}}{\tau_{n}^{*}}}\\=
M^{-}-M_{\infty}-M_{o}-\sum_{n=1}^{N}\frac{M_{-1,\tau_{n}}^{+}}{z-\tau_{n}}-\sum_{n=1}^{N}
\frac{M_{-1,-\frac{q_{0}^{2}}{\tau_{n}}}^{+}}{z-\frac{q_{0}^{2}}{\tau_{n}}}-
\sum_{n=1}^{N}\frac{M_{-1,\tau_{n}^{*}}^{-}}{z-\tau_{n}^{*}}-\sum_{n=1}^{N}
\frac{M_{-1,-\frac{q_{0}^{2}}{\tau_{n}^{*}}}^{-}}{z-\frac{q_{0}^{2}}{\tau_{n}^{*}}}
\end{split}
\end{align}
where $N=N_{1}+N_{2}+N_{3}$.
\end{thm}
The modified engenfunctions under the residue conditions can be given as follows,
\begin{align}
\begin{split}
m_{1}^{\pm}(z)=&\left(
              \begin{array}{c}
                \frac{\textbf{q}_{\pm}}{q_{0}} \\
                \frac{iq_{0}}{z} \\
              \end{array}
            \right)
+\sum_{n=1}^{N_{1}}\frac{\widehat{E}_{n}e^{-2i\theta_{1}(z_{n}^{*})}}{z-z_{n}^{*}}m^{-}_{3}(z_{n}^{*})+
\sum_{n=1}^{N_{1}}\frac{z}{iq_{0}}\frac{\widetilde{E}_{n}e^{-2i\theta_{1}
(-\frac{q_{0}^{2}}{z_{n}})}}{z-(-\frac{q_{0}^{2}}{z_{n}})}m^{+}_{1}(z_{n})+\\
&\sum_{n=1}^{N_{2}}\frac{F_{n}(\zeta_{n})e^{i(\theta_{2}-\theta_{1})(\zeta_{n})}}
{z-\zeta_{n}}m^{+}_{2}(\zeta_{n})+\sum_{n=1}^{N_{3}}\frac
{\widetilde{G}_{n}e^{i(\theta_{2}-\theta_{1})}}
{z-(-\frac{q_{0}^{2}}{\omega_{n}})}m^{+}_{2}(\omega_{n})-\frac{1}{2\pi i}\int_{\Sigma}\frac{M^{-}\widehat{L}(\xi)}{\xi-z}d\xi,
\end{split}
\end{align}
\begin{align}
\begin{split}
m_{2}^{\pm}(z)=&\left(
              \begin{array}{c}
                \frac{\textbf{q}_{\pm}^{\bot}}{q_{0}} \\
                \frac{iq_{0}}{z} \\
              \end{array}
            \right)
+\sum_{n=1}^{N_{2}}\left(\frac{\hat{F}_{n}e^{i(\theta_{1}-\theta_{2})(\zeta_{n}^{*})}}{z-\zeta_{n}^{*}}
+\frac{\zeta_{n}^{*}}{iq_{0}}\frac{\check{F}_{n}e^{-i(\theta_{1}+\theta_{2})
(-\frac{q_{0}^{2}}{\zeta_{n}^{*}})}}{z-(-\frac{q_{0}^{2}}{\zeta_{n}^{*}})}\right)m^{-}_{1}(\zeta_{n}^{*})
\\&+\sum_{n=1}^{N_{3}}\left(\frac{\widehat{G}_{3}(\omega_{n}^{*})e^{-i(\theta_{1}+\theta_{2})
(\omega_{n}^{*})}}{z-\omega_{n}^{*}}+\frac{\omega^{*}_{n}}{iq_{0}}\frac{\check{G}_{n}
e^{i(\theta_{1}-\theta_{2})(-\frac{q_{0}^{2}}{\omega_{n}^{*}})}}{z-(-\frac{q_{0}^{2}}{\omega_{n}^{*}})}\right)
m^{-}_{3}(\omega_{n}^{*})-\frac{1}{2\pi i}\int_{\Sigma}\frac{M^{-}\widehat{L}(\xi)}{\xi-z}d\xi,\\
m_{3}^{\pm}(z)=&\left(
              \begin{array}{c}
                -\frac{i\textbf{q}}{z} \\
                1 \\
              \end{array}
            \right)
+\sum_{n=1}^{N_{1}}\frac{E_{n}e^{2i\theta_{1}(z_{n})}}{z-z_{n}}m^{+}_{1}(z_{n})+\sum_{n=1}^{N_{1}}
\frac{z_{n}^{*}}{iq_{0}}\frac{\check{E}_{n}e^{2i\theta_{1}(-\frac{q_{0}^{2}}
{z_{n}^{*}})}}{z-(-\frac{q_{0}^{2}}{z_{n}^{*}})}m_{3}^{-}(z_{n}^{*})\\&+\sum_{n=1}^{N_{2}}
\frac{\widetilde{F}_{n}e^{i(\theta_{1}+\theta_{2})(-\frac{q_{0}^{2}}{\zeta_{n}})}}
{z-(-\frac{q_{0}^{2}}{\zeta_{n}})}m^{+}_{2}(\zeta_{n})+\sum_{n=1}^{N_{3}}\frac
{G_{n}e^{i(\theta_{1}+\theta_{2})(\omega_{n})}}{z-\omega_{n}}m^{+}_{2}(\omega_{n})
-\frac{1}{2\pi i}\int_{\Sigma}\frac{M^{-}\widehat{L}(\xi)}{\xi-z}d\xi.
\end{split}
\end{align}

Generally, once the solution of the RH problem is obtained, the asymptotic behavior of the eigenfunction can be compared with that of the direct scattering problem, and according to the norm constant and the scattering coefficient to reconstruct the potential.
\begin{thm}\label{7.2}
Taking $M=M_{1}^{+}$ and comparing the $1,3$ element , the reconstruction formula for the potential can be derived
\begin{align}
\begin{split}
q=q_{+}+&i\sum_{n=1}^{N_{1}}E_{n}e^{2i\theta_{1}(z_{n})}m^{+}_{11}(z_{n})+i\sum_{n=1}^{N_{1}}
\frac{z_{n}^{*}}{iq_{0}}\check{E}_{n}m_{13}^{-}(z_{n}^{*})e^{2i\theta_{1}(-\frac{q_{0}^{2}}
{z_{n}^{*}})}\\+&i\sum_{n=1}^{N_{2}}\widetilde{F}_{n}m^{+}_{12}(\zeta_{n})
e^{i(\theta_{1}+\theta_{2})(-\frac{q_{0}^{2}}{\zeta_{n}})}+i\sum_{n=1}^{N_{3}}G_{n}m^{+}_{12}
(\omega_{n})e^{i(\theta_{1}+\theta_{2})(\omega_{n})}.
\end{split}
\end{align}
\end{thm}
In what follows, the purpose is to determine the unknowns elements $m^{+}_{11}(z_{n})$, $m_{13}^{-}(z_{n}^{*})$, $m^{+}_{12}(\zeta_{n})$ and $m^{+}_{12}
(\omega_{n})$ appear in Therom\ref{7.2},  so one has
\begin{align}\label{26}
\begin{split}
m_{11}^{+}(z_{i^{'}})=&\frac{u_{+}}{q_{0}}
+\sum_{n=1}^{N_{1}}\frac{\widehat{E}_{n}e^{-2i\theta_{1}(z_{n}^{*})}}{z_{i^{'}}-z_{n}^{*}}m^{-}_{13}
(z_{n}^{*})+\sum_{n=1}^{N_{1}}\frac{z}{iq_{0}}\frac{\widetilde{E}_{n}e^{-2i\theta_{1}
(-\frac{q_{0}^{2}}{z_{n}})}}{z_{i^{'}}-(-\frac{q_{0}^{2}}{z_{n}})}m^{+}_{11}(z_{n})\\
&+\sum_{n=1}^{N_{2}}\frac{F_{n}(\zeta_{n})e^{i(\theta_{2}-\theta_{1})(\zeta_{n})}}
{z_{i^{'}}-\zeta_{n}}m^{+}_{12}(\zeta_{n})+\sum_{n=1}^{N_{3}}\frac
{\widetilde{G}_{n}e^{i(\theta_{2}-\theta_{1})(-\frac{q_{0}^{2}}{\omega_{n}})}}
{z_{i^{'}}-(-\frac{q_{0}^{2}}{\omega_{n}})}m^{+}_{12}(\omega_{n}),\\
m_{11}^{-}(\zeta_{j^{'}}^{*})=&\frac{u_{+}}{q_{0}}
+\sum_{n=1}^{N_{1}}\frac{\widehat{E}_{n}e^{-2i\theta_{1}(z_{n}^{*})}}
{\zeta_{j^{'}}^{*}-z_{n}^{*}}m^{-}_{13}(z_{n}^{*})+
\sum_{n=1}^{N_{1}}\frac{z}{iq_{0}}\frac{\widetilde{E}_{n}e^{-2i\theta_{1}
(-\frac{q_{0}^{2}}{z_{n}})}}{\zeta_{j^{'}}^{*}-(-\frac{q_{0}^{2}}{z_{n}})}m^{+}_{11}(z_{n})\\
&+\sum_{n=1}^{N_{2}}\frac{F_{n}(\zeta_{n})e^{i(\theta_{2}-\theta_{1})(\zeta_{n})}}
{\zeta_{j^{'}}^{*}-\zeta_{n}}m^{+}_{12}(\zeta_{n})+\sum_{n=1}^{N_{3}}\frac
{\widetilde{G}_{n}e^{i(\theta_{2}-\theta_{1})(-\frac{q_{0}^{2}}{\omega_{n}})}}
{\zeta_{j^{'}}^{*}-(-\frac{q_{0}^{2}}{\omega_{n}})}m^{+}_{12}(\omega_{n}),
\end{split}
\end{align}
\begin{align}
\begin{split}
m_{12}^{+}(\zeta_{j^{'}})=&\frac{v_{+}}{q_{0}}
+\sum_{n=1}^{N_{2}}\left(\frac{\hat{F}_{n}e^{i(\theta_{1}-\theta_{2})(\zeta_{n}^{*})}}
{\zeta_{j^{'}}-\zeta_{n}^{*}}
+\frac{\zeta_{n}^{*}}{iq_{0}}\frac{\check{F}_{n}e^{-i(\theta_{1}+\theta_{2})
(-\frac{q_{0}^{2}}{\zeta_{n}^{*}})}}{\zeta_{j^{'}}-(-\frac{q_{0}^{2}}{\zeta_{n}^{*}})}
\right)m^{-}_{11}(\zeta_{n}^{*})\\&+\sum_{n=1}^{N_{3}}\left(\frac{\widehat{G}_{n}(\omega_{n}^{*})e^{-i(\theta_{1}+\theta_{2})
(\omega_{n}^{*})}}{\zeta_{j^{'}}-\omega_{n}^{*}}+\frac{\omega^{*}_{n}}{iq_{0}}\frac{\check{G}_{n}
e^{i(\theta_{1}-\theta_{2})(-\frac{q_{0}^{2}}{\omega_{n}^{*}})}}{\zeta_{j^{'}}-(-\frac{q_{0}^{2}}{\omega_{n}^{*}})}\right)
m^{-}_{13}(\omega_{n}^{*}),
\end{split}
\end{align}
\begin{align}
\begin{split}
m_{12}^{+}(\omega_{\ell^{'}})=&\frac{v_{+}}{q_{0}}
+\sum_{n=1}^{N_{2}}\left(\frac{\hat{F}_{n}e^{i(\theta_{1}-\theta_{2})(\zeta_{n}^{*})}}
{\omega_{\ell^{'}}-\zeta_{n}^{*}}+\frac{\zeta_{n}^{*}}{iq_{0}}\frac
{\check{F}_{n}e^{-i(\theta_{1}+\theta_{2})(-\frac{q_{0}^{2}}{\zeta_{n}^{*}})}}
{\omega_{\ell^{'}}-(-\frac{q_{0}^{2}}{\zeta_{n}^{*}})}\right)m^{-}_{11}(\zeta_{n}^{*})
\\&+\sum_{n=1}^{N_{3}}\left(\frac{\widehat{G}_{n}(\omega_{n}^{*})e^{-i(\theta_{1}+\theta_{2})
(\omega_{n}^{*})}}{\omega_{\ell^{'}}-\omega_{n}^{*}}+\frac{\omega^{*}_{n}}{iq_{0}}\frac{\check{G}_{n}
e^{i(\theta_{1}-\theta_{2})(-\frac{q_{0}^{2}}{\omega_{n}^{*}})}}{\omega_{\ell^{'}}-(-\frac{q_{0}^{2}}{\omega_{n}^{*}})}\right)
m^{-}_{13}(\omega_{n}^{*}),
\end{split}
\end{align}
\begin{align}\label{27}
\begin{split}
m_{13}^{-}(z_{i^{'}}^{*})=&-\frac{iu_{+}}{q_{0}}
+\sum_{n=1}^{N_{1}}\frac{E_{n}e^{2i\theta_{1}(z_{n})}}{z_{i^{'}}^{*}-z_{n}}m^{+}_{11}(z_{n})+\sum_{n=1}^{N_{1}}
\frac{z_{n}^{*}}{iq_{0}}\frac{\check{E}_{n}e^{2i\theta_{1}(-\frac{q_{0}^{2}}
{z_{n}^{*}})}}{z_{i^{'}}^{*}-(-\frac{q_{0}^{2}}{z_{n}^{*}})}m_{13}^{-}(z_{n}^{*})\\&+\sum_{n=1}^{N_{2}}
\frac{\widetilde{F}_{n}e^{i(\theta_{1}+\theta_{2})(-\frac{q_{0}^{2}}{\zeta_{n}})}}
{z_{i^{'}}^{*}-(-\frac{q_{0}^{2}}{\zeta_{n}})}m^{+}_{12}(\zeta_{n})+\sum_{n=1}^{N_{3}}\frac
{G_{n}e^{i(\theta_{1}+\theta_{2})(\omega_{n})}}{z_{i^{'}}^{*}-\omega_{n}}m^{+}_{12}(\omega_{n}),
\end{split}
\end{align}
\begin{align}
\begin{split}
m_{13}^{+}(\omega_{\ell^{'}}^{*})=&-\frac{iu_{+}}{q_{0}}
+\sum_{n=1}^{N_{1}}\frac{E_{n}e^{2i\theta_{1}(z_{n})}}{\omega_{\ell^{'}}^{*}-z_{n}}m^{+}_{11}(z_{n})+\sum_{n=1}^{N_{1}}
\frac{z_{n}^{*}}{iq_{0}}\frac{\check{E}_{n}e^{2i\theta_{1}(-\frac{q_{0}^{2}}
{z_{n}^{*}})}}{\omega_{\ell^{'}}^{*}-(-\frac{q_{0}^{2}}{z_{n}^{*}})}m_{13}^{-}(z_{n}^{*})\\&+\sum_{n=1}^{N_{2}}
\frac{\widetilde{F}_{n}e^{i(\theta_{1}+\theta_{2})(-\frac{q_{0}^{2}}{\zeta_{n}})}}
{\omega_{\ell^{'}}^{*}-(-\frac{q_{0}^{2}}{\zeta_{n}})}m^{+}_{12}(\zeta_{n})+\sum_{n=1}^{N_{3}}\frac
{G_{n}e^{i(\theta_{1}+\theta_{2})(\omega_{n})}}{z_{i^{'}}^{*}-\omega_{n}}m^{+}_{12}(\omega_{n}).
\end{split}
\end{align}
In order to simplify the following operations, we introduce some new notations
\begin{align*}
&\Delta_{n}^{(1)}=\frac{\hat{F}_{n}e^{i(\theta_{1}-\theta_{2})(\zeta_{n}^{*})}}
{z-\zeta_{n}^{*}}+\frac{\zeta_{n}^{*}}{iq_{0}}
\frac{\check{F}_{n}e^{-i(\theta_{1}+\theta_{2})(-\frac{q_{0}^{2}}{\zeta_{n}^{*}})}}
{z-(-\frac{q_{0}^{2}}{\zeta_{n}^{*}})},\\
&\Delta_{n}^{(2)}=\frac{\hat{G}_{n}e^{-i(\theta_{2}+\theta_{1})(\omega_{n}^{*})}}{z-\omega_{n}^{*}}+
\frac{\omega^{*}_{n}}{iq_{0}}\frac{\check{G}_{n}e^{i(\theta_{1}-\theta_{2})
(-\frac{q_{0}^{2}}{\omega_{n}^{*}})}}{z-(-\frac{q_{0}^{2}}{\omega_{n}^{*}})},\\
&\Delta_{n}^{(3)}=\frac{z}{iq_{0}}\frac{\widetilde{E}_{n}e^{-2i\theta_{1}
(-\frac{q_{0}^{2}}{z_{n}})}}{z-(-\frac{q_{0}^{2}}{z_{n}})},~~
\Delta_{n}^{(4)}=\frac{\widehat{E}_{n}e^{-2i\theta_{1}(z_{n}^{*})}}{z-z_{n}^{*}},~~
\Delta_{n}^{(5)}=\frac{F_{n}e^{i(\theta_{2}-\theta_{1})(\zeta_{n})}}{z-\zeta_{n}},\\
&\Delta_{n}^{(6)}=\frac{\widetilde{G}_{n}e^{i(\theta_{2}-\theta_{1})
(-\frac{q_{0}^{2}}{\omega_{n}})}}{z-(-\frac{q_{0}^{2}}{\omega_{n}})},~~
\Delta_{n}^{(7)}=\frac{E_{n}e^{2i\theta_{1}(z_{n})}}{z-z_{n}},~~
\Delta_{n}^{(8)}=\frac{z_{n}^{*}}{iq_{0}}\frac{\check{E}_{n}
e^{2i\theta_{1}(-\frac{q_{0}^{2}}{z_{n}^{*}})}}{{z-(-\frac{q_{0}^{2}}{z_{n}^{*}})}},
\end{align*}
\begin{align*}
\Delta_{n}^{(9)}=\frac{\widetilde{F}_{n}e^{i(\theta_{1}+\theta_{2})(-\frac{q_{0}^{2}}{\zeta_{n}})}}
{z-(-\frac{q_{0}^{2}}{\zeta_{n}})},~~
\Delta_{n}^{(10)}=\frac{G_{n}e^{i(\theta_{1}+\theta_{2})(\omega_{n})}}{z-\omega_{n}}.
\end{align*}
Substituting $m_{11}^{-}(\xi_{n}^{*})$ and $m_{13}^{-}(\omega_{n}^{*})$ into $m_{12}^{+}(\xi_{j}^{'})$ and $m_{12}^{+}(\omega_{j})$, one has
\begin{align}\label{28}
\begin{split}
m_{12}^{+}(z)&=\frac{v_{+}}{q_{0}}-\sum_{n=1}^{N_{2}}\Delta_{n}^{(1)}(z)\frac{u_{+}}{q_{0}}-
\sum_{n=1}^{N_{3}}\Delta_{n}^{(2)}(z)\frac{iu_{+}}{q_{0}}+\sum_{n=1}^{N_{2}}\sum_{n^{'}=1}^{N_{1}}
\Delta_{n}^{(1)}(z)\Delta_{n^{'}}^{(4)}(\xi_{n}^{*})m_{13}^{-}(z_{n^{'}}^{*})\\&+
\sum_{n=1}^{N_{2}}\sum_{n^{'}=1}^{N_{1}}\frac{z}{iq_{0}}\Delta_{n}^{(1)}(z)\Delta_{n^{'}}^{(3)}(\xi_{n}^{*})
m_{11}^{+}(z_{n^{'}}^{*})+\sum_{n=1}^{N_{2}}\sum_{n^{'}=1}^{N_{2}}
\Delta_{n}^{(1)}(z)\Delta_{n^{'}}^{(5)}(\xi_{n}^{*})m_{12}^{+}(\xi_{n})
\\&+\sum_{n=1}^{N_{2}}
\sum_{n^{'}=1}^{N_{3}}\Delta_{n}^{(1)}(z)\Delta_{n^{'}}^{(6)}(\xi_{n}^{*})m_{12}^{+}(\omega_{n^{'}}^{*})
+\sum_{n=1}^{N_{3}}\sum_{n^{'}=1}^{N_{1}}
\Delta_{n}^{(2)}(z)\Delta_{n^{'}}^{(7)}(\omega_{n}^{*})m_{11}^{+}(z_{n^{'}})\\&+
\sum_{n=1}^{N_{3}}\sum_{n^{'}=1}^{N_{1}}\frac{z_{n}^{*}}{iq_{0}}
\Delta_{n}^{(2)}(z)\Delta_{n^{'}}^{(8)}(\omega_{n}^{*})m_{13}^{-}(z_{n^{'}}^{*})
+\sum_{n=1}^{N_{2}}\sum_{n^{'}=1}^{N_{2}}
\Delta_{n}^{(2)}(z)\Delta_{n^{'}}^{(9)}(\omega_{n}^{*})m_{12}^{+}(\xi_{n^{'}})
\\&+\sum_{n=1}^{N_{2}}\sum_{n^{'}=1}^{N_{3}}
\Delta_{n}^{(2)}(z)\Delta_{n^{'}}^{(10)}(\omega_{n}^{*})m_{12}^{+}(\omega_{n^{'}}),
\end{split}
\end{align}
where $z=\xi_{n^{'}}$ and $z=\omega_{n^{'}}$. To this end, combining with \eqref{26}, \eqref{27} and \eqref{28}, one obtains a linear  system. In what follows, to get the brief expression of the reflectionless potentials, we would like to define the vector $\mathbf{x}=(x_{1}, x_{2},\cdots, x_{2N_{1}+N_{2}+N_{3}})$,
\begin{align}
\begin{split}
\mathbf{x_{k}}=\left\{\begin{aligned}
&m_{11}^{+}(z_{k}),\qquad k=1,\cdots,N_{1},\\
&m_{13}^{-}(z_{k-N_{1}}^{*}),\qquad k=N_{1}+1,\cdots,2N_{1},\\
&m_{12}^{+}(\xi_{k-2N_{1}}),\qquad k=2N_{1}+1,\cdots,2N_{1}+N_{2},\\
&m_{12}^{+}(\omega_{k-2N_{1}-N_{2}}),\qquad k=2N_{1}+N_{2}+1,\cdots,2N_{1}+N_{2}+N_{3}.
\end{aligned}\right.
\end{split}
\end{align}
\begin{align}
\begin{split}
\textbf{b}_{k}=\left\{\begin{aligned}
&\frac{u_{+}}{q_{0}},\qquad k=1,\cdots,N_{1},\\
&-\frac{iu_{+}}{q_{0}},\qquad k=N_{1}+1,\cdots,2N_{1},\\
&\frac{v_{+}}{q_{0}}-\sum_{n=1}^{N_{2}}\Delta_{n}^{(1)}(\xi_{k-2N_{1}})\frac{u_{+}}{q_{0}}-
\sum_{n=1}^{N_{3}}\Delta_{n}^{(2)}(\xi_{k-2N_{1}})\frac{iu_{+}}{q_{0}},\qquad k=2N_{1}+1,\cdots,2N_{1}+N_{2},\\
&\frac{v_{+}}{q_{0}}-\sum_{n=1}^{N_{2}}\Delta_{n}^{(1)}(\omega_{k-2N_{1}-2N_{2}})\frac{u_{+}}{q_{0}}-
\sum_{n=1}^{N_{3}}\Delta_{n}^{(2)}(\omega_{k-2N_{1}-2N_{2}})\frac{iu_{+}}{q_{0}}, k=2N_{1}+N_{2}+1,\cdots,2N_{1}+N_{2}N_{3}.
\end{aligned}\right.
\end{split}
\end{align}
\begin{align}
\begin{split}
y_{k}=\left\{\begin{aligned}
&-iE_{k}e^{2i\theta_{1}(z_{k})},\qquad k=1,\cdots,N_{1},\\
&-\frac{z_{k-N_{1}}^{*}}{q_{0}}\check{E}_{k-N_{1}}e^{2i\theta_{1}(\frac{-q_{0}^{2}}{z_{k-N_{1}}^{*}})},\qquad k=N_{1}+1,\cdots,2N_{1},\\
&-i\widetilde{F}_{k-2N_{1}}e^{i(\theta_{1}+\theta_{2})(\frac{-q_{0}^{2}}{\zeta_{k-2N_{1}}})},\qquad k=2N_{1}+1,\cdots,2N_{1}+N_{2},\\
&-iG_{k-2N_{1}-N_{2}}e^{i(\theta_{1}+\theta_{2})(\omega_{k-2N_{1}-N_{2}})},
k=2N_{1}+N_{2}+1,\cdots,2N_{1}+N_{2}+N_{3}.
\end{aligned}\right.
\end{split}
\end{align}
The matrix $F$ can be defined as follows:\\
For $i,j=1, \ldots, N_{1}$, we have
\begin{align}
F_{ij}=\Delta_{j}^{(3)}(z_{i}).
\end{align}
For $i=1, \ldots, N_{1}$, $j=N_{1}+1, \ldots, 2N_{1}$, we have
\begin{align}
F_{ij}=\Delta_{j-N_{1}}^{(4)}(z_{i}).
\end{align}
For $i=1, \ldots, N_{1}$, $j=2N_{1}+1, \ldots, N_{2}$, we have
\begin{align}
F_{ij}=\Delta_{j-2N_{1}}^{(5)}(z_{i}).
\end{align}
For $i=1, \ldots, N_{1}$, $j=2N_{1}+N_{2}+1, \ldots, 2N_{1}+N_{2}+N_{3}$, we have
\begin{align}
F_{ij}=\Delta_{j-2N_{1}-N_{2}}^{(6)}(z_{i}).
\end{align}
For $i=N_{1}+1, \ldots, 2N_{1}$, $j=1, \ldots, N_{1}$, we have
\begin{align}
F_{ij}=\Delta_{j}^{(7)}(z_{i-N_{1}}^{*}).
\end{align}
For $i=N_{1}+1, \ldots, 2N_{1}$, $j=N_{1}+1, \ldots, 2N_{1}$, we have
\begin{align}
F_{ij}=\Delta_{j-N_{1}}^{(8)}(z_{i-N_{1}}^{*}).
\end{align}
For $i=N_{1}+1, \ldots, 2N_{1}$, $j=2N_{1}+1, \ldots, N_{2}$, we have
\begin{align}
F_{ij}=\Delta_{j-2N_{1}}^{(9)}(z_{i-N_{1}}^{*}).
\end{align}
For $i=N_{1}+1, \ldots, 2N_{1}$, $j=2N_{1}+N_{2}, \ldots, 2N_{1}+N_{2}+N_{3}$, we have
\begin{align}
F_{ij}=\Delta_{j-2N_{1}-N_{2}}^{(10)}(z_{i-N_{1}}^{*}).
\end{align}
For $i=2N_{1}+1, \ldots, 2N_{1}+N_{2}$, $j=1, \ldots, N_{1}$, we have
\begin{align}
F_{ij}=\sum_{n=1}^{N_{2}}\Delta_{n}^{(1)}(z_{i-2N_{1}})\Delta_{j}^{(3)}(\xi_{n}^{*})+
\sum_{n=1}^{N_{3}}\Delta_{n}^{(2)}(z_{i-2N_{1}})\Delta_{j}^{(7)}(\omega_{n}^{*}).
\end{align}
For $i=2N_{1}+1, \ldots, 2N_{1}+N_{2}$, $j=N_{1}+1, \ldots, 2N_{1}$, we have
\begin{align}
F_{ij}=\sum_{n=1}^{N_{2}}\Delta_{n}^{(1)}(z_{i-2N_{1}})\Delta_{j-N_{1}}^{(4)}(\xi_{n}^{*})+
\sum_{n=1}^{N_{3}}\Delta_{n}^{(2)}(z_{i-2N_{1}})\Delta_{j-N_{1}}^{(8)}(\omega_{n}^{*}).
\end{align}
For $i=2N_{1}+1, \ldots, 2N_{1}+N_{2}$, $j=2N_{1}+1, \ldots, N_{2}$, we have
\begin{align}
F_{ij}=\sum_{n=1}^{N_{2}}\Delta_{n}^{(1)}(z_{i-2N_{1}})\Delta_{j-2N_{1}}^{(5)}(\xi_{n}^{*})+
\sum_{n=1}^{N_{3}}\Delta_{n}^{(2)}(z_{i-2N_{1}})\Delta_{j-2N_{1}}^{(9)}(\omega_{n}^{*}).
\end{align}
For $i=2N_{1}+1, \ldots, 2N_{1}+N_{2}$, $j=2N_{1}+N_{2}+1, \ldots, 2N_{1}+N_{2}+N_{3}$, we have
\begin{align}
F_{ij}=\sum_{n=1}^{N_{2}}\Delta_{n}^{(1)}(z_{i-2N_{1}})\Delta_{j-2N_{1}-N_{2}}^{(6)}(\xi_{n}^{*})+
\sum_{n=1}^{N_{3}}\Delta_{n}^{(2)}(z_{i-2N_{1}})\Delta_{j-2N_{1}-N_{2}}^{(10)}(\omega_{n}^{*}).
\end{align}
For $i=2N_{1}+N_{2}+1, \ldots, 2N_{1}+N_{2}+N_{3}$, $j=1, \ldots, N_{1}$, we have
\begin{align}
F_{ij}=\sum_{n=1}^{N_{2}}\Delta_{n}^{(1)}(z_{i-2N_{1}-N_{2}})\Delta_{j}^{(3)}(\xi_{n}^{*})+
\sum_{n=1}^{N_{3}}\Delta_{n}^{(2)}(z_{i-2N_{1}-N_{2}})\Delta_{j}^{(7)}(\omega_{n}^{*}).
\end{align}
For $i=2N_{1}+N_{2}+1, \ldots, 2N_{1}+N_{2}+N_{3}$, $j=N_{1}+1, \ldots, 2N_{1}$, we have
\begin{align}
F_{ij}=\sum_{n=1}^{N_{2}}\Delta_{n}^{(1)}(z_{i-2N_{1}-N_{2}})\Delta_{j-N_{1}}^{(4)}(\xi_{n}^{*})+
\sum_{n=1}^{N_{3}}\Delta_{n}^{(2)}(z_{i-2N_{1}-N_{2}})\Delta_{j-N_{1}}^{(8)}(\omega_{n}^{*}).
\end{align}
For $i=2N_{1}+N_{2}+1, \ldots, 2N_{1}+N_{2}+N_{3}$, $j=2N_{1}+1, \ldots, N_{2}$, we have
\begin{align}
F_{ij}=\sum_{n=1}^{N_{2}}\Delta_{n}^{(1)}(z_{i-2N_{1}-N_{2}})\Delta_{j-2N_{1}}^{(5)}(\xi_{n}^{*})+
\sum_{n=1}^{N_{3}}\Delta_{n}^{(2)}(z_{i-2N_{1}-N_{2}})\Delta_{j-2N_{1}}^{(9)}(\omega_{n}^{*}).
\end{align}
For $i=2N_{1}+N_{2}+1, \ldots, 2N_{1}+N_{2}+N_{3}$, $j=2N_{1}+N_{2}+1, \ldots, 2N_{1}+N_{2}+N_{3}$, we have
\begin{align}
F_{ij}=\sum_{n=1}^{N_{2}}\Delta_{n}^{(1)}(z_{i-2N_{1}-N_{2}})\Delta_{j-2N_{1}-N_{2}}^{(6)}(\xi_{n}^{*})+
\sum_{n=1}^{N_{3}}\Delta_{n}^{(2)}(z_{i-2N_{1}-N_{2}})\Delta_{j-2N_{1}-N_{2}}^{(10)}(\omega_{n}^{*}).
\end{align}
The solution of the system \eqref{QQ1} reads
\begin{align}
\mathbf{x_{k}}=\frac{\det \mathbf{W_{k}}^{aug}}{\det \mathbf{W}},\quad k=1,2,\ldots, 2N_{1}+N_{2}+N_{3},
\end{align}
where $\mathbf{W_{k}}^{aug}=(\mathbf{W_{1}}, \ldots, \mathbf{W_{k-1}}, \mathbf{b}, \mathbf{W_{k+1}}, \ldots, \mathbf{W_{2N_{1}+N_{2}+N_{3}}})$.
Therefore, substituting the above $\mathbf{x_{k}}$ into  the reconstruction formula, the brief expression for the potential can be obtained by
\begin{align}
q(x,t)=\frac{\det \mathbf{W}^{aug}}{\det \mathbf{W}},
\end{align}
where
\begin{align}
\mathbf{W}^{aug}=\left(
                   \begin{array}{cc}
                     q_{+} & \mathbf{y}^{T} \\
                    \mathbf{ b} & \mathbf{W} \\
                   \end{array}
                 \right),
\end{align}
$\mathbf{b}=(b_{1}, \ldots, b_{2N_{1}+N_{2}+N_{3}})$, and $\mathbf{y}=(y_{1}, \ldots, y_{2N_{1}+N_{2}+N_{3}})^{T}$.
\section{Soliton solutions}
In this section, we mainly study several cases under the condition of single soliton solution, including $N_{1}=1$, $N_{2}=1$ and $N_{3}=1$.
\subsection{Single soliton solutions for $N_{1}=1$}
In this subsection, we assume that the eigenvalues $\tau_{n}$ is the first kind eigenvalues, which implies that $N_{1}=1$ and $N_{2}=N_{3}=0$. Let $\mathbf{q}_{+}=(1,~~1)^{T}$, $E_{1}=e^{\alpha+i\beta}$, $z=\varrho e^{i\delta},(0<\delta<\pi)$. The soliton solution of the cmKdV equation can be derived
\begin{align}
q(x,t)=\frac{\det\left(
                   \begin{array}{ccc}
                     q_{+} & y_{1} & y_{2} \\
                     b_{1} & 1-F_{11} & -F_{12} \\
                     b_{2} & -F_{21} & 1-F_{22} \\
                   \end{array}
                 \right)
}{\det\left(
                   \begin{array}{cc}
                     1-F_{11} & -F_{12} \\
                     -F_{21} & 1-F_{22} \\
                   \end{array}
                 \right)
},
\end{align}
where\\
\begin{align*}
&y_{1}=-iE_{1}e^{2i\theta_{1}(z_{1})}, ~ y_{2}=-\frac{z_{1}}{q_{0}}\check{E}_{1}e^{2i\theta_{1}(-\frac{q_{0}^{2}}{z_{1}})},~
b_{1}=\frac{u_{+}}{q_{0}}, b_{2}=-\frac{iu_{+}}{q_{0}},~\\ &F_{11}=\Delta_{1}^{(3)}(z_{1}),~
F_{12}=\Delta_{1}^{(4)}(z_{1}), F_{21}=\Delta_{1}^{(7)}(z_{1}^{*}), ~ F_{22}=\Delta_{1}^{(8)}(z_{1}^{*}),\\
&\Delta_{1}^{(3)}(z_{1})=\frac{\widetilde{E}_{1}e^{-2i\theta_{1}(-\frac{q_{0}^{2}}{z_{1}})}}
{z_{1}-(-\frac{q_{0}^{2}}{z_{1}})}, ~ \Delta_{1}^{(4)}(z_{1})=\frac{\hat{E}_{n}e^{-2i\theta_{1}(z_{1}^{*})}}{z_{1}-z_{1}^{*}},\\
&\Delta_{1}^{(7)}(z_{1}^{*})=\frac{E_{1}e^{2i\theta_{1}(z_{1})}}{z_{1}^{*}-z_{1}},~
\Delta_{1}^{(8)}(z_{1}^{*})=\frac{\check{E}_{1}e^{2i\theta_{1}(-\frac{q_{0}^{2}}{z_{1}^{*}})}}
{z_{1}^{*}-(-\frac{q_{0}^{2}}{z_{1}^{*}})}.
\end{align*}

~\qquad\quad{\includegraphics[width=5cm,height=4.9cm,angle=0]{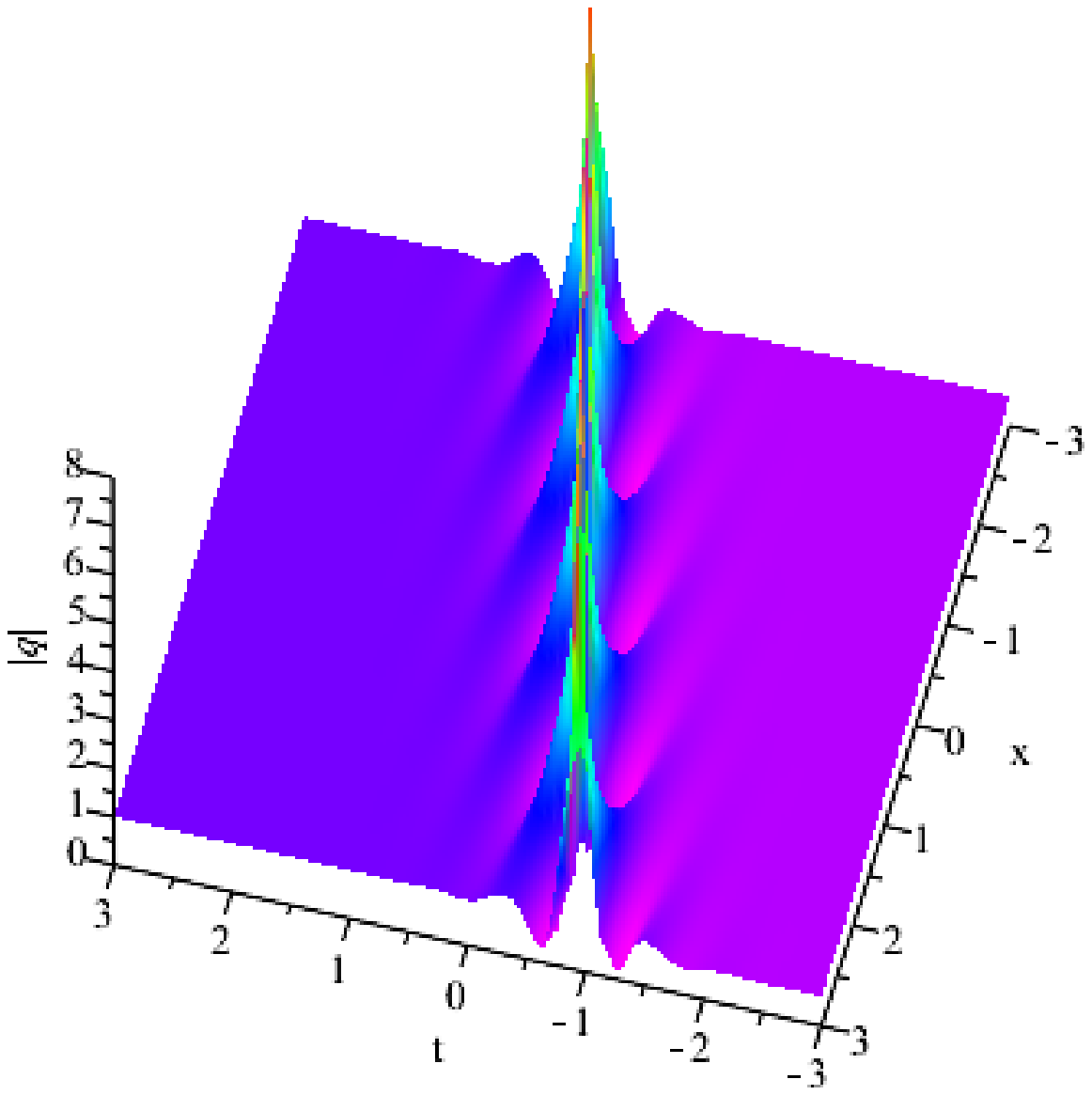}}
~~~\quad\qquad{\rotatebox{0}{\includegraphics[width=5cm,height=4.9cm,angle=0]{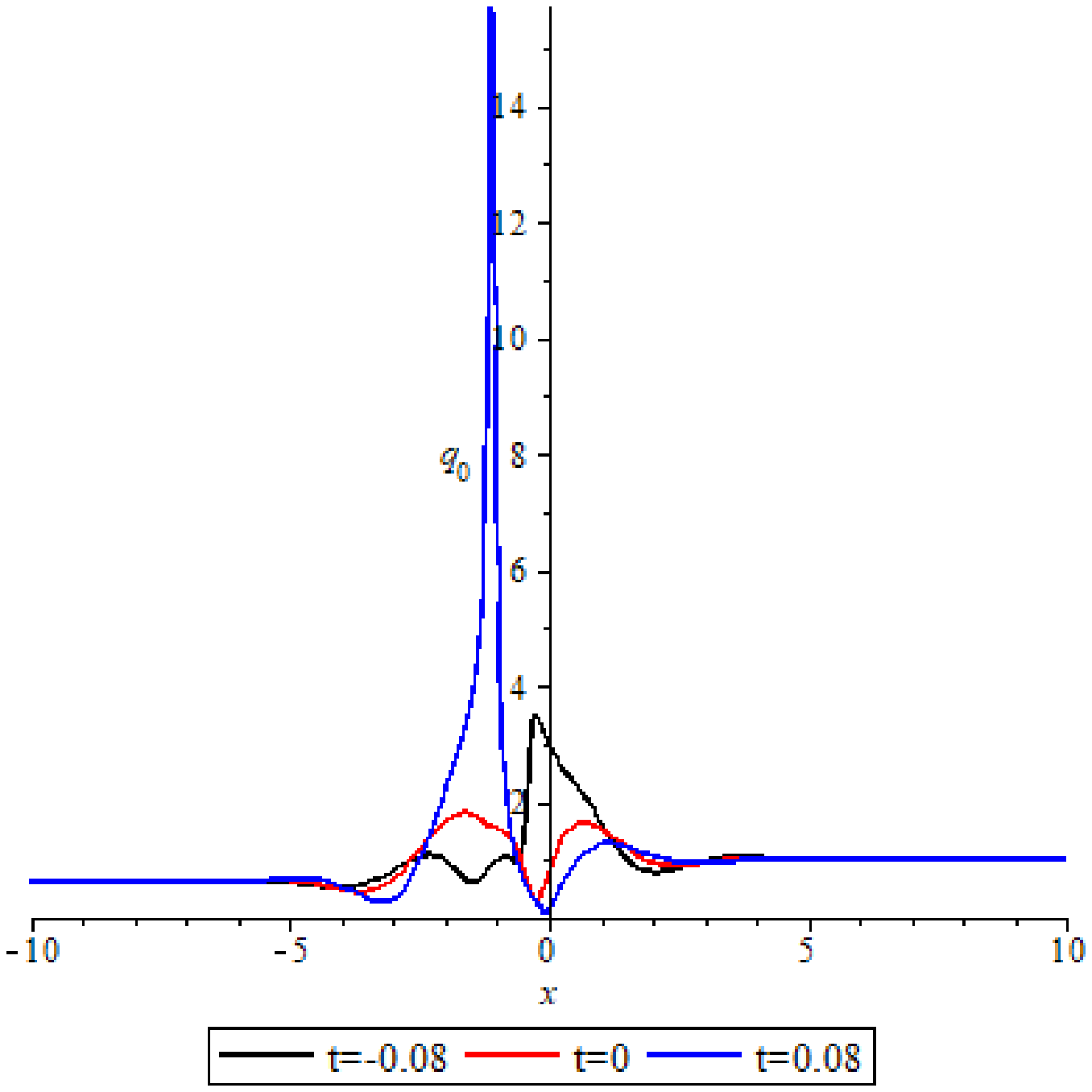}}}

$~~~~~~\qquad\qquad\qquad\qquad(\textbf{a})\quad\qquad\qquad\qquad\qquad\qquad\qquad
~~~~~~~~\qquad(\textbf{b})$

~\qquad{\includegraphics[width=5cm,height=4.9cm,angle=0]{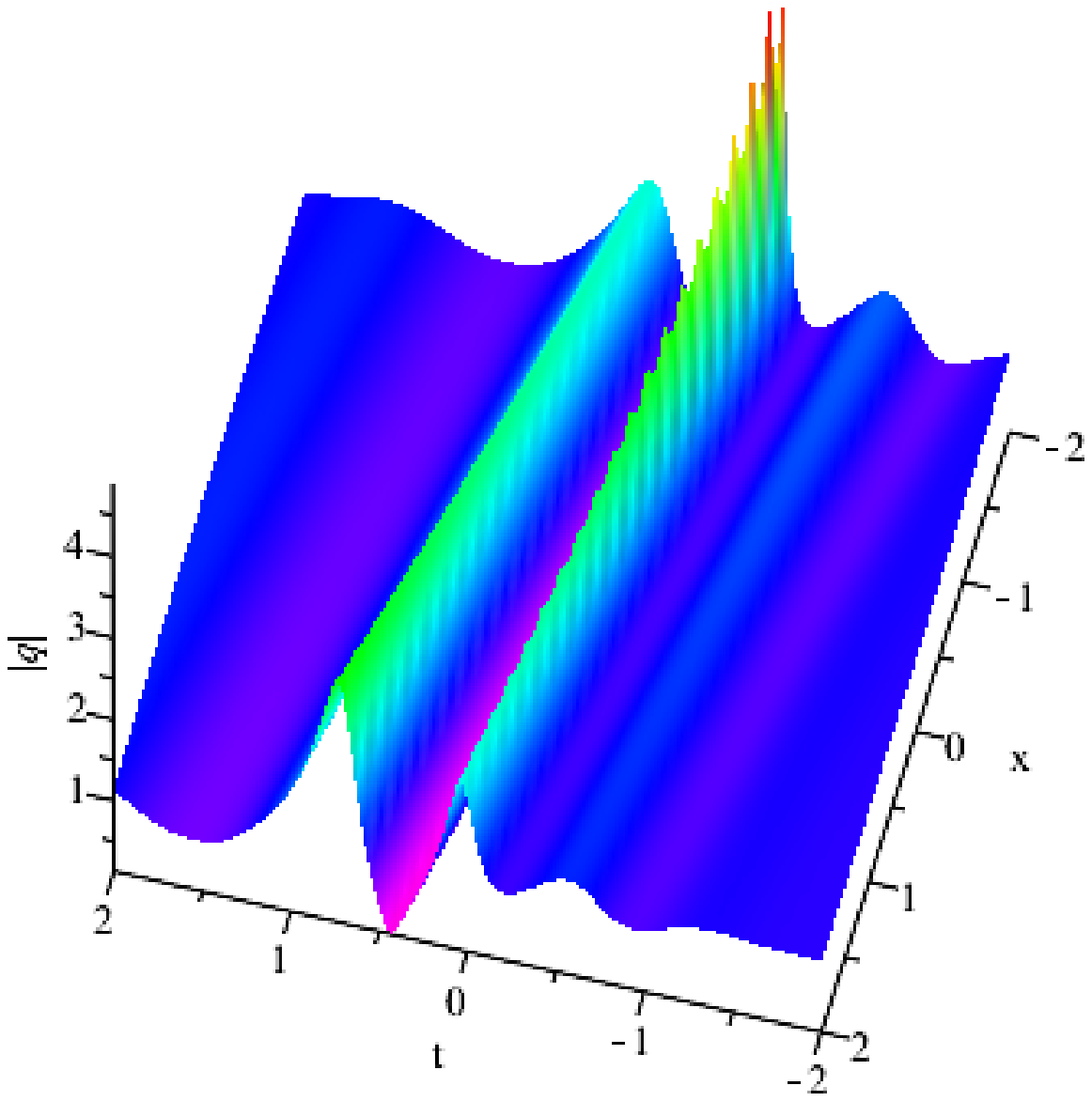}}
~~~\quad\qquad{\rotatebox{0}{\includegraphics[width=5cm,height=4.9cm,angle=0]{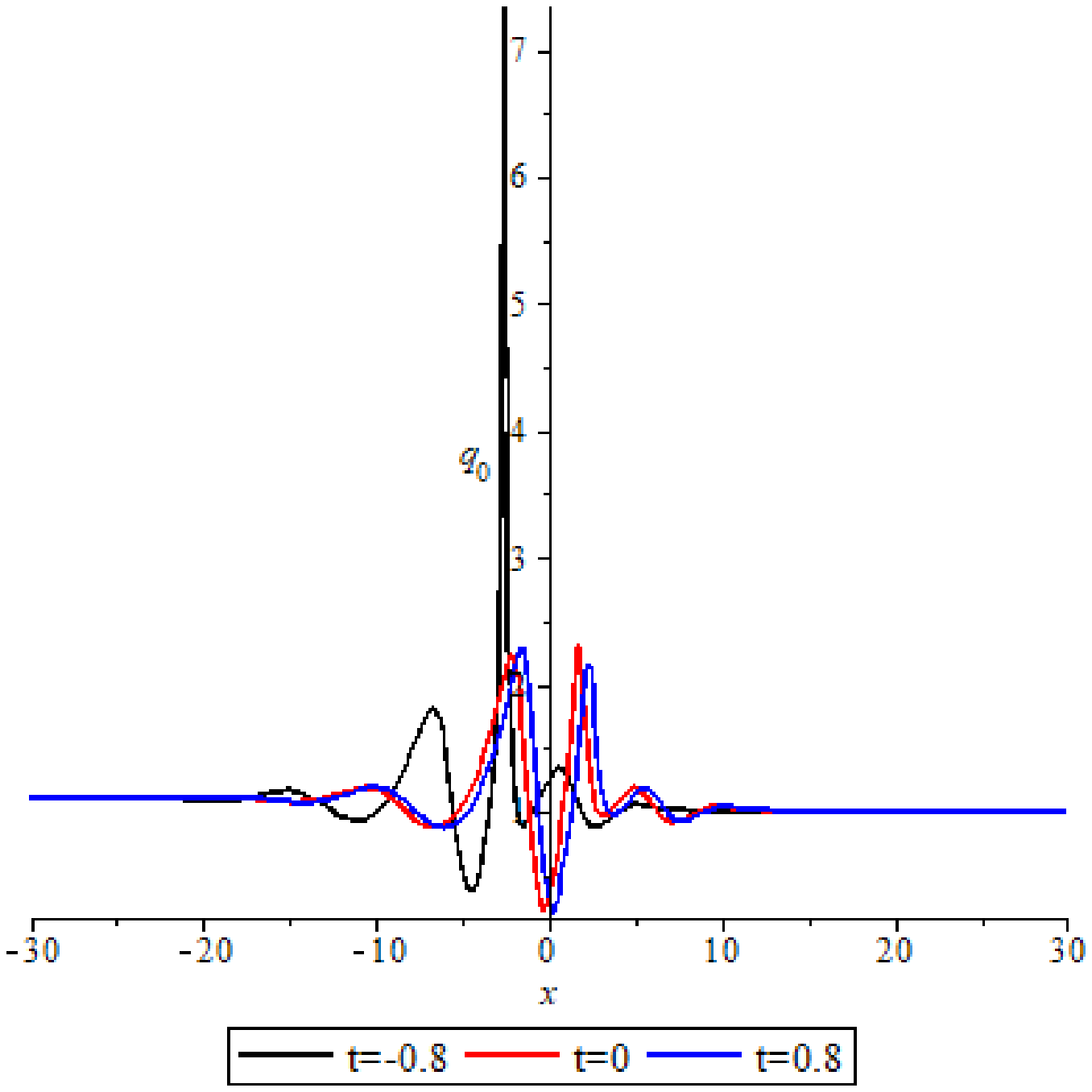}}}

$~~~~\qquad\qquad\qquad\qquad(\textbf{c})\quad\qquad\qquad\qquad\qquad\qquad\qquad
~~~~~~~~\qquad(\textbf{d})$

\noindent { \small \textbf{Figure 3.} $\textbf{(a)(c)}$ The solution \eqref{e} of the equation \eqref{QQ1}. $\textbf{(b)(d)}$ display the propagation behavior of
solutions at different times. $\rho_{1}=2$, $\rho_{2}=1.15$, $\delta_{1}=\frac{2\pi}{5}$, $\delta_{2}=\frac{15\pi}{24}$.}

An example of such a solution is shown in Fig. 3. The Fig. $\textbf{(a)}$  obviously is a breathing solution and a bright soliton solution. It propagates periodically along the $x$-axis, but aperiodically along the $t$-axis. Fig. $\textbf{(c)}$ is a bimodal bright soliton solution, and we can observe that the amplitude on the right side is higher than that on the left side.
\subsection{Single soliton solutions for $N_{2}=1$}
In this subsection, we assume that the eigenvalues $\tau_{n}$ is the second kind eigenvalues, which implies that $N_{2}=1$ and $N_{1}=N_{3}=0$. Let $\mathbf{q}_{+}=(1,~~1)^{T}$, $E_{1}=e^{\alpha+i\beta}$, $z=\varrho e^{i\delta},(0<\delta<\pi)$. The soliton solution of the cmKdV equation can be derived
\begin{align}\label{e}
q(x,t)=\frac{\det\left(
                   \begin{array}{ccc}
                     q_{+} & y_{1}  \\
                     b_{1} & 1-F_{11}  \\
                   \end{array}
                 \right)
}{1-F_{11}},
\end{align}
where
\begin{align*}
&y_{1}=-i\widetilde{F}_{1}e^{i(\theta_{1}+\theta_{2})(-\frac{q_{0}^{2}}{\zeta_{1}})},~~b_{1}=
\frac{v_{+}}{q_{0}}-\frac{u_{+}}{q_{0}}\Delta_{1}^{(1)}(\zeta_{1}),~
F_{11}=\Delta_{1}^{(1)}(\zeta_{1})\Delta_{1}^{(5)}(\zeta_{1}^{*}),\\
&\Delta_{1}^{(1)}(\zeta_{1})=\frac{F_{1}e^{i(\theta_{1}+\theta_{2})(\zeta_{1}^{*})}}
{\zeta_{1}-\zeta_{1}^{*}}+\frac{\zeta_{1}^{*}}{iq_{0}}
\frac{\check{F}_{1}e^{-i(\theta_{1}+\theta_{2})(-\frac{q_{0}^{2}}{\zeta_{1}^{*}})}}
{\zeta_{1}-(-\frac{q_{0}^{2}}{\zeta_{1}^{*}})},
\Delta_{1}^{(5)}(\zeta_{1}^{*})=\frac{F_{1}e^{i(\theta_{1}+\theta_{2})(\zeta_{1})}}{\zeta_{1}^{*}-\zeta_{1}}.
\end{align*}

{\rotatebox{0}{\includegraphics[width=3.3cm,height=3.3cm,angle=0]{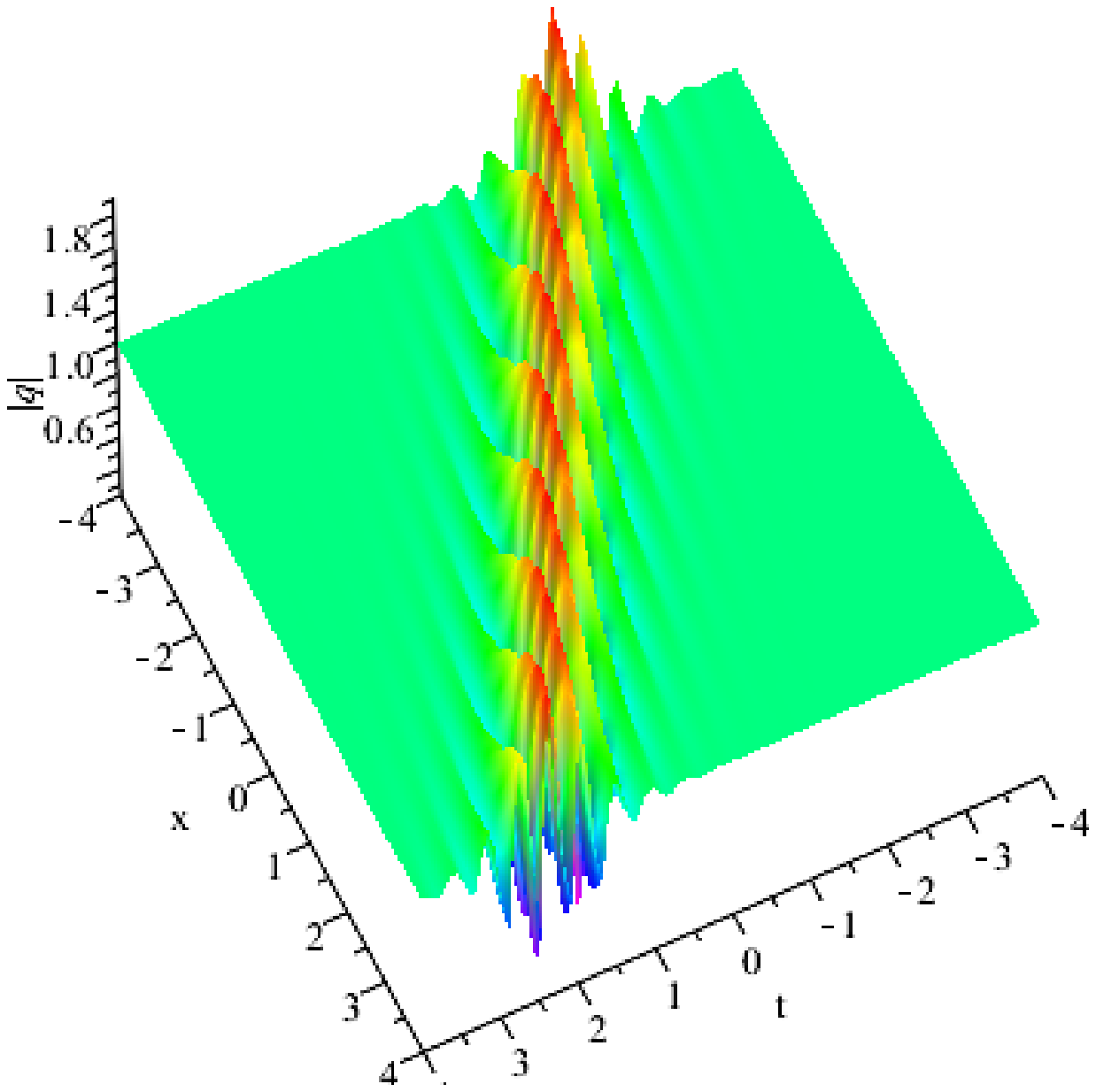}}}
\quad{\rotatebox{0}{\includegraphics[width=3.3cm,height=3.3cm,angle=0]{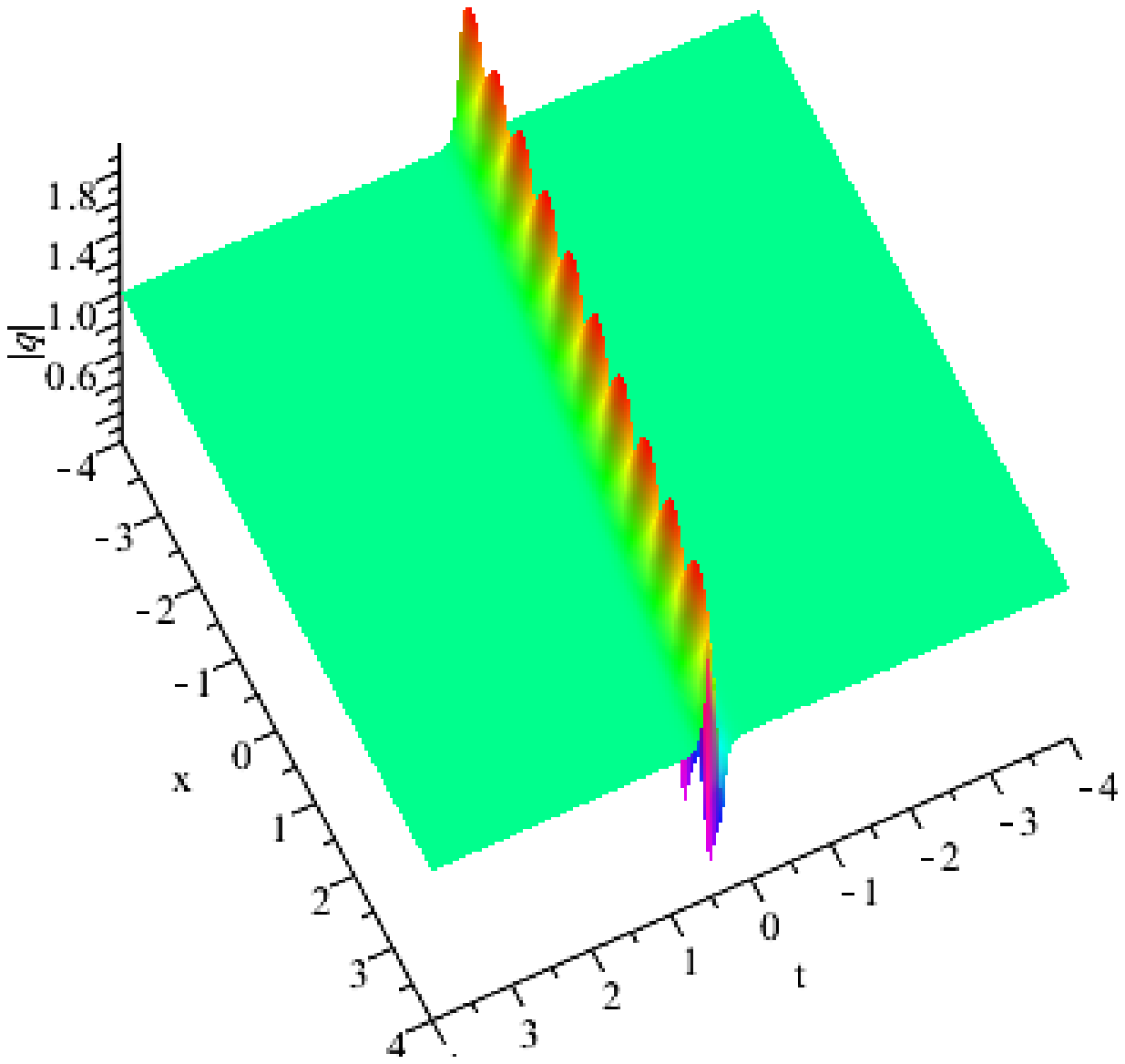}}}
\quad\quad{\rotatebox{0}{\includegraphics[width=3.3cm,height=3.3cm,angle=0]{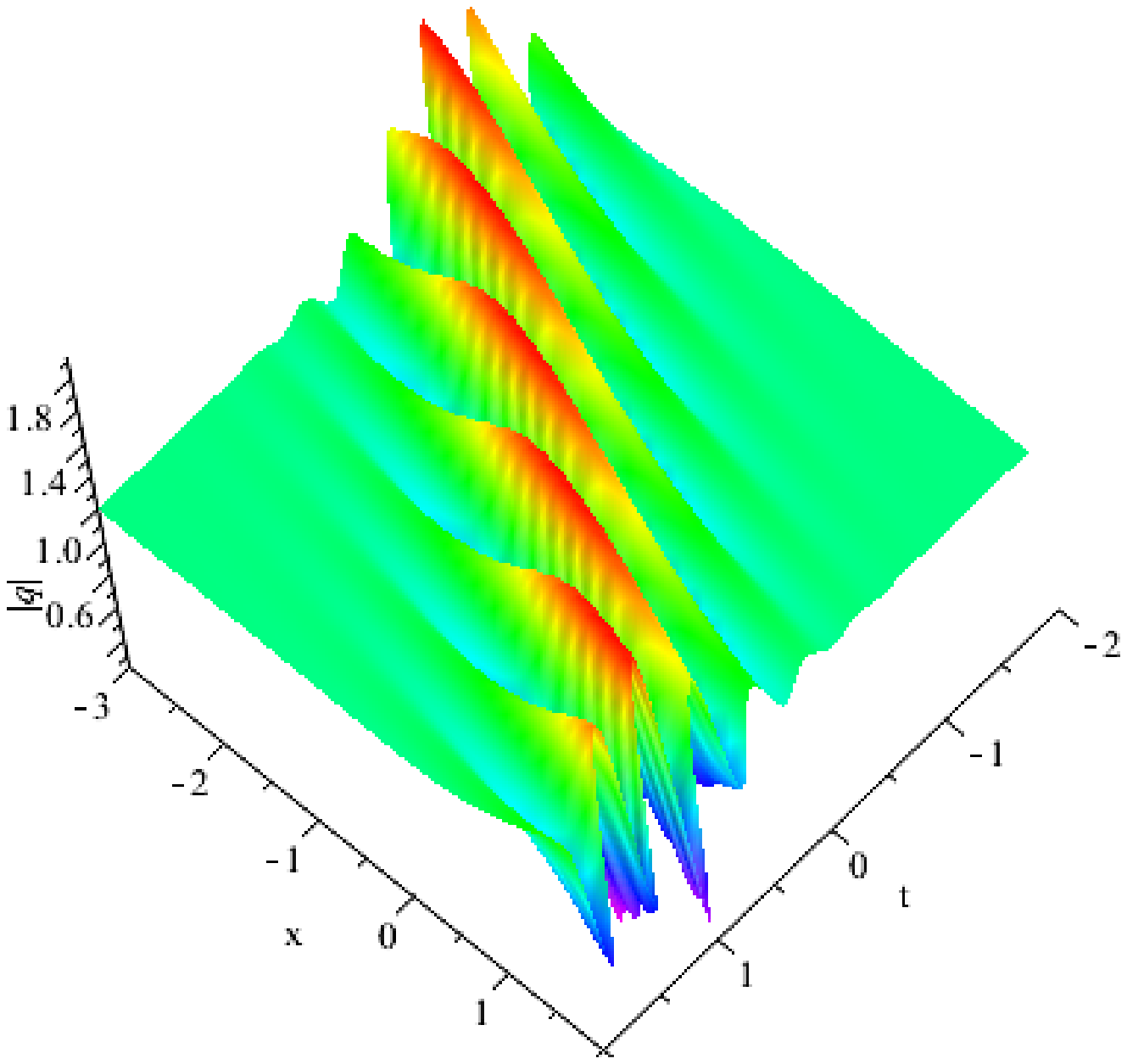}}}

$\qquad\qquad~~(\textbf{a})\quad \ \qquad\qquad\qquad\quad\quad~~(\textbf{b})~~\qquad\qquad\qquad\qquad\qquad(\textbf{c})$

{\rotatebox{0}{\includegraphics[width=3.3cm,height=3.3cm,angle=0]{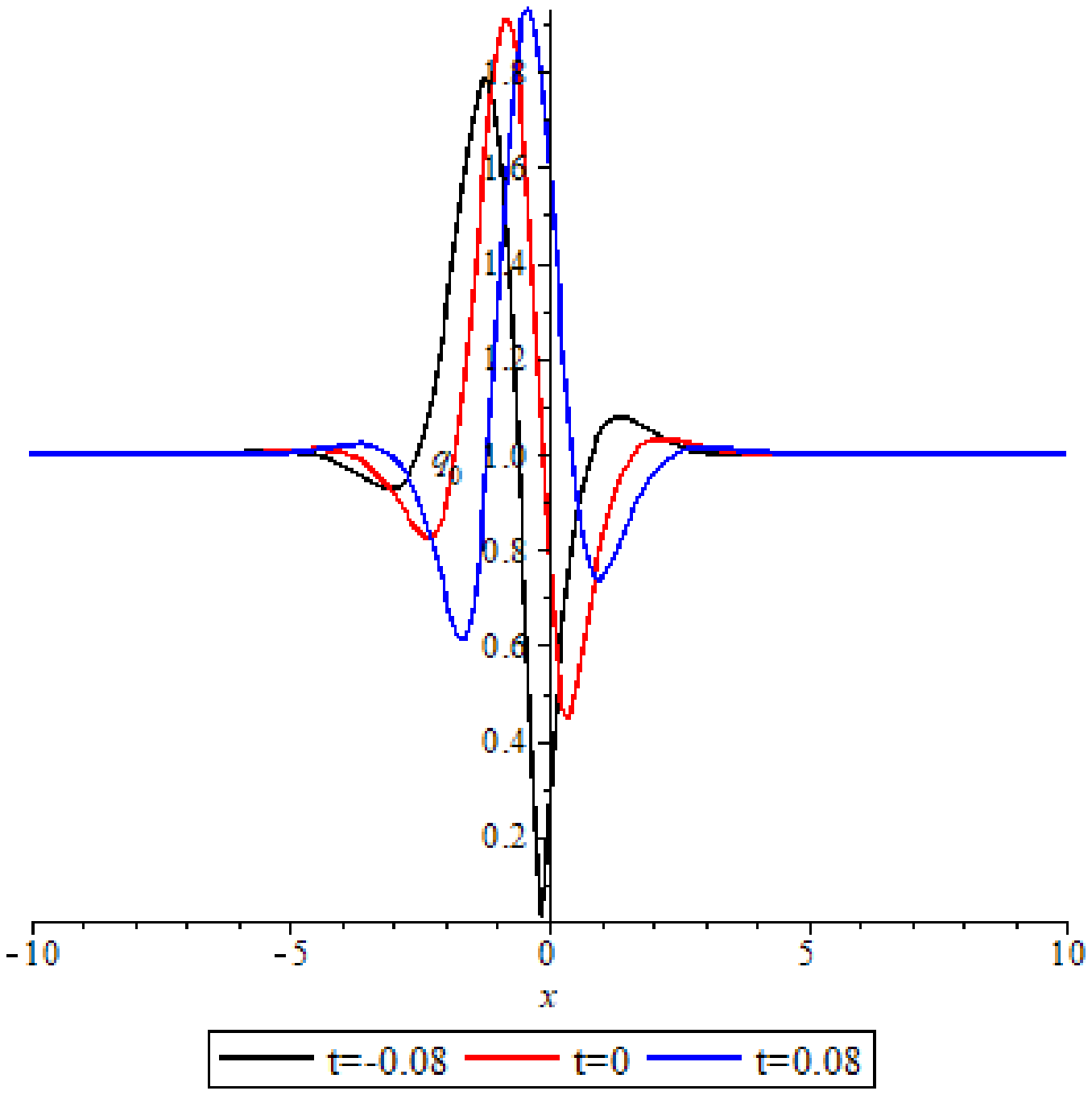}}}
\quad{\rotatebox{0}{\includegraphics[width=3.3cm,height=3.3cm,angle=0]{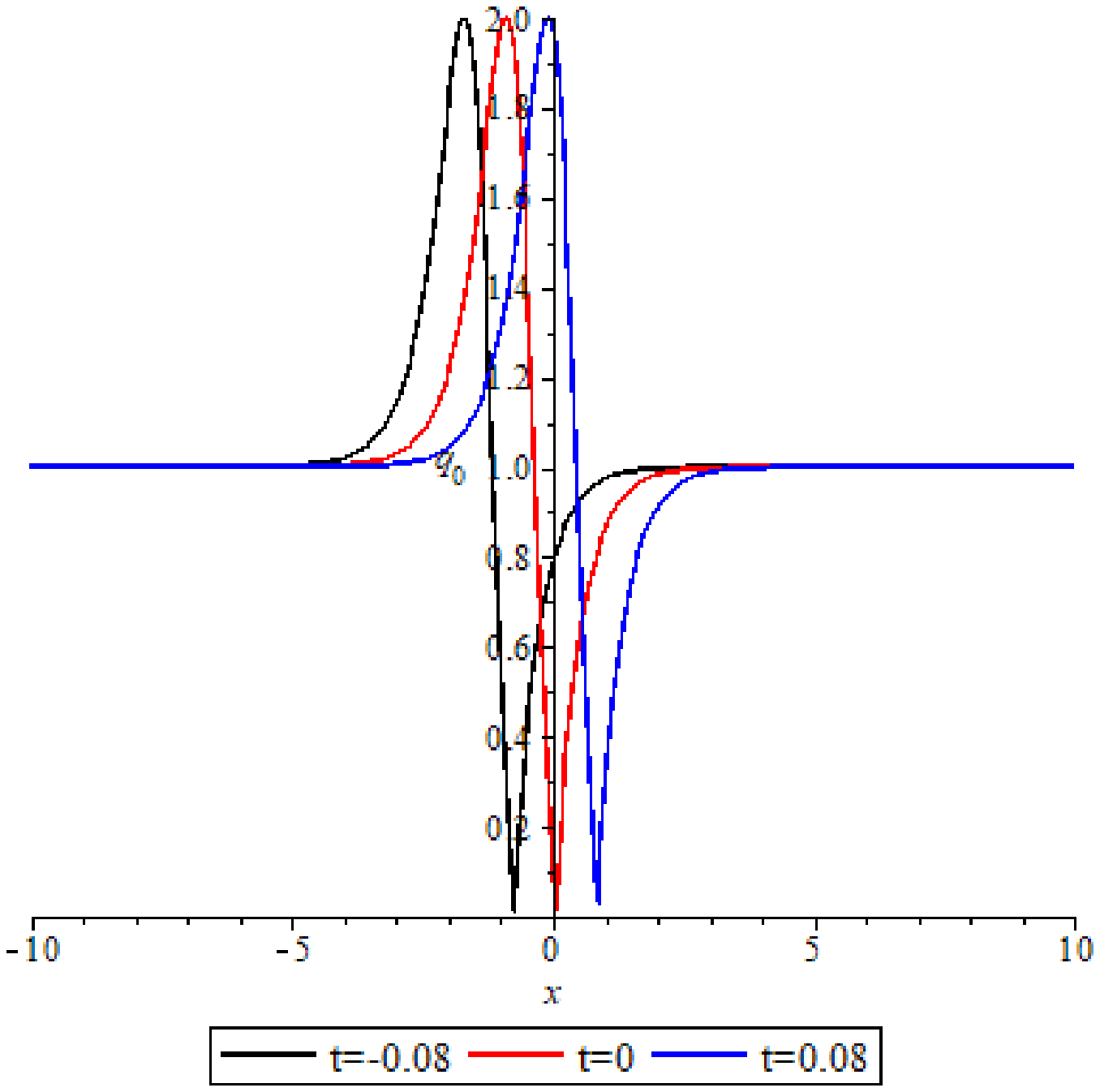}}}
\quad\quad{\rotatebox{0}{\includegraphics[width=3.3cm,height=3.3cm,angle=0]{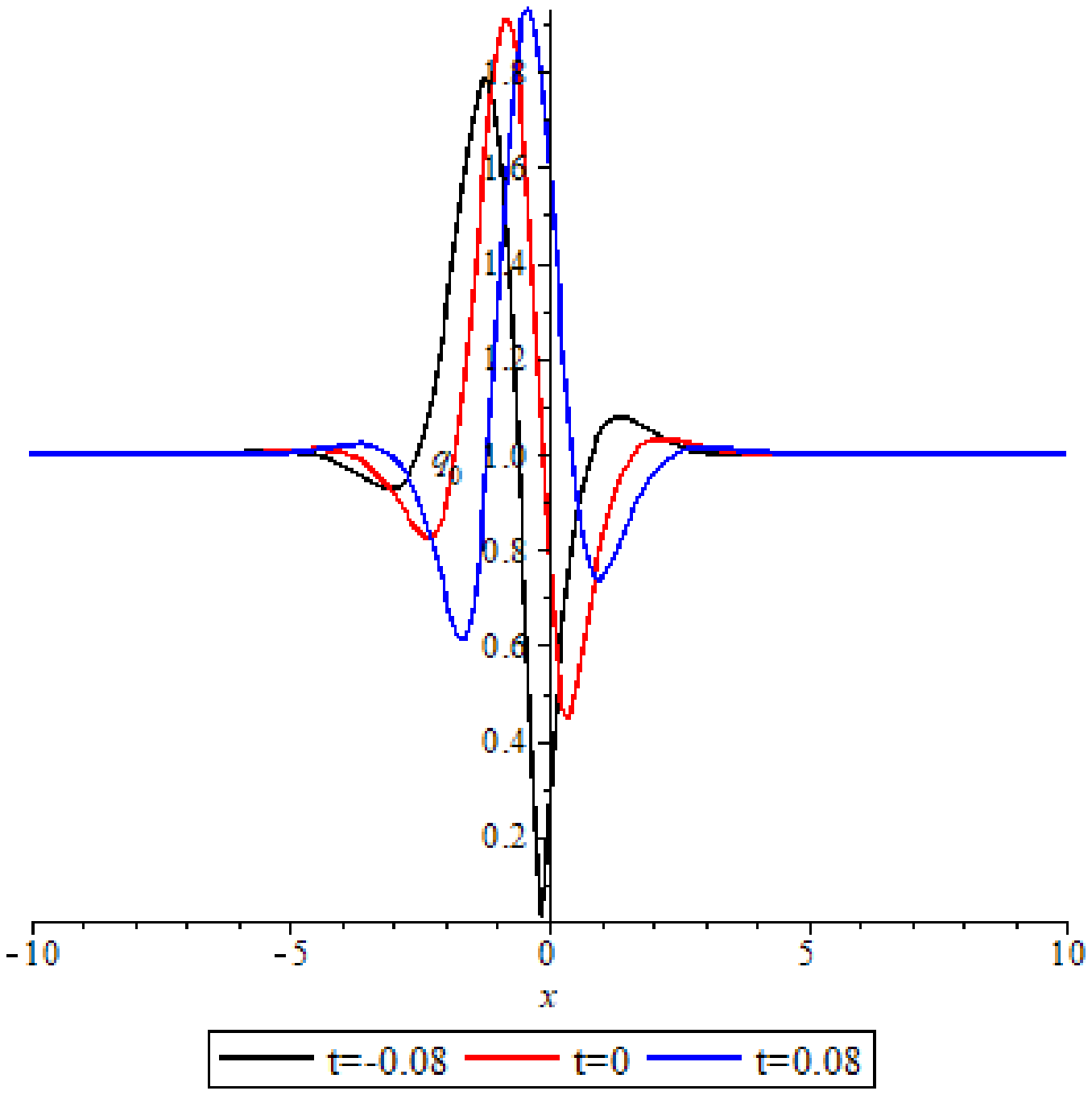}}}

$\qquad\qquad~~(\textbf{d})\quad \ \qquad\qquad\qquad\quad~~~\quad(\textbf{e})~~~~~~\quad\qquad\qquad\qquad\qquad(\textbf{f})$ 

\noindent { \small \textbf{Figure 4.} $\textbf{(a)(b)(c)}$ The solution \eqref{e} of the equation \eqref{QQ1}. $\textbf{(d)(e)(f)}$ display the propagation behavior of
solutions at different times. $\rho=1$, $\delta_{1}=\frac{\pi}{4}$, $\delta_{2}=\frac{\pi}{2}$, $\delta_{3}=\frac{3\pi}{4}$.}

{\rotatebox{0}{\includegraphics[width=3.3cm,height=3.3cm,angle=0]{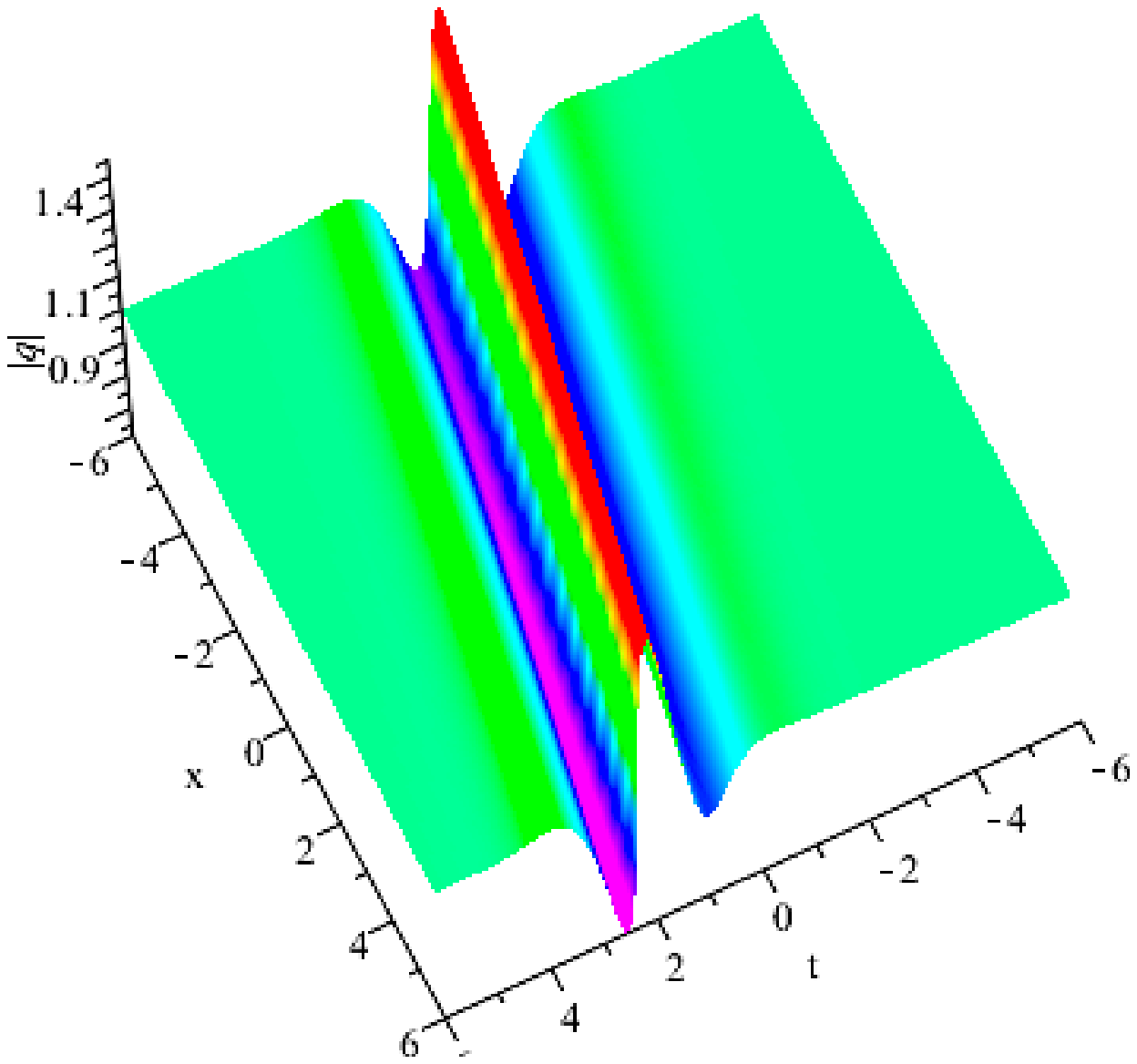}}}
\quad{\rotatebox{0}{\includegraphics[width=3.3cm,height=3.3cm,angle=0]{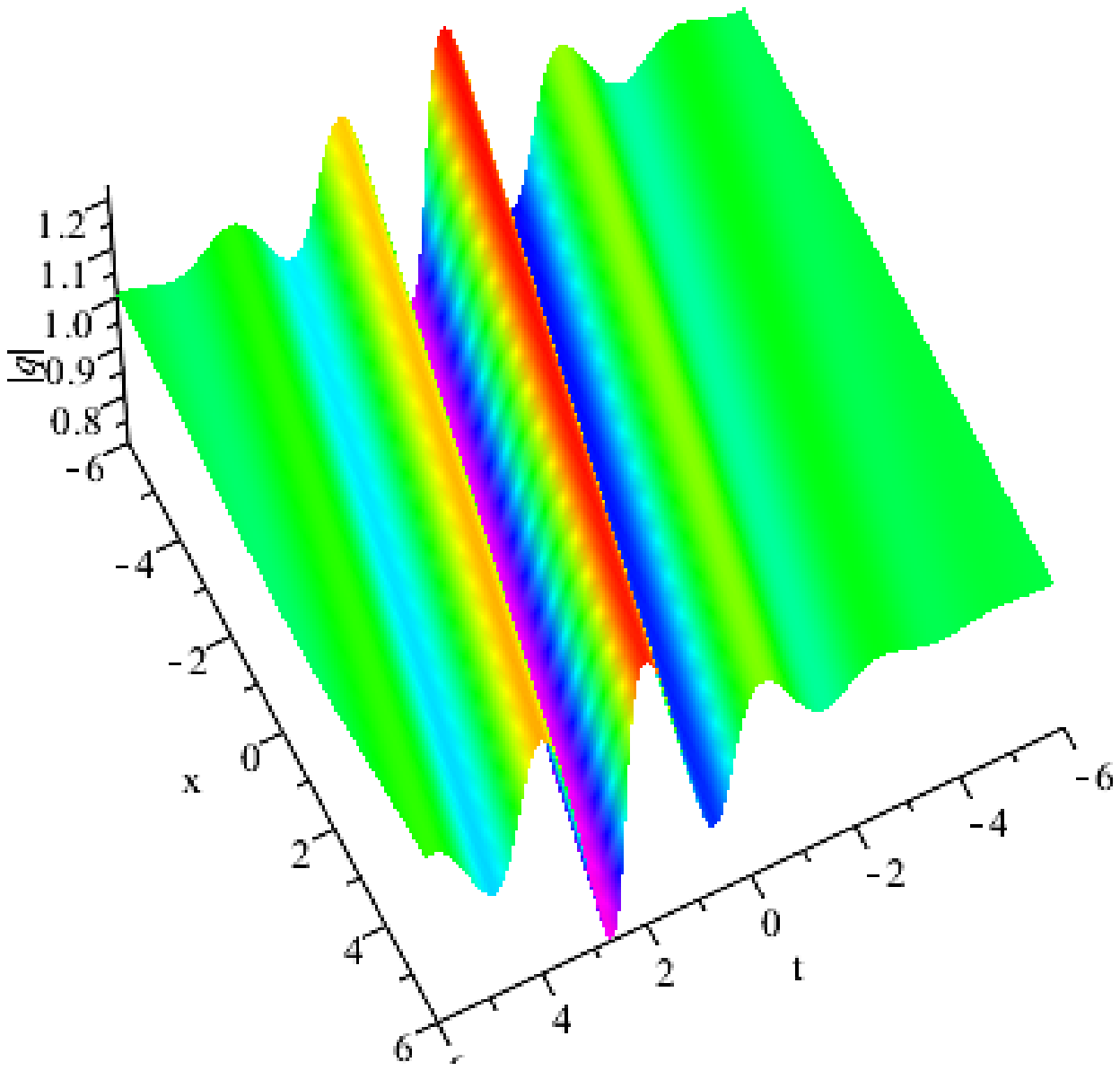}}}
\quad\quad{\rotatebox{0}{\includegraphics[width=3.3cm,height=3.3cm,angle=0]{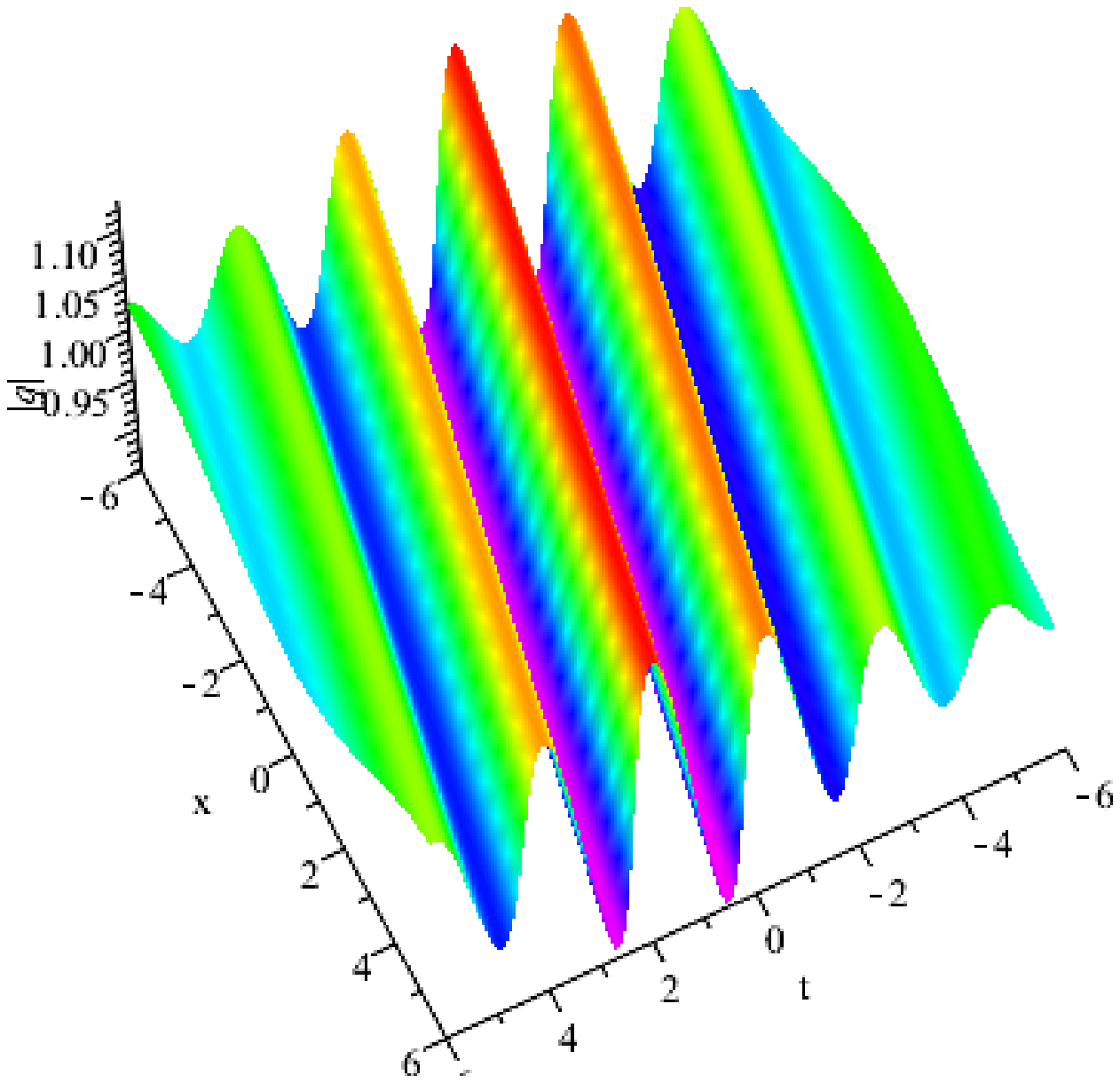}}}

$\qquad\qquad~~(\textbf{a})\quad \ \qquad\qquad\qquad\quad\quad~~(\textbf{b})~~\qquad\qquad\qquad\qquad\qquad(\textbf{c})$

{\rotatebox{0}{\includegraphics[width=3.3cm,height=3.3cm,angle=0]{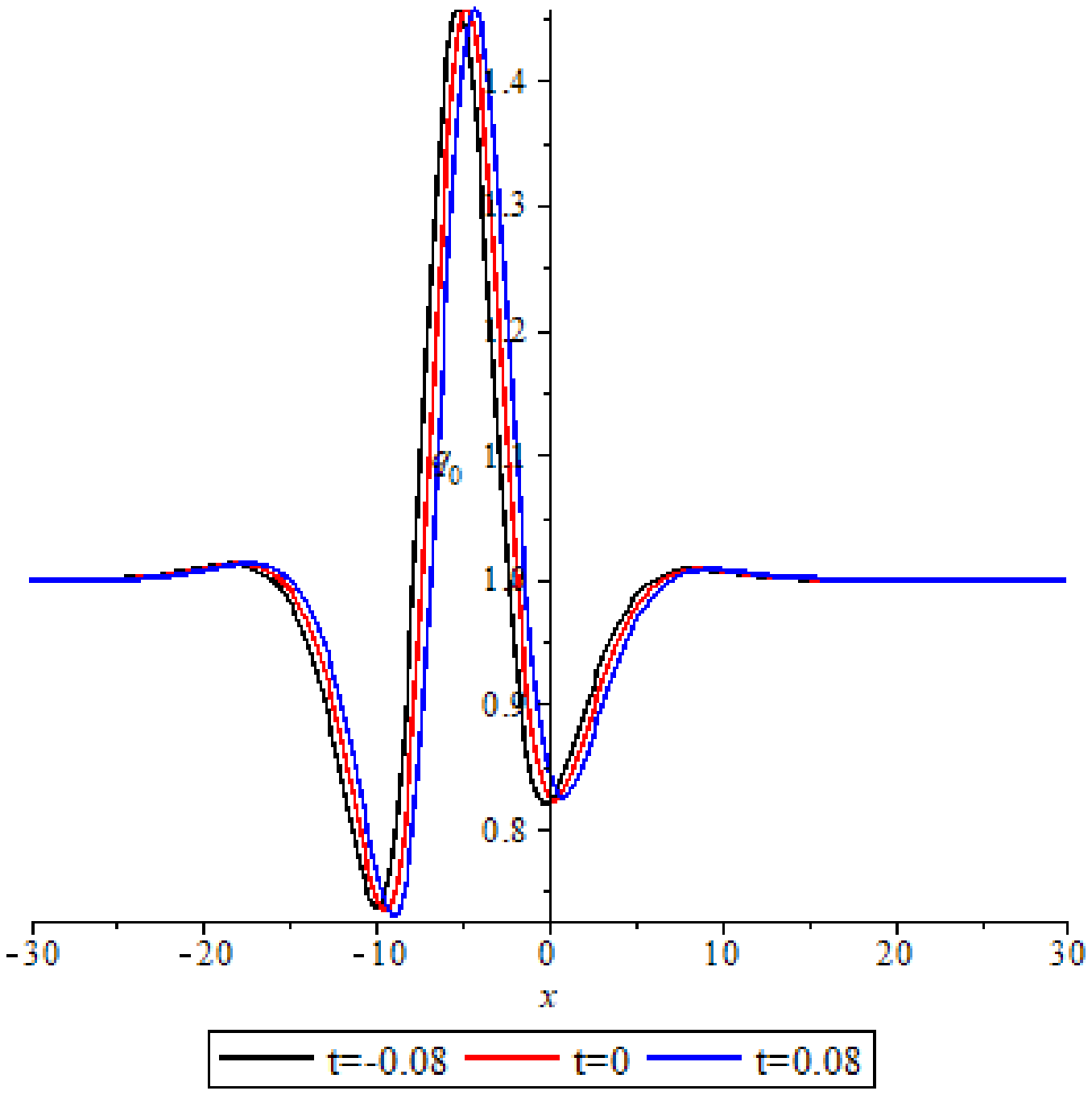}}}
\quad{\rotatebox{0}{\includegraphics[width=3.3cm,height=3.3cm,angle=0]{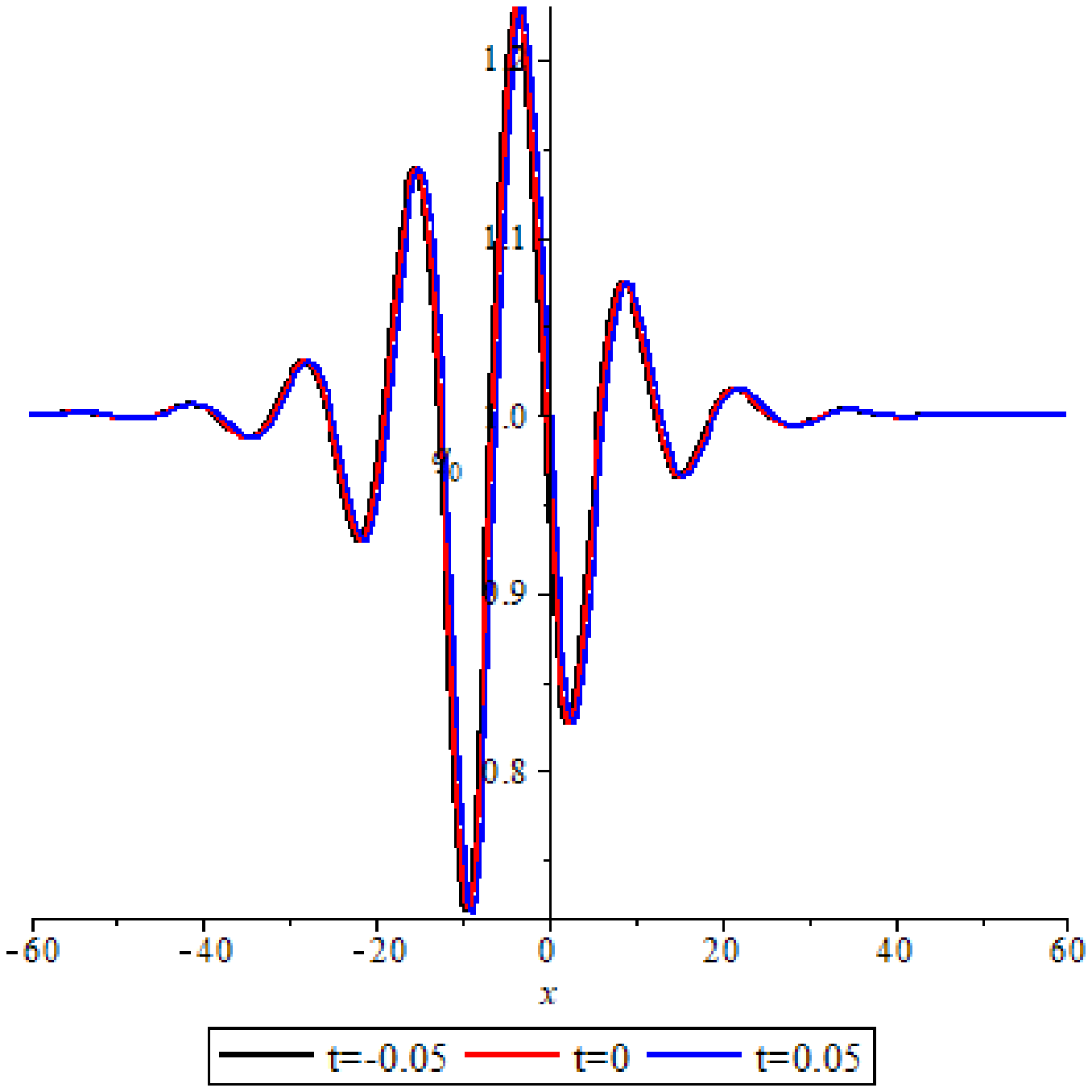}}}
\quad\quad{\rotatebox{0}{\includegraphics[width=3.3cm,height=3.3cm,angle=0]{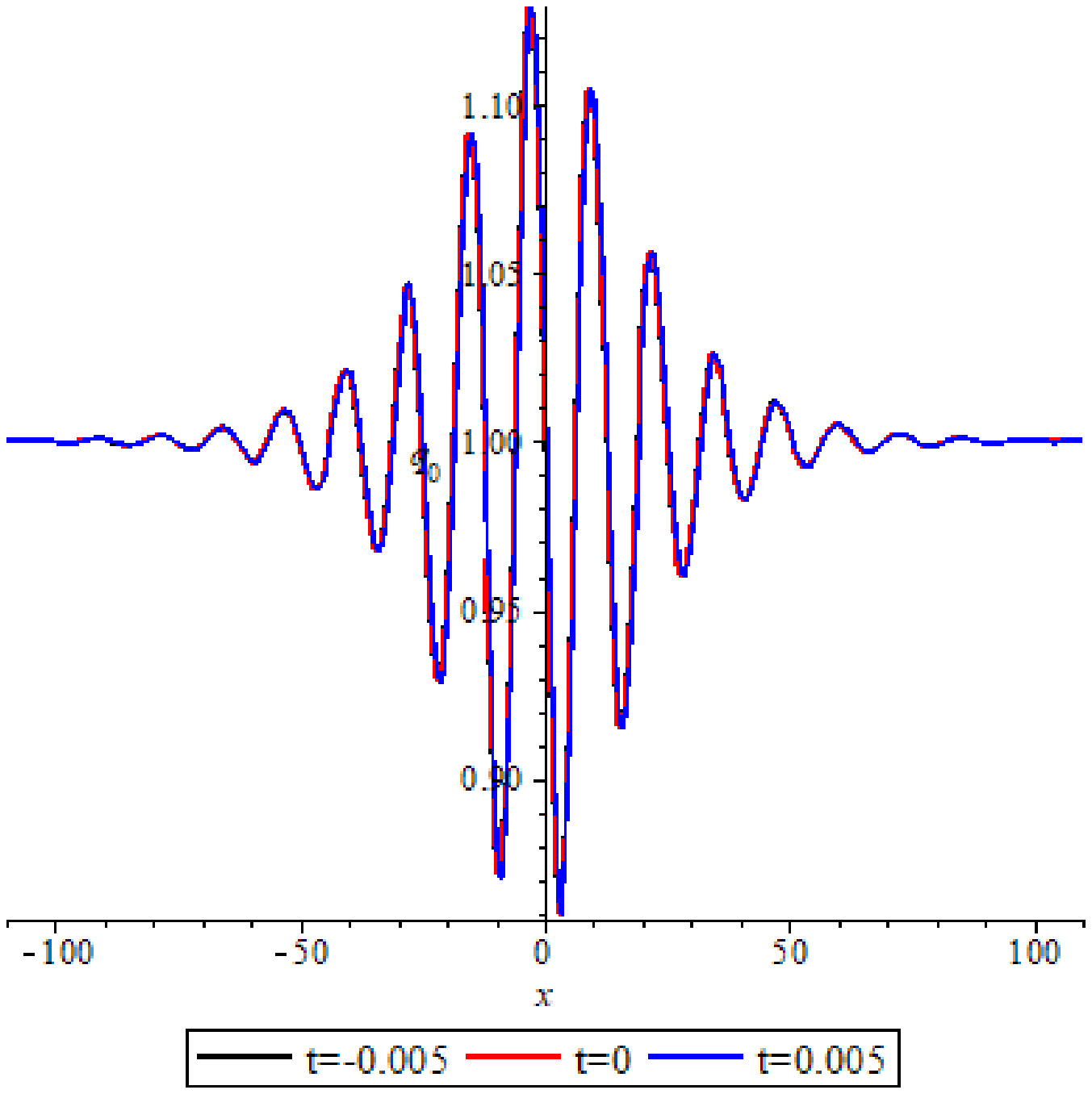}}}

$\qquad\qquad~~(\textbf{d})\quad \ \qquad\qquad\qquad\quad~~~\quad(\textbf{e})~~~~~~\quad\qquad\qquad\qquad\qquad(\textbf{f})$

\noindent { \small \textbf{Figure 5.} $\textbf{(a)(b)(c)}$ The solution \eqref{e} of the equation \eqref{QQ1}. $\textbf{(d)(e)(f)}$ display the propagation behavior of
solutions at different times. $\rho=4$, $\delta_{1}=\frac{\pi}{4}$, $\delta_{2}=\frac{\pi}{12}$, $\delta_{3}=\frac{3\pi}{24}$.}

{\rotatebox{0}{\includegraphics[width=3.3cm,height=3.3cm,angle=0]{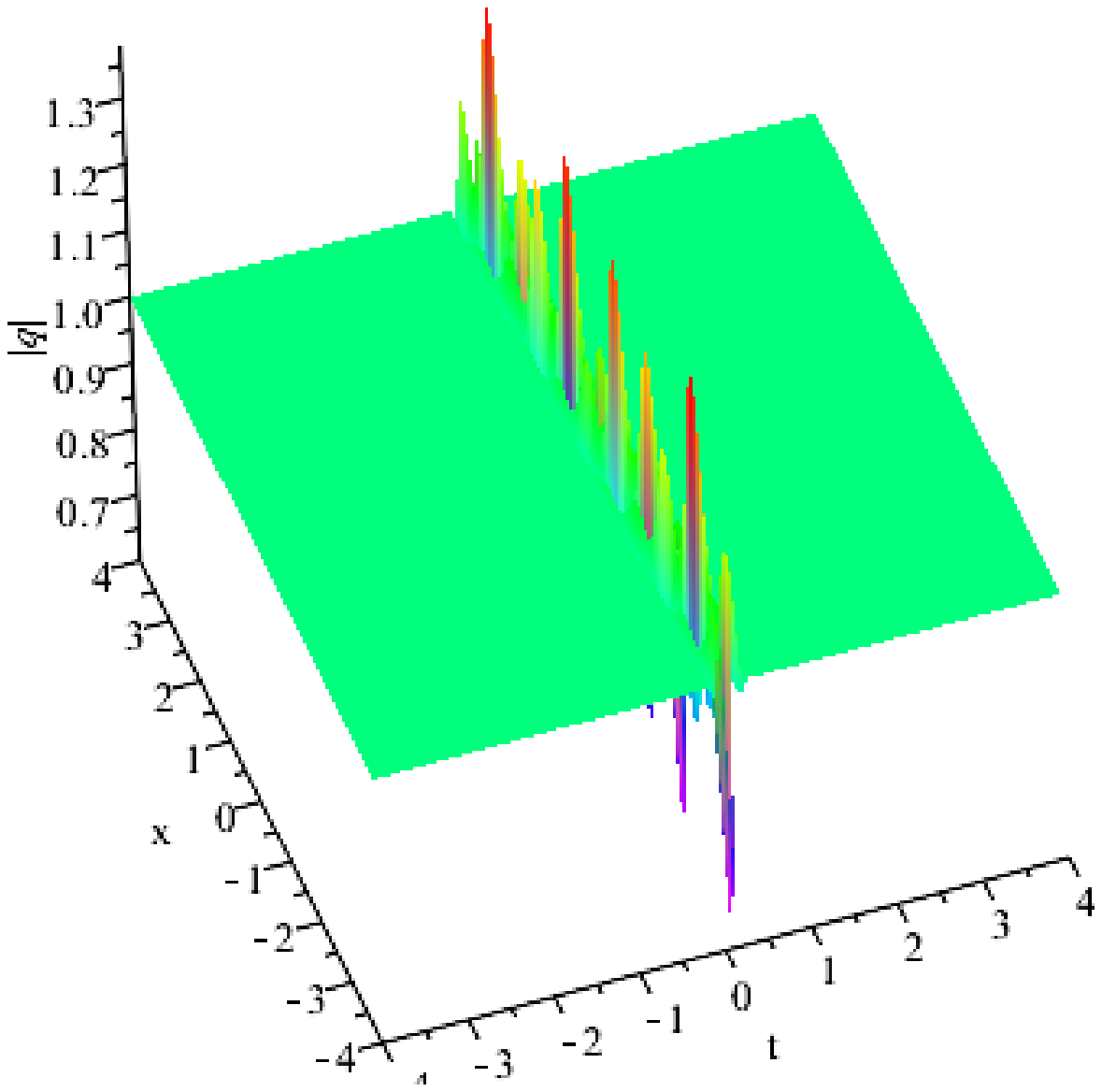}}}
\quad{\rotatebox{0}{\includegraphics[width=3.3cm,height=3.3cm,angle=0]{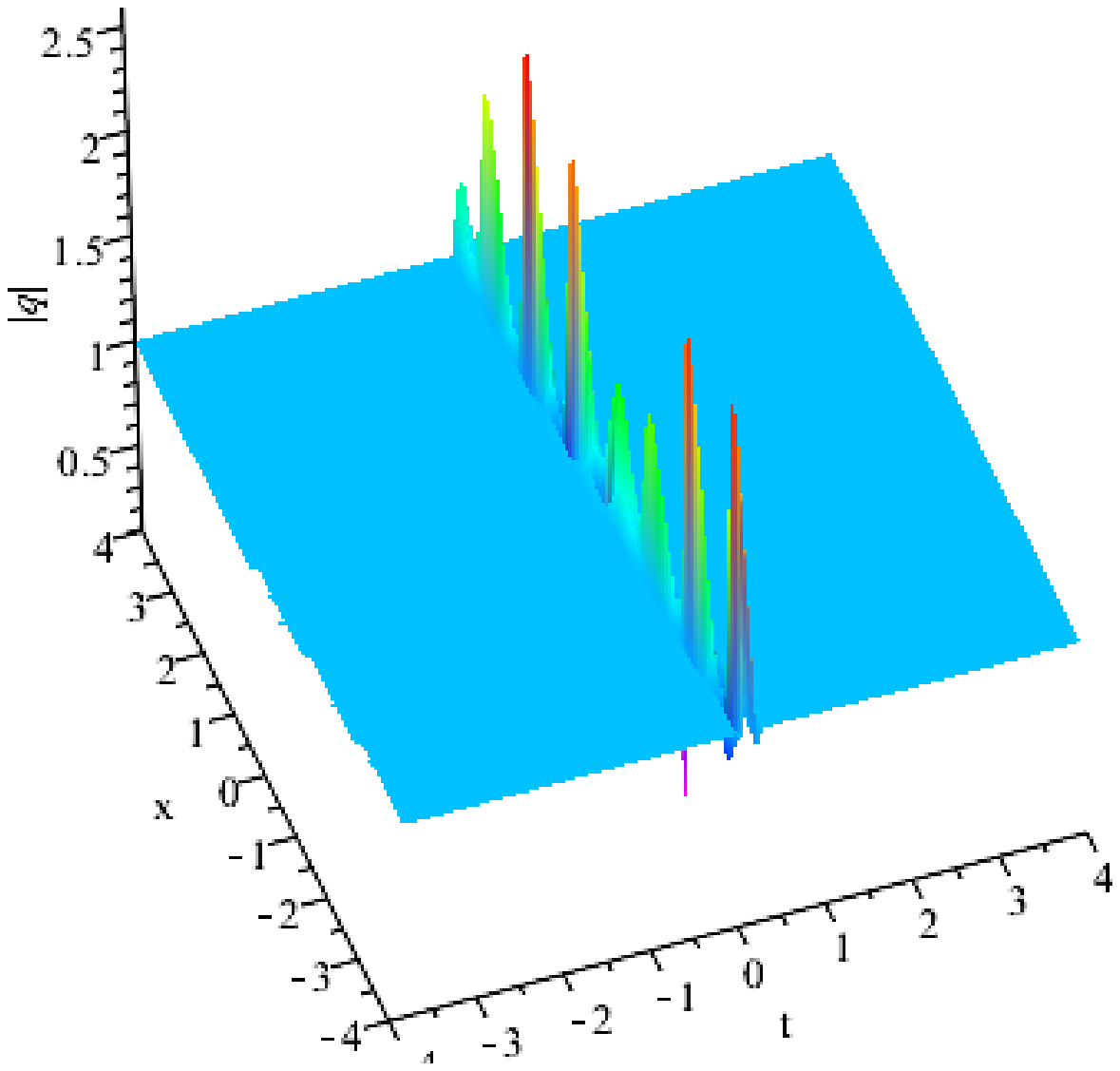}}}
\quad\quad{\rotatebox{0}{\includegraphics[width=3.3cm,height=3.3cm,angle=0]{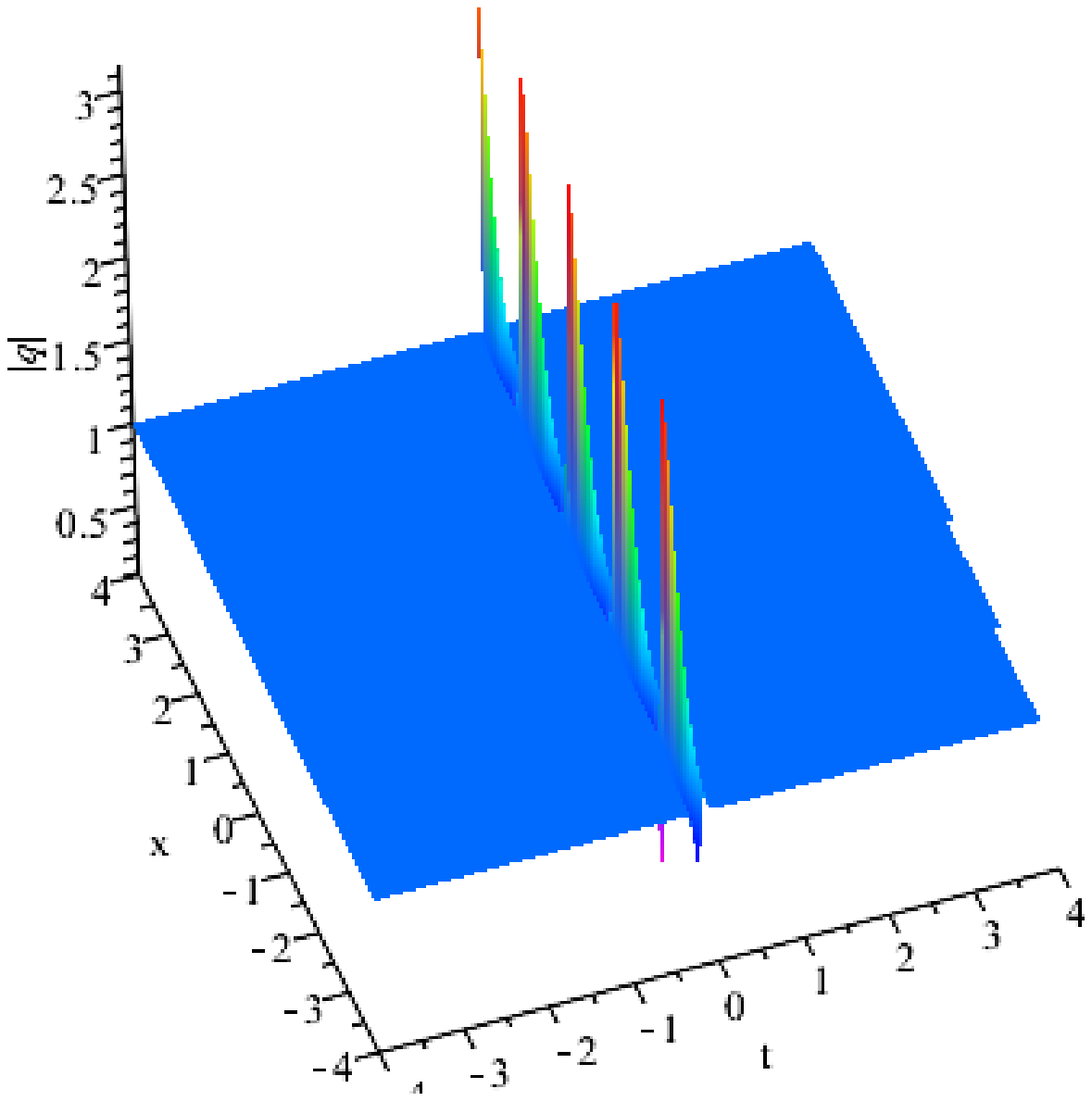}}}

$\qquad\qquad~~(\textbf{a})\quad \ \qquad\qquad\qquad\quad\quad~~(\textbf{b})~~\qquad\qquad\qquad\qquad\qquad(\textbf{c})$

{\rotatebox{0}{\includegraphics[width=3.3cm,height=3.3cm,angle=0]{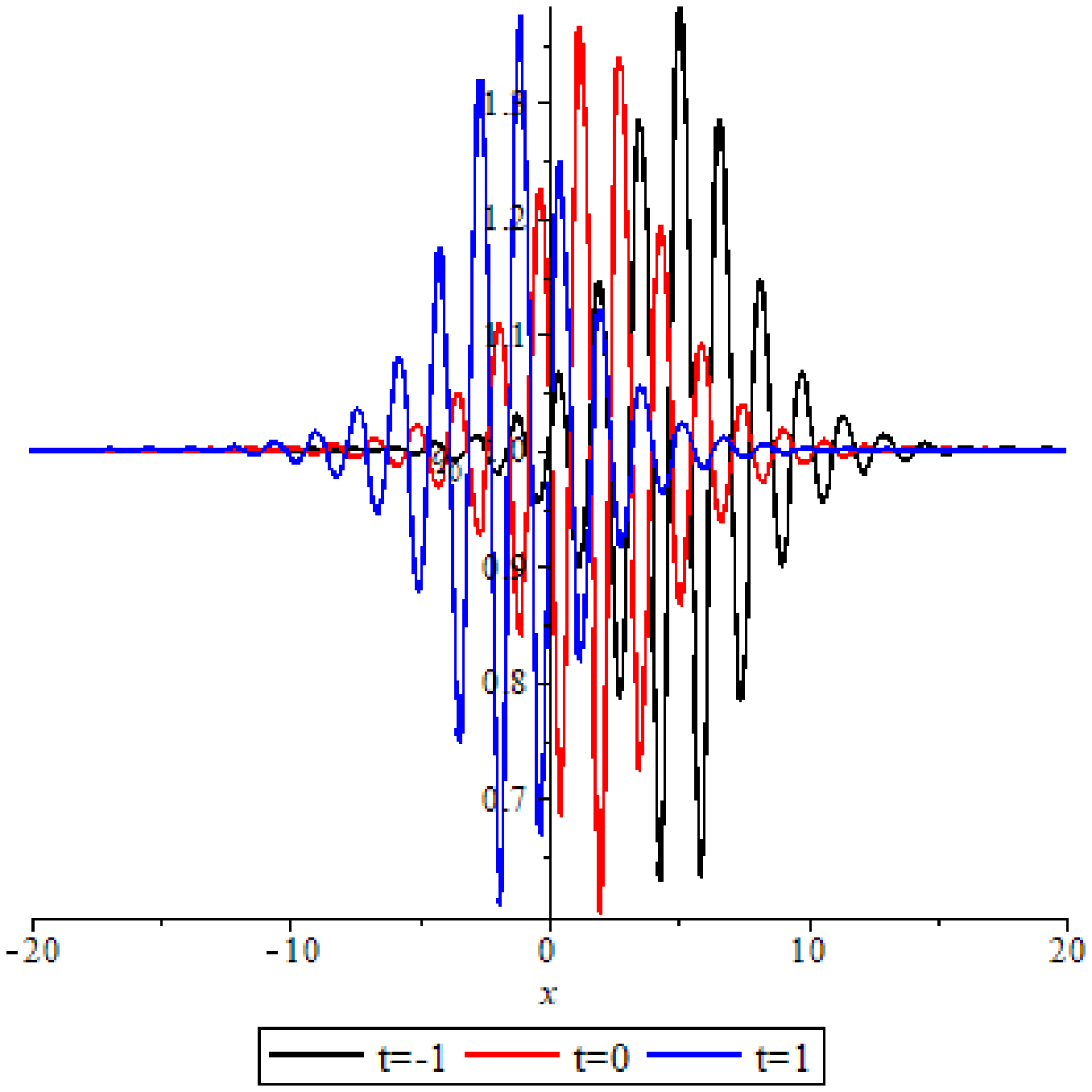}}}
\quad{\rotatebox{0}{\includegraphics[width=3.3cm,height=3.3cm,angle=0]{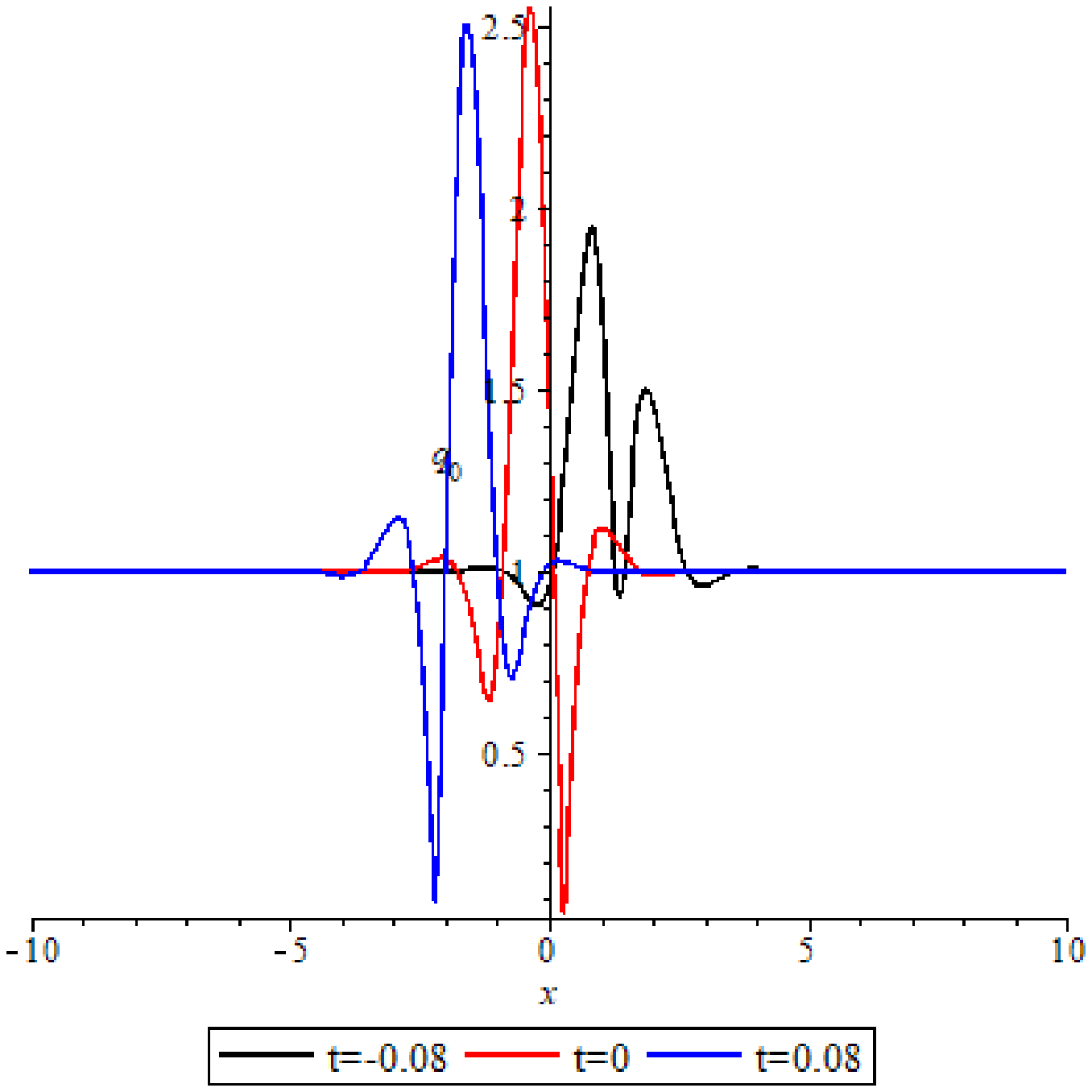}}}
\quad\quad{\rotatebox{0}{\includegraphics[width=3.3cm,height=3.3cm,angle=0]{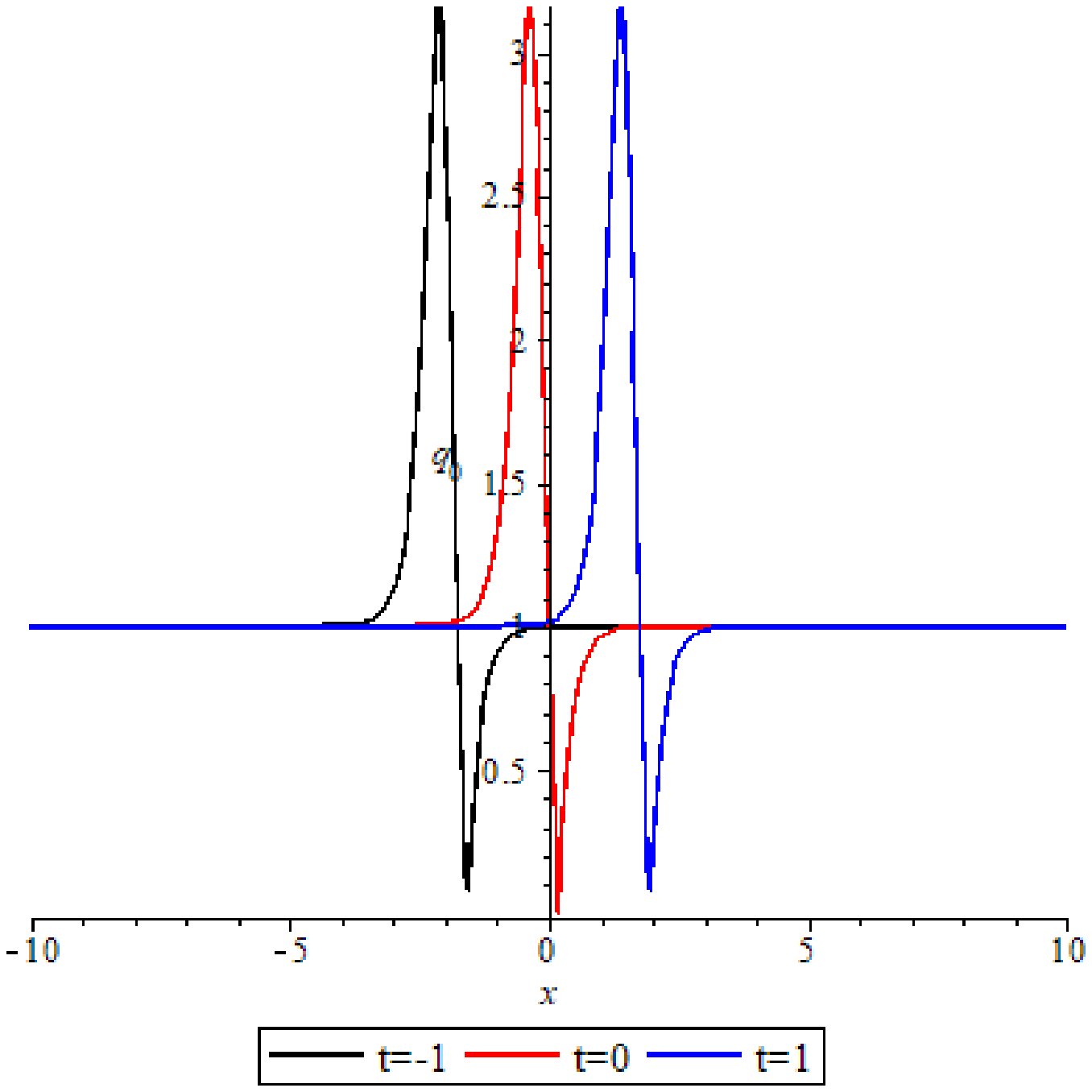}}}

$\qquad\qquad~~(\textbf{d})\quad \ \qquad\qquad\qquad\quad~\quad(\textbf{e})~~~~~~\quad\qquad\qquad\qquad\qquad(\textbf{f})$ 

\noindent { \small \textbf{Figure 6.} $\textbf{(a)(b)(c)}$ The solution \eqref{e} of the equation \eqref{QQ1}. $\textbf{(d)(e)(f)}$ display the propagation behavior of
solutions at different times. $\rho=\frac{1}{2}$, $\delta_{1}=\frac{\pi}{24}$, $\delta_{2}=\frac{5\pi}{24}$, $\delta_{3}=\frac{\pi}{2}$.}

It can be seen from Fig. 4 and Fig. 6 that the potential function $q$ propagates periodically along the space axis, but is aperiodic in the time axis. From a visual point of view, it is like generating several parallel breathers under the background of a cycle. As $\delta$ increases, the slope of the breather's propagation direction in the x-t plane increases. When $\delta=\frac{\pi}{2}$, the slope reaches its maximum value. When the value of $\delta$ continues to increase, the slope gradually decreases. Compared with Fig. 6, no matter what value $\delta$ takes, the slope is always the same.  As shown in Fig. 5, the shape (peak shape and number) of multimodal solitons also changes during propagation. From the figure, we can find that with the increasing value of $\delta$, the amplitude of the wave becomes smaller and smaller, but the shape remains unchanged.

\subsection{Single soliton solutions for $N_{3}=1$}
In this subsection, we assume that the eigenvalues $\tau_{n}$ is the third kind eigenvalues, which implies that $N_{3}=1$ and $N_{1}=N_{2}=0$. Let $\mathbf{q}_{+}=(1,~~1)^{T}$, $G_{1}=e^{\alpha+i\beta}$, $z=\varrho e^{i\delta},(0<\delta<\pi)$. The soliton solution of the cmKdV equation can be derived
\begin{align}
q(x,t)=\frac{\det\left(
                   \begin{array}{ccc}
                     q_{+} & y_{1}  \\
                     b_{1} & 1-F_{11}  \\
                   \end{array}
                 \right)
}{1-F_{11}},
\end{align}
where
\begin{align*}
&y_{1}=-iG_{1}e^{i(\theta_{1}+\theta_{2})(\omega_{1})},~~b_{1}=
\frac{v_{+}}{q_{0}}-\frac{iu_{+}}{q_{0}}\Delta_{1}^{(2)}(\omega_{1}),
F_{11}=\Delta_{1}^{(2)}(\omega_{1})\Delta_{1}^{(10)}(\omega_{1}^{*}),\\
&\Delta_{1}^{(2)}(\omega_{1})=\frac{\hat{G}_{1}e^{-i(\theta_{1}+\theta_{2})(\omega_{1}^{*})}}
{\omega_{1}-\omega_{1}^{*}}+\frac{\omega_{1}^{*}}{iq_{0}}
\frac{\check{G}_{1}e^{i(\theta_{1}-\theta_{2})(-\frac{q_{0}^{2}}{\omega_{1}^{*}})}}
{\omega_{1}-(-\frac{q_{0}^{2}}{\omega_{1}^{*}})},\\
&\Delta_{1}^{(10)}(\omega_{1}^{*})=\frac{G_{1}e^{i(\theta_{1}+\theta_{2})(\omega_{1})}}
{\omega_{1}^{*}-\omega_{1}}.
\end{align*}

%
%
%
%
\section{Conclusion}
In this work, the RH method is used to study the cmKdV equation  with $3\times3$ spectrum problem under NZBCs and various soliton solutions have been obtained, such as bounded solutions, breathing solutions and singular solutions. Firstly, in order to avoid the existence of multi-valued functions, a suitable Riemann surface has been introduced, then a new  independent variable is introduced to avoid the discussion on Riemann surface. Instead, the initial spectral parameter $k$ is transformed into a new spectral parameter $z$.

It is worth noting that compared with the nonzero boundary value problem of the $2\times2$ spectral problem, the following difficulties will be encountered in the discussion of the $3\times3$ spectral problem. Firstly, the second column of eigenfunctions is not analytic in the given regions; secondly, exponential oscillation term appears in the constructed RH problem; the third is that four jumping conditions will be generated when constructing the appropriate RH problem. Accordingly, when dealing with these problems, firstly we have introduced the auxiliary eigenfunctions to  solve the defects of analytic properties in spectrum analysis. In order to eliminate the oscillation term in the RH problem, we have further introduced the modified auxiliary eigenfunctions, which are very important in the construction of the RH problem. Different from the $2\times2$ non-zero boundary value problem, another key point is the distribution of eigenvalues. Here, in order to characterize the discrete spectrum, we introduce four analytic $3\times3$ matrices in a given region, and then produce three different types of eigenvalues.

 In particular, we have shown that  for the three symmetries of eigenfunction, scattering coefficient and auxiliary eigenfunction, this is the key to the characterization of discrete spectrum. We also point out that the discrete spectrum produces three types of discrete eigenvalues according to the symmetry of scattering data, and each eigenvalue corresponds to different types of soliton solutions. Furthermore, the dynamic propagation behaviors of soliton solutions have been obtained.
\section*{Acknowledgments}
This work was supported by the National Natural Science Foundation of China under Grant No. 11975306, the Natural Science Foundation of Jiangsu Province under Grant No. BK20181351, the Six Talent Peaks Project in Jiangsu Province under Grant No. JY-059, and the Fundamental Research Fund for the Central Universities under the Grant Nos. 2019ZDPY07 and 2019QNA35.

\end{document}